\documentclass{emulateapj}
\usepackage{colordvi}
\usepackage{epsfig}
\usepackage{graphicx}
\usepackage{amssymb}
\usepackage{lscape}
\newcommand{\mic}{~$\mu$m}

\newcommand{\spitzer}{{\it Spitzer}}
\newcommand{\HST}{{\it HST}}

\begin{document}

\title{UV-to-FIR analysis of {\it Spitzer}/IRAC sources in the Extended Groth Strip I: 
Multi-wavelength photometry and spectral energy distributions}
\author{Guillermo Barro\altaffilmark{1}, P.G. P\'erez-Gonz\'alez\altaffilmark{1,2}, J. Gallego\altaffilmark{1}, M. L. N. Ashby\altaffilmark{3}, M. Kajisawa\altaffilmark{4}, S. Miyazaki\altaffilmark{5}, V. Villar\altaffilmark{1}, T. Yamada\altaffilmark{4}, J. Zamorano\altaffilmark{1}}
\altaffiltext{1}{Departamento de Astrof\'{\i}sica, Facultad de CC. F\'{\i}sicas,
Universidad Complutense de Madrid, E-28040 Madrid, Spain}
\altaffiltext{2}{Associate Astronomer at Steward Observatory, The University of Arizona}
\altaffiltext{3}{Harvard-Smithsonian Center for Astrophysics, 60 Garden St., Cambridge, MA 02138}
\altaffiltext{4}{Astronomical Institute, Tohoku University, Aramaki, Aoba, Sendai 980–8578, Japan}
\altaffiltext{5}{National Astronomical Observatory of Japan, Mitaka, Tokyo 181-8588, Japan}
\slugcomment{Last edited: \today}
\date{Submitted: \today}
\label{firstpage}
\begin{abstract}
  We present an IRAC 3.6+4.5~$\mu$m selected catalog in the Extended
  Groth Strip (EGS) containing photometry from the ultraviolet to the
  far-infrared and stellar parameters derived from the analysis of the
  multi-wavelength data. In this paper, we describe the method used to
  build coherent spectral energy distributions (SEDs) for all the
  sources. In a companion paper, we analyze those SEDs to obtain
  robust estimations of stellar parameters such as photometric
  redshifts, stellar masses, and star formation rates. The catalog
  comprises 76,936 sources with [3.6]$\leq$23.75~mag (85\%
  completeness level of the IRAC survey in the EGS) over
  0.48~deg$^{2}$. For approximately 16$\%$ of this sample, we are able
  to deconvolve the IRAC data to obtain robust fluxes for the multiple
  counterparts found in ground-based optical images. Typically, the
  SEDs of the IRAC sources in our catalog count with more than 15
  photometric data points, spanning from the ultraviolet wavelengths
  probed by GALEX to the far-infrared observed by {\it Spitzer}, and
  going through ground- and space-based optical and near-infrared data
  taken with 2-8 meter class telescopes.  Approximately 95\% and 90\%
  of all IRAC sources are detected in the deepest optical and
  near-infrared bands. These fractions are reduced to 85\% and 70\%
  for SNR$>$5 detections in each band. Only 10\% of the sources in the
  catalog have optical spectroscopy and redshift estimations. Almost
  20\% and 2\% of the sources are detected by MIPS at 24 and
  70~$\mu$m, respectively. We also cross-correlate our catalog with
  public X-ray and radio catalogs.  Finally, we present the {\it
    Rainbow Navigator} public web-interface utility, designed to
  browse all the data products resulting from this work, including
  images, spectra, photometry, and stellar parameters.
\end{abstract}
\keywords{
galaxies: starburst --- galaxies: photometry --- galaxies: high-redshift --- infrared: galaxies.}

\section{Introduction}\label{intro}

Multi-wavelength observations of blank fields provide a fertile ground
for studies of the evolution of galaxies from the early Universe.
Indeed, in the past decade we have advanced amazingly in our knowledge
about the formation of galaxies thanks to deep field imaging and
spectroscopic surveys. The extraordinary success of these surveys is
sustained by the coordinated effort of several telescope facilities,
institutions, and research groups that gather large collections of
multi-wavelength photometry and spectroscopy, providing the entire
scientific community with a vast pool of data to analyze. Remarkable
examples of this kind of projects are the Hubble Deep Field (HDF)
observations \citep{1996AJ....112.1335W}, the Classifying Objects by
Medium- Band Observations project
\citep[][COMBO17]{2001A&A...377..442W}, Great Observatories Optical
Deep Survey \citep[][GOODS]{2004ApJ...600L..93G}, VIMOS-VLT Deep
Survey \citep[][VVDS]{2005A&A...439..845L}, All-wavelength Extended
Groth strip International Survey \citep[][AEGIS]{2007ApJ...660L...1D},
or Cosmic Evolution Survey \citep[][COSMOS]{2007ApJS..172....1S}.

Nevertheless, the full scientific exploitation of these surveys
unavoidably needs a consistent merging of the data coming from
heterogeneous sources (with different depths and resolutions) to build
catalogs of galaxies characterized with panchromatic photometry and
spectroscopy. Although a substantial effort has been devoted to create
homogeneously processed multi-wavelength catalogs for the most
important fields (e.g., Chandra Deep Field South,
\citealt{2001A&A...377..442W}; Hubble Deep Field North,
\citealt{2003ApJ...592..728S}, \citealt{2004AJ....127..180C}; EGS,
\citealt{2004ApJ...617..765C}, \citealt{2006A&A...457..841I}), many of
these catalogs are selected in the optical bands (i.e., rest-frame
ultraviolet -UV- at high redshift), and lack near-infrared (NIR)
imaging (an important dataset for the studies of galaxy populations at
high redshift) reaching depths that match the optical observations.
Fortunately, the proliferation of deep and wide NIR surveys has
supported the publication of an increasing number of multi-band
samples selected in the $K$-band (\citealt{2006A&A...449..951G},
\citealt{2007AJ....134.1103Q}; \citealt{2008ApJ...682..985W}) or
IRAC-bands \citep[][hereafter, PG05 and PG08]{2008MNRAS.386..697R,
  2005ApJ...630...82P, 2008ApJ...675..234P}. A detailed UV-to-NIR
coverage of the spectral energy distribution (SED) improves the
estimates of important stellar parameters, such as the mass, age,
extinction, or star formation rate (see, e.g.,
\citealt{2008A&A...491..713W}; \citealt{2007ApJS..173..267S}).
Moreover, in order to obtain the most reliable stellar mass estimates
for z$\sim$2-3 galaxies, and to distinguish young dusty starburst from
quiescent galaxies at these redshifts, we need to obtain data probing
the rest-frame NIR for these populations
(\citealt{2006ApJ...651..120B}; \citealt{2007A&A...474..443P};
\citealt{2007ApJ...655...51W}; \citealt{2009ApJ...691.1879W};
\citealt{2009ApJ...701.1839M}). This became possible with the launch
of {\it Spitzer} and the use of one of its instruments, IRAC
\citep{2004ApJS..154...39F}, which covers the 3.6-to-8.0\mic\,
spectral range.

The power of multi-band catalogs is significantly enhanced when
optical and NIR data are complemented with mid-IR ($>$5$\mu$m) to
radio fluxes, such as those from {\it Spitzer}/MIPS, SCUBA, VLA or
{\it Herschel} surveys. These data directly probe the emission of the
dust component of galaxies (\citealt{2009A&A...504..751S};
\citealt{2008ApJ...682..985W}). Actively star-forming galaxies harbor
large amounts of dust, which cause that a fraction of the UV emission,
directly related to the ongoing star formation, is extincted and
re-emitted in the IR. Thus, modeling the IR emission offers not only
a complementary approach to estimate the star formation rate (SFR) of
a galaxy but also an improved measurement of the intrinsic UV
extinction (\citealt{2006ApJ...653.1004R};
\citealt{2007ApJ...670..156D}; \citealt{2007ApJS..173..267S};
\citealt{2007ApJ...670..279I}). The IR approach to studies of the
star formation becomes particularly relevant at higher redshifts,
where the number of luminous infrared galaxies (LIRGs, whose
integrated IR luminosity is L(IR)$>$10$^{11}$L$_{\odot}$) and their
contribution to the cosmic SFR density increase significantly
(Chary\&Elbaz 2001, PG05, \citealt{2007ApJ...660...97C}).

The downside to having an exceptional data pool available in many
regions of the sky is that the data quality is largely heterogeneous.
Unfortunately, the high-redshift community still lacks the existence
of a unified database that facilitates the access to the multiple
datasets and resources, similar to local extragalactic databases such
as the NASA Extragalactic Database (NED) or the Sloan Digitalized Sky
Survey (SDSS; \citealt{2000AJ....120.1579Y}) database.

The purpose of this work is to present a NIR selected sample of
galaxies with well sampled spectral energy distributions, and analyze
their properties maximally benefiting from the panchromatic data. To
do this, we have built an IRAC-3.6+4.5$\mu$m selected photometric and
spectroscopic catalog including data from X-ray to radio wavelengths
for 76,936 galaxies at 0$<$z$\lesssim$4 in the Extended Groth
Strip (EGS).

The EGS has been intensively observed as a part of the All-Wavelength
Extended Groth Strip International Survey (AEGIS;
\citealt{2007ApJ...660L...1D}) collaboration in order to assemble an
exceptional multi-wavelength dataset, including deep optical imaging
from the CFHTLS, {\it HST} coverage in two bands, UV data from GALEX,
and mid-IR and far-IR photometry from {\it Spitzer}. In addition, the
EGS is the key field for the DEEP2 survey, one the largest and deepest
spectroscopic surveys to date \citep{2003SPIE.4834..161D}, with more
than 10,000 optical spectra down to R$\sim$24. This vast dataset
converts the EGS in one of the main fields for the study of galaxy
evolution at different epochs of the lifetime of the Universe. Our
goal is get advantage of this impressive multi-wavelength data
collection and use it to build UV-to-FIR SEDs, whose analysis will
allow us to obtain estimations of interesting parameters, such as the
photometric redshifts, stellar masses, and SFRs.

The photometric catalog along with the photometric redshifts and the
inferred stellar parameters are intended to become a multi-purpose
resource useful for many different scientific goals. Some of them will
be presented in forthcoming papers. We make all the catalogs publicly
available through our website and through a dedicated a web-interface,
dubbed {\it Rainbow Navigator}, conceived to facilitate the access to
the data, but also to serve as a permanent repository for updates in
these catalogs, or similar catalogs in other cosmological fields
(e.g., those presented in PG08 for the GOODS fields).

In this paper, we concentrate on the description of the dataset and
the methods developed to measure the merged photometry for the IRAC
sample in the EGS. We also analyze the multi-band properties of the
sample to understand the main properties of the {\it Spitzer} surveys.
All the multi-wavelength photometry is released in a public database,
conceived to allow the astronomical community to access all the
results from our work and use them for their own purposes. In a
companion paper (hereafter Paper II), we will present our methodology
to fit the SEDs presented in this paper, and to estimate photometric
redshifts, stellar masses, and SFRs out of them. We will also assess
the quality of the inferred parameters, analyzing in detail their
intrinsic systematic and random uncertainties.

The outline of this paper follows. In $\S$~\ref{alldata} we present
the available datasets that we have compiled for this paper. In
$\S$~\ref{mergedcat} we present the techniques developed to extract
the IRAC 3.6+4.5~$\mu$m selected sample and we discuss the properties
of the IRAC photometry of the sample. In $\S$~\ref{merged_technique}
we present the methods developed to build the merged multi-band
photometric catalog. In $\S$~\ref{merged_properties} we describe in
detail the photometric properties and reliability of the catalog. In
$\S$~\ref{dataacess} we describe the format of the published catalogs,
the database built to allow an easy access and handling of those,
baptized as the {\it Rainbow} Cosmological Surveys Database, and the
publicly available web-interface to surf the database, {\it Rainbow
  Navigator}.

Throughout this paper we use AB magnitudes. We adopt the cosmology
$H_{0}=70$ km$^{-1}$s$^{-1}$Mpc$^{-1}$, $\Omega_{m}=0.3$ and
$\Omega_{\lambda}=0.7$.

\section{Data description}\label{alldata}

The Extended Groth Strip (EGS; $\alpha$~=~$14^{h}17^{m}$,
$\delta$~=~+52$^{\circ}$30') is one of the most targeted cosmological
deep fields. Noticeably, a comprehensive panchromatic dataset has been
compiled in this field within the AEGIS collaboration
\citep{2007ApJ...660L...1D}.

The sample of galaxies studied in this paper is based on an
IRAC-3.6+4.5$\mu m$ selection. This choice obeys to several reasons.
First, the IRAC bands are specially tailored to probe the rest-frame
near-infrared (NIR) fluxes of distant galaxies, thus being the perfect
tool for studies of massive galaxies at high-z (e.g.,
\citealt{2007A&A...470...21R}; \citealt{2009A&A...500..705M}).
Second, the quality, depth and ubiquity of the IRAC observations in
the so-called cosmological fields favors the assembly of coherent flux
limited catalogs over large cosmological volumes (e.g., PG08), which
is the cornerstone of the observational cosmology.  The IRAC bands
offer an alternative to the less efficient NIR ground-based surveys,
and provide a starting point to consistently anchor surveys at longer
wavelengths ({\it Spitzer}/MIPS, {\it Herschel}, ALMA).  Despite its
lower spatial resolution (FWHM$\sim$2'';
\citealt{2004ApJS..154...39F}) compared to optical/NIR ground-based
surveys, the image quality of IRAC is very stable and several authors
have been able to perform deblending techniques successfully merging
IRAC catalogs into their panchromatic datasets (PG05,
\citealt{2006A&A...449..951G}, PG08, \citealt{2009ApJ...690.1236I}).

\begin{figure*}
\centering
\includegraphics[width=18cm,angle=0.]{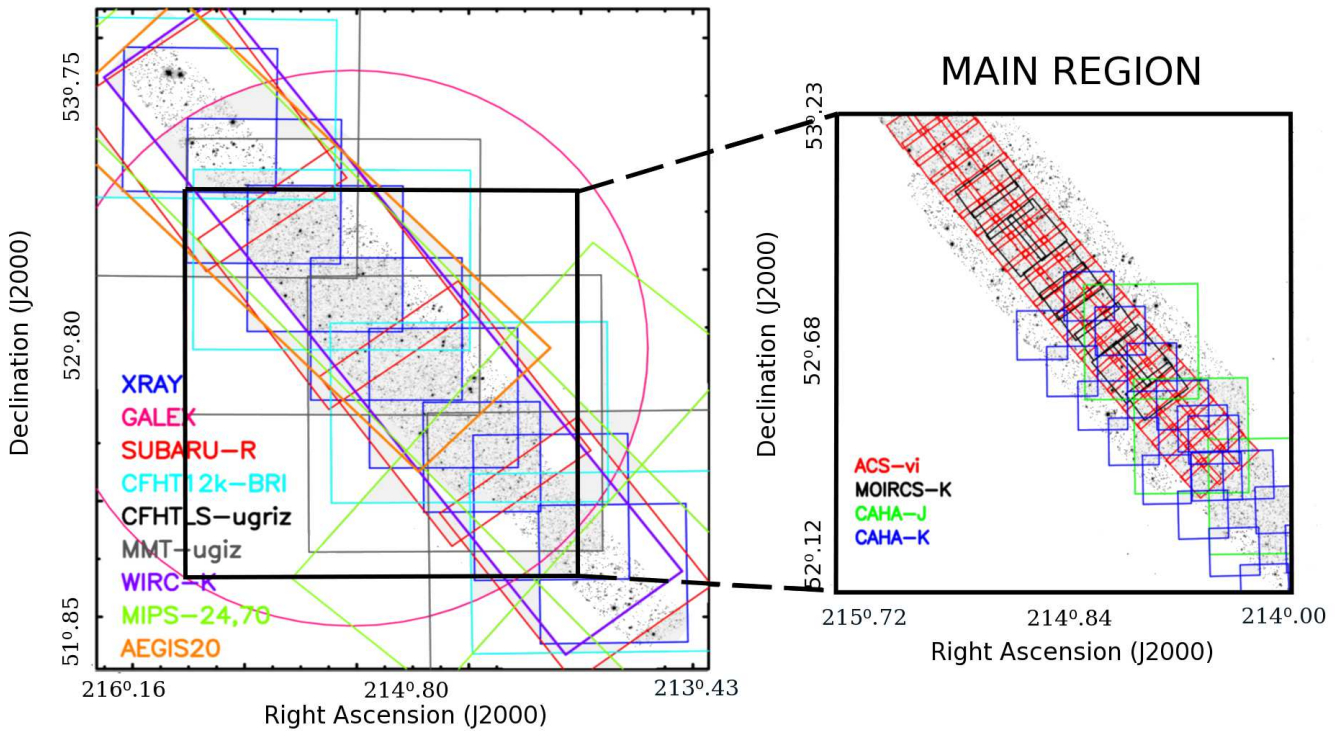}
\caption{\label{layout} Layout of all the imaging surveys covering the
  EGS.  The background grey-scale image shows the IRAC 3.6~$\mu$m
  mosaic. The black square depicts the CFHTLS pointing.  The
  intersection of both footprints defines the region of higher
  photometric coverage (magnified on the right panel), which is
  referred as ``main region'' throughout the text (see
  \S\ref{merged_properties}) and contains the highest resolution HST
  data.  The cyan and red rectangles show the CFHT12k-$BRI$ and SUBARU
  $R$-band coverage, respectively.  The gray and purple squares show
  the MMT-$u'giz$ and {\it Chandra}/ACIS pointings, respectively.  The
  magenta circle shows the GALEX survey footprint. The green
  rectangles show the {\it Spitzer}/MIPS coverage. The orange
  rectangle shows VLA-20cm coverage. In the main region, magnified on
  the right side of the figure, the red tiles depict the \HST/ACS
  footprints (the NICMOS images overlap with this area).  Black
  rectangles show the MOIRCS $K_{s}$-band imaging, and the green and
  purple lines delimit the CAHA $J$- and $K_{s}$-band surveys.}
\end{figure*}

In the rest of this section, we describe the multi-wavelength datasets
in EGS compiled for this paper to characterize the IRAC selected
sample.  Table~\ref{multiwav_data} summarizes the main characteristics
of these datasets, including the depth, the area and the image
quality.  Figure \ref{layout} shows the footprints of the various
surveys, highlighting the area with higher band coverage.

\subsection{X-ray data}

The EGS region has been observed in the X-ray (0.5-10 keV) using {\it
  Chandra/ACIS} during two observation cycles in 2002 and 2005. The
AEGIS-X survey covers an area of 0.67~deg$^{2}$ in 8 pointings
($\sim$17'x17' each), completely overlapping with the region covered
by IRAC. The nominal exposure time of the frames is 200~ks per pixel,
reaching limiting fluxes of 5.3$\times$10$^{-17}$ and
3.8$\times$10$^{-16}$ erg~cm$^{2}$~s${-1}$ in the soft (0.5-2 keV) and
hard (2-10 keV) bands, respectively. In this work, we use the data
reduction and point source catalogs published in \citet[][see also
\citealt{2005MNRAS.356..568N} for a first version of the
catalogs]{2009ApJS..180..102L}. The two-band (soft and hard) merged
catalog comprises 1325 sources with $<$1.5\% spurious detections.  The
authors identified optical and NIR counterparts for 1013 and 830
sources, respectively, from the CFHT legacy survey optical catalog
(see \S~\ref{CFHT}), and the {\it Spitzer}/IRAC catalog of
\citet{2008ApJS..177..431B}.  The cross-match is based on maximum
likelihood method with a search radius of 2\arcsec\ (more than a
factor of 1.5 the {\it rms} of their astrometric accuracy).

\subsection{Ultraviolet data}

The \textit{Galaxy Evolution Explorer} (GALEX,
\citealt{2005ApJ...619L...1M}) observed the EGS over three consecutive
years providing deep UV data in two channels (FUV at 153~nm, and NUV
at 231~nm) on a 1.13~deg$^{2}$ circular area around
$\alpha$~=~$14^{h}20^{m}$, $\delta$~=~+52$^{\circ}$47'. The total
exposure time for the composite stacks are 58~ks and 120~ks in the FUV
and NUV filters, respectively. The approximate limiting magnitude in
both bands is $\sim$25.1~mag.

\subsection{Optical data}

\subsubsection{CFHTLS/CFHT12K}
\label{CFHT}

A 1$^{\circ}\times1^{\circ}$ square region centered on the coordinates
$\alpha$~=~$14^{h}19^{m}27^{s}$, $\delta$~=~+52$^{\circ}$40'56'' was
observed within the Canada-France Hawaii Telescope (CFHT) Legacy
Survey (CFHTLS; sector D3). Using the MEGACAM camera on the CFHT,
imaging data were obtained in five broadband filters over the
wavelength range 350~nm$<$$\lambda$$<$940~nm:
$u^{*},~g',~r',~i',~z'$. The overall exposure times ranges from 4h to
40h (for the $u^{*}$ and $z'$ bands, respectively), reaching limiting
magnitudes between 26-27~mag. The reduced images have gone through
several releases. The data used in this paper are part of the CFHTLS
'T0004' release produced at the TERAPIX data center (Gwyn et al. 2010,
in preparation\footnotemark[2]).  Although the total mosaicked area
covers 1~deg$^{2}$, the overlap with the IRAC observations is just
0.35~deg$^{2}$.
\footnotetext[2]{http://www.astro.uvic.ca/grads/gwyn/cfhtls/index.html}

In addition to the CFHTLS data, shallower images were acquired in the
$B$, $R$ and $I$ bands over a larger area, using the wide-field 12K
mosaic camera on CFHT. These observations were intended to provide
optical coverage for the DEEP2 spectroscopic survey, extending through
the whole EGS in 4 different pointings. The observing times were 3-6h,
to achieve a limiting magnitude of $\sim$25~mag. The photometric
catalogs were published within the DEEP2 DR1
\citep{2004ApJ...617..765C}, and the raw (non-reduced) images can be
retrieved from the CADC archive\footnotemark[3]. We downloaded the raw
and calibrations files from the archive to perform our own data
reduction and cataloging. This reduction was carried out with the
IRAF\footnotemark[4] task \textit{mscred}. The photometric calibration
was performed using the $BRI$ public catalogs. The average dispersion
of the photometry comparison is smaller than 0.03~mag.
\footnotetext[3]{http://www4.cadc-ccda.hia-iha.nrc-cnrc.gc.ca/cadcbin/cfht/wdbi.cgi/cfht/quick/form}
\footnotetext[4]{http://iraf.noao.edu/}

\subsubsection{MMT/Megacam}

The EGS was observed with the Megacam camera
\citep{2006sda..conf..337M} on the MMT in the $u^\prime g i z$ bands
during several campaigns from 2006 to 2009.  Four pointings of the
24\arcmin$\times$24\arcmin Megacam field of view were arranged to
cover the full extension of the 2 degrees long IRAC mosaic.

The MMT/Megacam imaging data were reduced with a combination of
standard IRAF/{\tt mscred} routines and custom software.  Final
mosaics were calibrated against the coextensive SDSS photometry with
appropriate color corrections.  The 5$\sigma$ limiting magnitudes
within a 2\arcsec\ diameter aperture for point sources varies among
the four different pointings but are approximately 26.5, 27.2, 26.0,
and 26.0~mag in the $u^\prime, g, i$, and $z$
bands, respectively.  The effective seeing in the final mosaics is
roughly 1\arcsec\ FWHM or slightly better in some bands.

\subsubsection{\HST/ACS}

As a part of a GO program (PI: Davis), the \HST\, Advanced Camera for
Surveys (ACS) acquired deep imaging of EGS in two optical bands:
$F606W$ and $F814W$ (hereafter the $V_{606}$ and $i_{814}$ images).
The ACS survey covers an area of $\sim$710.9~arcmin$^{2}$
(10.1$\arcmin$$\times$70.5$\arcmin$) on 63 contiguous tiles following
the direction of the IRAC mosaic, covering approximately 50$\%$ of the
total area surveyed by IRAC, and $\sim80\%$ of the area overlapping
with the CFHTLS observations. The fully reduced, drizzled frames and
calibration products were released by the AEGIS Team
\citep{2006astro.ph..2088L}. Each science image has an approximate
exposure time of $\sim$0.6~h per pixel and a limiting magnitude of
$\sim$28~mag.

\subsubsection{Subaru Suprime-Cam}

Ground based $R$-band imaging of the EGS was carried out with the
Subaru Telescope as a part of the Subaru Suprime-Cam Weak-Lensing
Survey \citep{2007ApJ...669..714M}. The Suprime-Cam field of view is
0.25~deg$^{2}$ in size, covering the whole IRAC map in four pointings.
The total exposure time was 30 minutes for each pointing, taken in
four 7.5 minute exposures, in a dithering pattern with $\sim$1\arcmin\
spacing. The approximate limiting magnitude is $R$$=$26~mag. These
data were downloaded from the SMOKA database and reduced with the
Suprime-Cam pipeline (SDFRED, v1.0).

\subsection{Near Infrared data}
\subsubsection{\HST/NICMOS}

Simultaneously to the ACS observations, the Near-Infrared Camera and
Multi-Object Spectrometer (NICMOS) covered parallel fields in the
$F110W$ and $F160W$ bands (hereafter $J_{110}$ and $H_{160}$ bands,
respectively), with a similar exposure time and a limiting magnitude
of 23.5 and 24.2, respectively.  The observations were designed to
maximize the overlap between both \HST\, surveys. Virtually all (58
out of the 63) NICMOS frames lie within the area covered by the ACS
primary imaging. However, the smaller field of view of NICMOS
($\sim$1'$\times$1') leads to a total NIR coverage of only
0.0128~deg$^{2}$, with a 90$\%$ overlap with the area covered by ACS
but less than a $4\%$ with that covered by IRAC.

\subsubsection{Subaru MOIRCS}

In addition to the optical imaging, NIR observations of the EGS were
also obtained with the Multi-Object InfraRed Camera and Spectrograph
(MOIRCS) on 5 nights during April - May 2006 (PIs: Fukugita, Yamada)
and a complementary run on Jun 25, 2007. The dataset comprises 11
pointings covering a total of 0.09~deg$^{2}$ oriented along the
original strip within $52.5^{\circ}<$$\delta$$<53.0^{\circ}$ and
completely overlapping with the \HST-ACS and CFHTLS imaging. The
median exposure time per frame is $\sim$1h for an approximate limiting
magnitude of $K_{s}$$\sim$23-24~mag. The data was reduced using
dedicated scripts developed by the MOIRCS team involving the IRAF task
MSCRED, plus an additional de-fringing process (see e.g.,
\citealt{2009ApJ...702.1393K}).

\subsubsection{Palomar and Calar Alto imaging}

Given the importance of having a continuous band coverage for any kind
of study regarding galaxy populations \citep{2007ApJ...655...51W}, we
have incorporated in our data compilation the POWIR NIR catalog
\citep{2006ApJ...651..120B}. These data were acquired between
September 2002 and October 2005 using the WIRC camera in the Palomar
5m telescope. The total surveyed area in the EGS field is
2165~$\mathrm{arcmin}^{2}$ (0.6~$\deg^{2}$) in the $K$-band and
$\sim$1/3 of that area in the $J$-band. The approximate limiting
magnitudes are $K\sim22.9$ and $J\sim21.9$. Note that no images were
publicly available for this dataset, so we only use the catalogs.

In addition, we also make use of the NIR imaging obtained by the
Galaxy evolution and Young Assembly (GOYA\footnotemark[7]) project.
Two photometric campaigns were carried out to obtain NIR data of the
original Groth Strip $\alpha$=14$^{h}$17$^{m}$43$^{s}$, $\delta$=
52$^{\circ}$28'41'' \citep{2003ApJ...595...71C} and flanking fields
\citep{2009A&A...494...63B}. Here we make use of the $K$-band images
of the flanking fields, observed with the $\Omega'$ instrument in the
3.5m telescope at Calar Alto Spanish-German Astronomical Center
(CAHA). The frames cover a total area of $\sim$0.24~deg$^{2}$ to a
limiting magnitude of $K_{s}$$\sim$20.7.

Finally, we have also included data in the $J$-band from the narrow
band survey of H$\alpha$ emitters described in
\citet{2008ApJ...677..169V}. Three 15'x15' pointings centered at
$\alpha$=14$^{h}$17$^{m}$31$^{s}$, $\delta$=52$^{\circ}$28'11'',
$\alpha$=14$^{h}$17$^{m}$31$^{s}$, $\delta$=52$^{\circ}$28'11'' and
$\alpha$=14$^{h}$18$^{m}$14$^{s}$, $\delta$=52$^{\circ}$42'15'' were
observed in CAHA using the $\Omega2k$ instrument in the 3.5m telescope.
The combined pointings cover an area of 0.19~deg$^{2}$ to a limiting
magnitude of $J=22.9$.
\footnotetext[7]{http://www.astro.ufl.edu/GOYA/home.html}

\subsection{Mid-to-far IR data}
\subsubsection{{\it Spitzer}/IRAC}

Our sample is drawn from {\it Spitzer} near/mid-IR data obtained as
part of the Guaranteed Time Observations (GTO, PI: Fazio) and
presented in \citet{2008ApJS..177..431B}. We also included additional
data from the GO program with ID \#41023 (PI: K.Nandra). The GTO IRAC
imaging data at 3.6, 4.5, 5.8, and 8.0~$\mu m$ were obtained over two
epochs (December 2003 and June/July 2004). The dataset comprises 52
different pointings that cover a 2$^{\circ}\times$10' strip with
approximately the same depth.  To achieve this homogeneous coverage
and due to scheduling issues, the width of the mosaic is slightly
variable along the strip, ranging from 10' to 17'. The average
exposure time per pixel is 2.5~h (9100s) in the four channels.  An
area of 1440 arcmin$^{2}$ was observed for 1900s, 930 arcmin$^{2}$ for
9100s and $\sim100$arcmin$^{2}$ for $>$11500s
\citep{2008ApJS..177..431B}.  The additional GO data are located in
two strips of width $\sim$3.5\arcmin flanking the original strip and
covering the declination range
52.35$^{\circ}$$<$$\delta$$<$53.25$^{\circ}$. All the data were
reduced with the general {\it Spitzer pipeline}, which provides Basic
Calibrated Data, and then mosaicked with {\it Mopex} using a pixel
scale half of the original ($\sim$0.61 arcsec~pixel$^{-1}$). The
details of the image quality are discussed in
\S~\ref{merged_completeness}.

\subsubsection{{\it Spitzer}/MIPS 24~$\mu m$ and 70~$\mu m$}

Complementary to the IRAC observations, MIR and FIR observations were
also obtained with the Multiband Imaging Photometer for
\spitzer~(MIPS; \citealt{2004ApJS..154...25R}) as part of the GTO and
the Far-Infrared Deep Extragalactic Legacy Survey (FIDEL). For this
paper, we use the whole GTO+FIDEL dataset, reduced and mosaicked with
the {\it Spitzer} pipeline and MOPEX+GeRT software. The surveyed area
covers approximately the entire $2^{\circ}\times10'$ strip, being
slightly wider on the upper and lower edges. The MIPS mosaic overlaps
with the deepest part of the IRAC observations. The mean exposure time
at 24~$\mu m$ is $\sim$7200~s per pixel, while for the 70~$\mu m$
channel it is approximately 3800~s. The approximate limiting fluxes
are 60~$\mu$Jy and 3.5~mJy, respectively.

\subsection{Radio data}

A radio survey at 1.4~GHz (20cm) of the Northern half of the EGS
($\sim$50\% of the IRAC mosaic) was conducted with the Very Large
Array (VLA) in its B configuration during 2003-2005. The AEGIS20
survey covers 0.73~deg$^{2}$ down to 130~$\mu$Jy~beam$^{-1}$ including
a smaller region of 0.04~deg$^{2}$ with a 50~$\mu$Jy detection limit
(5$\sigma$). The data reduction and the source catalog, comprising
1123 sources, were presented in \citet{2007ApJ...660L..77I}.

\subsection{Keck Optical spectra}

EGS has also been the target of an exhaustive and unique spectroscopic
follow up. As one of the DEEP2 fields, optical multi-object
spectroscopy has been carried out from 2003 to 2005 with the Deep
Imaging Multi-Object Spectrograph (DEIMOS;
\citealt{2003SPIE.4841.1657F}) on the Keck II telescope. The
observations cover the spectral range 640$<$$\lambda$$<$910~nm with a
resolution of 0.14~nm. The DEEP2 DR3\footnotemark[6] contains
spectroscopic redshifts for 13,867 sources at 0$<$z$<$1.4 with a
median redshift z$=$0.75. The targets were selected from the CFHT12k
BRI images (within 1.31~deg$^{2}$) in the magnitude range
18.5$\leq$$R$$\leq$24.1. The spectroscopic redshifts for around 70\%
of the sources are labeled with a high quality flag (values of 3 and
4, meaning $>$95\% success rate). Lower quality flags are considered
unreliable and will be excluded from our analysis here and in Paper
II.  \footnotetext[6]{http://deep.berkeley.edu/DR3/dr3.primer.html}

We complemented the DEEP2 spectroscopy of z$<$1.4 galaxies with
redshifts for z$\sim3$ sources from the Lyman Break Galaxy (LBG)
survey of \citet{2003ApJ...592..728S}.  This survey covers a total
area of 0.38~deg$^{2}$ divided in several fields, one of them centered
in the EGS. The observations in EGS consist of a single 15'$\times$15'
mask centered at $\alpha$=14$^{h}$17$^{m}$43$^{s}$,
$\delta$=52$^{\circ}$28'48'' observed with the Low Resolution Imaging
Spectrometer (LRIS; \citealt{1995PASP..107..375O}) on Keck. The
spectra cover the 400-700~nm range with a median resolution of
0.75~nm.  The targets were pre-selected based on the LBG color-color
criteria \citep{1996ApJ...462L..17S} including only candidates
brighter than $R$$=$25.5.  The EGS catalog contains a total of 334 LBG
candidates in the surveyed area. Out of them, 193 are
spectroscopically confirmed to be at z$\sim$3. Unfortunately, the
overlap with the IRAC frame is not complete (and some of the galaxies
are extremely faint in the IRAC bands), and we were only able to
identify 243 (72\% of the spectroscopic sample) LBGs in our
3.6+4.5\mic\ selected catalog (we give more details on these sources
in Paper II).

\setlength{\tabcolsep}{0.03in} 
\begin{deluxetable*}{lcccccl}
\tabletypesize{\scriptsize}
\tablewidth{0pt}
\tablecaption{\label{multiwav_data}Properties of the dataset in the EGS}
\tablehead{
\colhead{Band} & \colhead{$\lambda_{\mathrm{eff}}$} & \colhead{m$_{\mathrm{lim}}$[AB]} & \colhead{FWHM} & \colhead{Area} & \colhead{Surf. Dens.($\times$10$^{3}$)}& \colhead{Source}\\
\colhead{(1)}&\colhead{(2)}&\colhead{(3)}&\colhead{(4)}&\colhead{(5)}&\colhead{(6)}&\colhead{(7)}}
\startdata
Hard X-Ray$\dag$&  0.31~nm (2-10~keV)   &3.8~$\times$10$^{-16}$erg cm$^{2}$s$^{-1}$  &0.5-6\arcsec&  0.67&1.102 & {\it Chandra}/ACIS; \citet{2009ApJS..180..102L}\\
Soft X-Ray$\dag$&  1.24~nm (0.5-2keV)  &5.3~$\times$10$^{-17}$erg cm$^{2}$s$^{-1}$  &0.5-4\arcsec& 0.67&1.540& {\it Chandra}/ACIS; \citet{2009ApJS..180..102L}\\
FUV.............&  153.9~nm & 25.6 &5.5\arcsec& 1.13 & 10.5  &\textit{GALEX} GTO \\
NUV.............&  231.6~nm & 25.6 &5.5\arcsec& 1.13 & 24.7  &\textit{GALEX} GTO \\
$u'$............&  362.5~nm & 26.1 &1.0\arcsec& 0.77   & 148.2 &MMT/Megacam \\
$u^{*}$.........&  381.1~nm & 25.7 &0.9\arcsec& 1      & 152.5 &CFHTLS/MegaCam \\
B...............&  439.0~nm & 25.7 &1.2\arcsec& 1.31   & 101.7 &CFHT-12k \\
$g$............ &  481.4~nm & 26.7 &1.3\arcsec& 0.77   & 203.8 &MMT/MegaCam \\
$g'$............&  486.3~nm & 26.5 &0.9\arcsec& 1      & 163.4 &CFHTLS/MegaCam \\
$V_{606}$........& 591.3~nm & 26.9 &0.2\arcsec& 0.197  & 440.7 &\textit{HST}/ACS \\
$r'$............&  625.8~nm & 26.3 &0.8\arcsec& 1      & 363.9 &CFHTLS/MegaCam \\
R...............&  651.8~nm & 26.1 &0.7\arcsec& 1      & 220.0 &Subaru/SuprimeCam\\
R...............&  660.1~nm & 25.3 &1.0\arcsec& 1.31   & 144.3 &CFHT-12k \\
$i'$............&  769.0~nm & 25.9 &0.8\arcsec& 1      & 341.1 &CFHTLS/MegaCam \\
$i$.............&  781.5~nm & 25.3 &1.0\arcsec& 0.77   & 275.8 &MMT/MegaCam \\
I...............&  813.2~nm & 24.9 &1.1\arcsec& 1.31   & 117.0 &CFHT-12k \\
$i_{814}$........&  807.3~nm & 26.1 &0.2\arcsec& 0.197  & 452.1 &\textit{HST}/ACS \\
$z'$............&  887.1~nm & 24.7 &0.8\arcsec& 1      & 179.0 &CFHTLS/Megacam \\
$z$.............&  907.0~nm & 25.3 &1.2\arcsec& 0.77   & 214.0 &MMT/Megacam \\
$J_{110}$......&  1.10~$\mu$m  & 23.5 &0.7\arcsec& 0.0128 & 252.0&\textit{HST}/NICMOS \\
$J$.............&  1.21~$\mu$m  & 22.9 &1.0\arcsec& 0.195  & 67.0 &CAHA-$\Omega2k$ \\
$H_{160}$......&  1.59~$\mu$m  & 24.2 &0.8\arcsec& 0.0128 & 252.0&\textit{HST}/NICMOS \\
$J\dag$.........&  1.24~$\mu$m  & 21.9 &1\arcsec  & 0.30   & 30.0 &Palomar-WIRC; \citet{2006ApJ...651..120B}\\
$K_{s}$............&  2.11~$\mu$m  & 20.7 &1.5\arcsec& 0.20   & 18.6 &CAHA-$\Omega'$ \\
$K_{s}$.............&  2.15~$\mu$m  & 23.7 &0.6\arcsec& 0.09   & 124.0&Subaru MOIRCS\\
$K\dag$.........&  2.16~$\mu$m  & 22.9 &1\arcsec  & 0.70   & 34.0 &Palomar-WIRC; \citet{2006ApJ...651..120B}\\
IRAC-3.6........&  3.6~$\mu$m &  23.9&2.1\arcsec  &0.48 & 315.0 &\textit{Spitzer} GTO \\
IRAC-4.5........&  4.5~$\mu$m &  23.9&2.1\arcsec  &0.48 & 274.4 &\textit{Spitzer} GTO \\
IRAC-5.8........&  5.8~$\mu$m &  22.3&2.2\arcsec  &0.48 & 129.7 &\textit{Spitzer} GTO \\
IRAC-8.0........&  8.0~$\mu$m &  22.3&2.2\arcsec  &0.48 & 115.2 &\textit{Spitzer} GTO \\
MIPS-24.........&  23.7~$\mu$m &  19.5 (60~$\mu$Jy)& 5\arcsec  & 0.79 & 30.0&\textit{Spitzer} GTO \\
MIPS-70.........&  71.4~$\mu$m &  15 (3.5~mJy)     & 19\arcsec  & 0.69 & 6.0 &\textit{Spitzer} GTO \\
Radio 20cm$\dag$&    20~cm     &  100~$\mu$Jy beam$^{-1}$  & 4.2\arcsec & 0.73 & 1.538 & VLA; \citet{2007ApJ...660L..77I}\\
R,redshift$\dag$......& 640-910~nm & 24.1 & - & 1.31 & 10.343 & DEEP2
\enddata
\tablecomments{\\
\dag Data drawn from a catalog. \\
Col(1) Name of the observing band.\\
Col(2) Effective wavelength of the filter calculated by convolving the Vega spectrum \citep{1994AJ....108.1931C} with the transmission curve of the filter+detector.\\
Col(3) Limiting AB magnitude (except for the X-ray catalogs) of the image estimated as the magnitude of a SNR$=$5 detection (see \S\ref{merged_technique} for details on the measurement
of the photometric errors).\\
Col(4) Median FWHM of the PSF in arcseconds measured in a large number of stars (see \S~\ref{STAR}).\\
Col(5): Area covered by the observations in deg$^{2}$.\\
Col(6): Source density per square degree up to the limiting magnitude given in Col(3).\\
Col(7): Source from where the data were obtained.}
\end{deluxetable*}

\section{Description of the sample selection}\label{mergedcat}

The dataset described in the previous Section was used to obtain
UV-to-FIR Spectral Energy Distributions (SEDs) for all the sources
detected in the EGS IRAC survey. This merged photometric catalog was
built following the procedure described in PG05 and PG08. Here we review
all the basic steps of the method, emphasizing the
improvements introduced for this paper concerning the extraction of
the IRAC catalog and the band merging procedure.

\subsection{The IRAC-3.6$\mu m$+4.5$\mu m$ selection}
\label{irac_catalog}

The source detection in the IRAC data was carried out separately in
the 3.6~$\mu$m and 4.5~$\mu$m images using SExtractor
\citep{1996A&AS..117..393B}. The complementary detection in the
slightly shallower 4.5$\mu m$-band helps to alleviate the source
confusion problems arising from the PSF size and the remarkable depth
of the IRAC data. Both catalogs were cross-matched using a 1\arcsec\,
search radius to remove repeated sources. This produces a master
IRAC-selected catalog containing the sources detected in any of the
two channels.  Eventually, most sources are simultaneously detected in
both channels.

The average survey depth is remarkably homogeneous across the strip,
$t_{exp}\sim$10ks, with the exception of two small areas with lower
exposure at the top ($\delta$$>$53.525$^{\circ}$) and bottom
($\delta$$<$52.025$^{\circ}$) of the mosaic. We took into account the
lower exposure times near the edges of the images by defining two
different areas: a shallower region with exposure time shorter than
3800 seconds (N(frames)$\leq$20), and a deeper region covering the
majority of the strip. The detection was carried out with different
SExtractor parameters in each region, using a more conservative
configuration for the shallower region. Then, we used this (more
restrictive) catalog to purge some low significance detections in the
other catalog within an overlapping area between them
(N(frames)=18-25). The purged catalog restricted to the area with
N(frames)$>$20 constitutes our master photometric catalog, and it
covers an area of 0.50~deg$^{2}$.

After the detection of sources, we removed spurious sources in the
wings of bright stars (where the PSF shows bright knots). For that
purpose, first we made a preliminary detection of star-like sources
based on the IRAC color-color criteria of \citet[][see
Section~\ref{STAR}]{2004ApJS..154...48E}.  Then, we eliminated
detections within the typical distance where the contamination from
the star is significant (typically $\sim$9\arcsec\ for the typical
star magnitudes and depth of the EGS observations).  After masking the
regions around stars, the total area covered by the catalog is
0.48~deg$^{2}$.

Aperture photometry for all 3.6~$\mu$m+4.5~$\mu$m detected sources was
measured with our own dedicated software (which takes into account
pixel fractions appropriately; see PG08) in all four IRAC images,
previously registered to the same WCS. The flux is measured in the
four IRAC bands simultaneously. If a source is undetected in the
shallower bands (i.e., [5.8] and [8.0]) we still measure an upper
limit flux as 3 times the {\it rms} of the sky. The majority of the
IRAC sources are unresolved in the 3.6~$\mu$m image
(FWHM~$\sim$2\arcsec). However, most of them are not point-like, but
slightly extended. Consequently, PSF fitting is not effective and
photometry is best measured with small circular apertures (PG08;
\citealt{2008ApJS..177..431B}; \citealt{2008ApJ...682..985W};
\citealt{2009ApJ...690.1236I}).  The flux measurement in all bands was
carried out at the positions specified in the IRAC master catalog. We
used a 2\arcsec\, radius aperture and applied aperture corrections for
each band derived from the point spread function (PSF) growth curves.
The values of the correction are
[0.32$\pm$0.03,0.36$\pm$0.02,0.53$\pm$0.02,0.65$\pm$0.02]~mag at
[3.6,4.5,5.8,8.0]~$\mu$m. The errors account for the typical WCS
alignment uncertainties.  For a small number of extended sources
($\sim$2$\%$ of the total catalog, and 75\% of them presenting
[3.6]$<$22.3), the 2\arcsec\ aperture tend to underestimate the total
magnitudes (by more than a 10\%). These sources are typically bright
nearby galaxies whose \citet{1980ApJS...43..305K} radius is larger
than $\sim$4.5\arcsec\ .  The flux measurement for these sources was
performed in larger apertures enclosing the full object and applying
the extended source aperture corrections given in the {\it
  Spitzer}/IRAC cookbook.

The uncertainties in the IRAC photometry were computed taking into
account the contributions from the sky emission, the readout noise,
the photon counting statistics, the uncertainties in the aperture
corrections, and a 2\% uncertainty from the zero point absolute
calibration \citep{2005PASP..117..978R}. We did not assume the
uncertainties resulting from SExtractor flux measurement. Instead, we
used a more realistic method to determine the background noise that
takes into account the effects of pixel-to-pixel signal correlation.
This procedure has also been applied to measure the photometric
uncertainties in all the bands and is outlined in
\S\ref{merged_errors}. Nevertheless, a straightforward comparison to
the SExtractor errors indicates that the noise correlation does not
introduce a significant contribution to the flux uncertainties at
bright magnitudes, leading to a median increment $\leq$0.02~mag up to
[3.6,4.5]$\sim$23.75.

A detailed comparison of our IRAC catalog of the EGS survey to the one
published by \citet{2008ApJS..177..431B} is presented in
\S~\ref{merged_comparison}, including a discussion on the source
confusion levels.

\subsection{Completeness and limiting magnitude of the IRAC catalog}
\label{merged_completeness}

\begin{figure*}
\includegraphics[width=8.5cm,angle=0.]{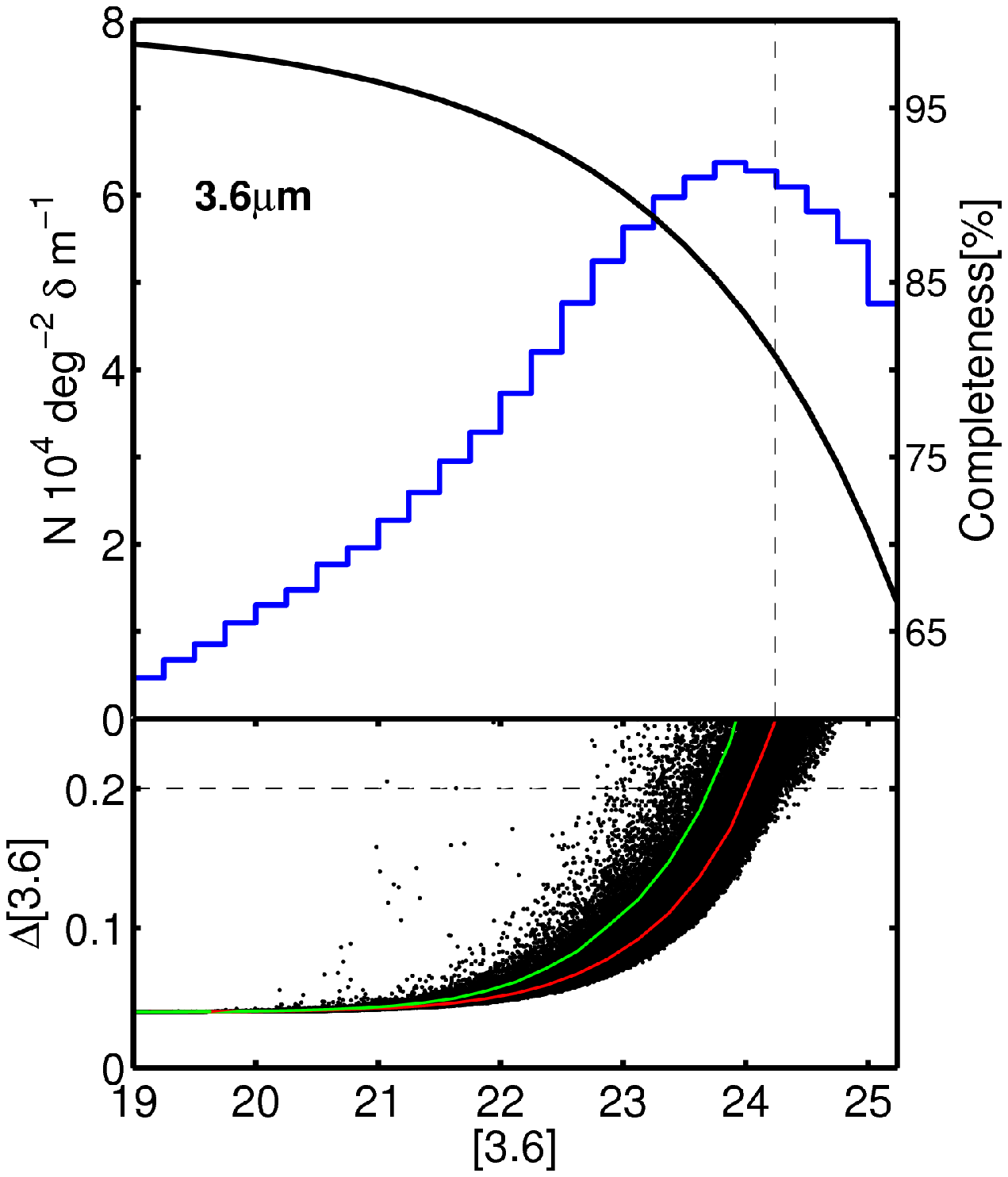}
\hspace{0.2cm}
\includegraphics[width=8.5cm,angle=0.]{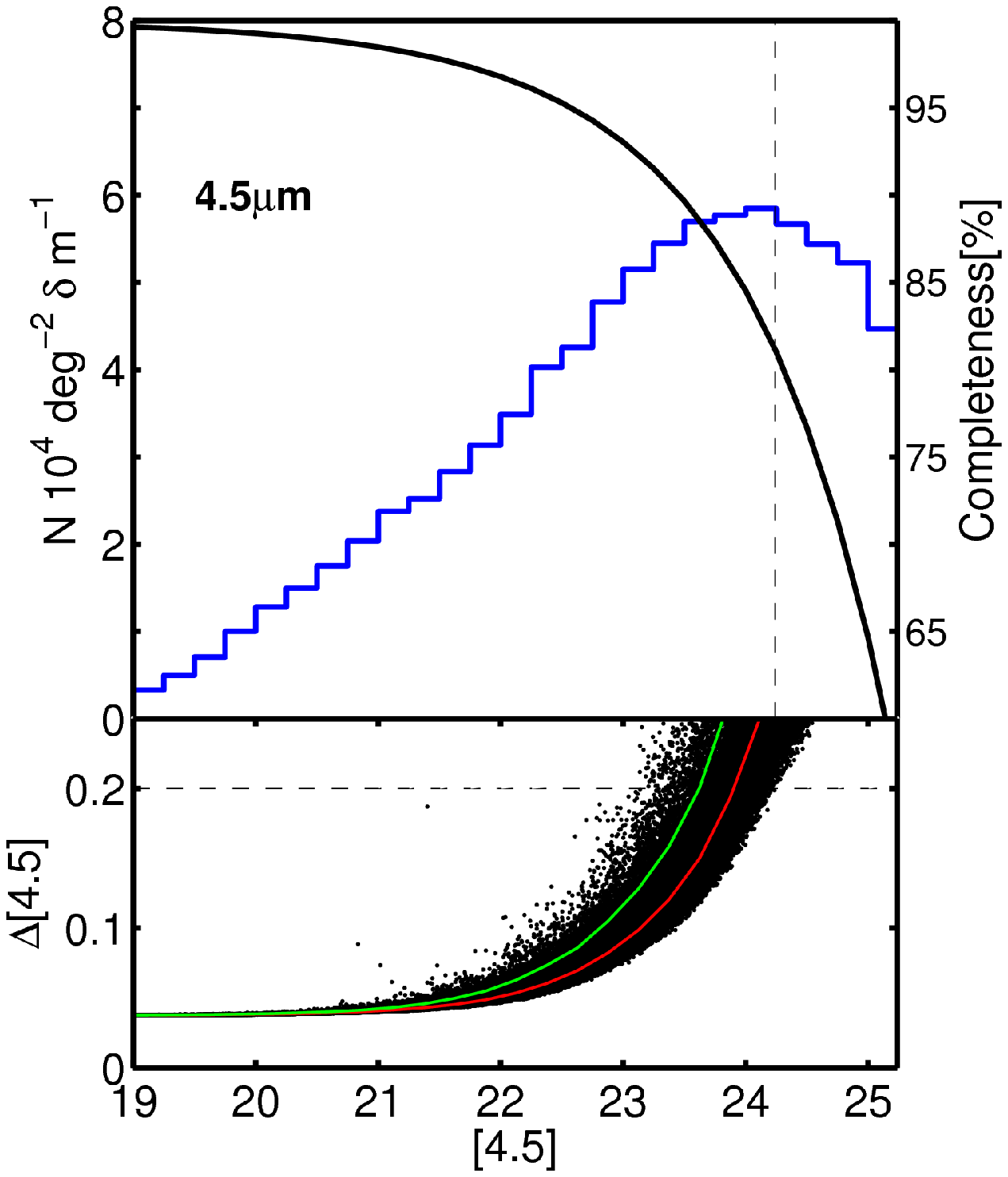}\\
\includegraphics[width=8.5cm,angle=0.]{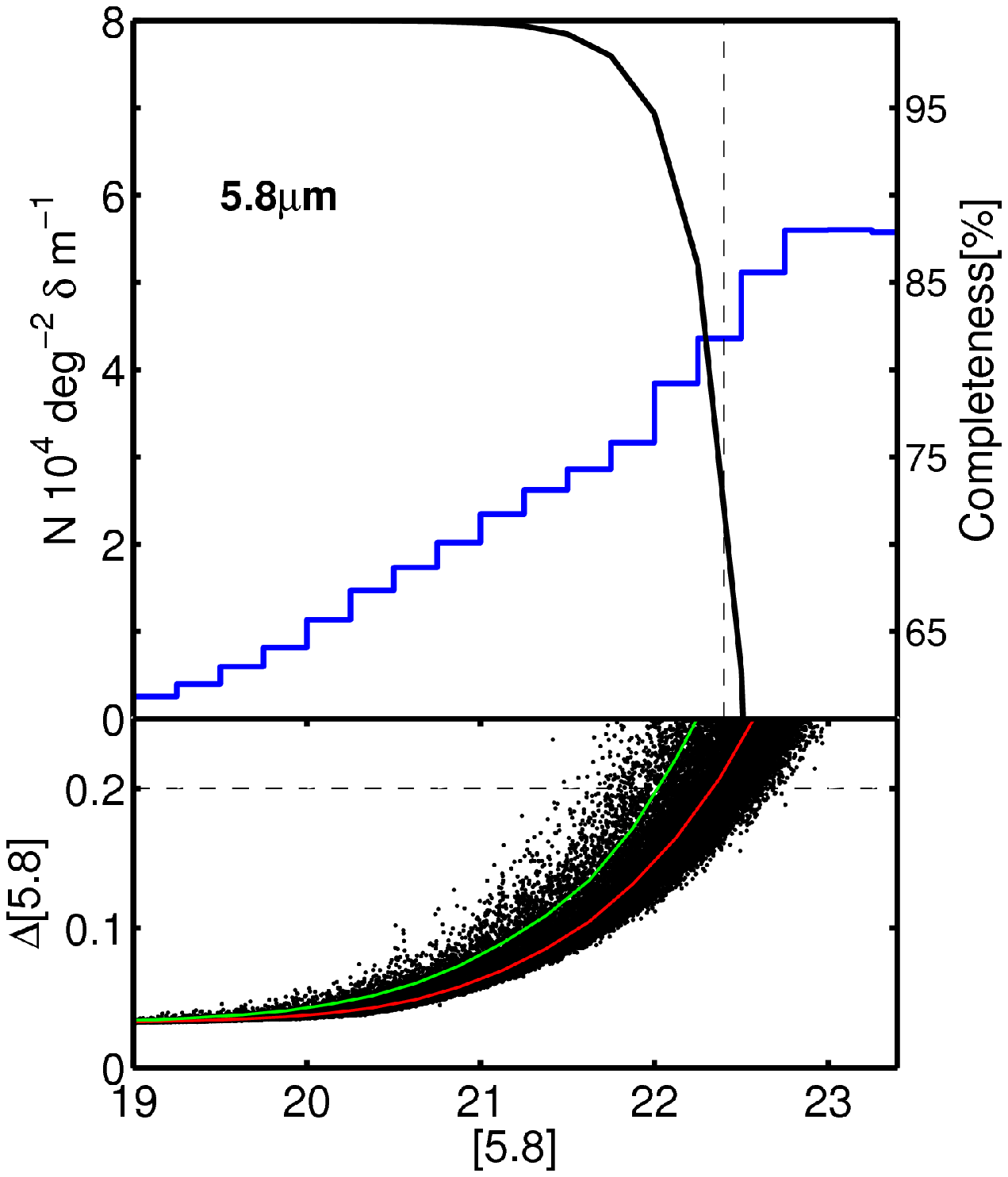}
\hspace{0.2cm}
\includegraphics[width=8.5cm,angle=0.]{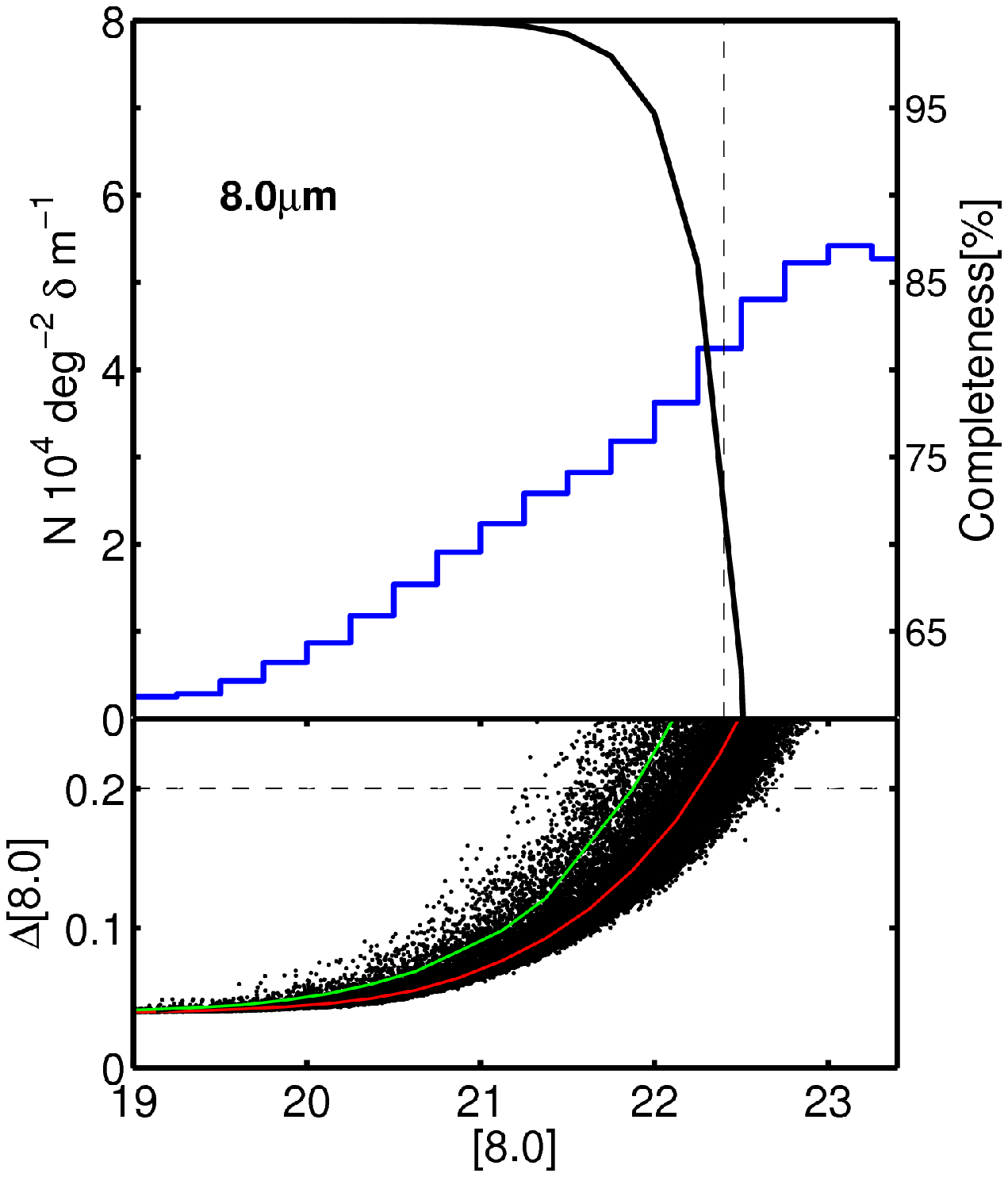}
\caption{\label{completeness} The upper panel of each quadrant shows
  the histogram of IRAC magnitudes for the sources in our sample
  selected in the IRAC-3.6$\mu m$+4.5$\mu m$ data.  The histograms are
  shown up to the 2$\sigma$ limiting magnitude in each band. The
  blacks continuous curve depicts the detection efficiency as
  estimated from recovery rate of simulated punctual sources. The
  vertical dashed lines represent 5 times the median sky-{\it rms}
  (5$\sigma$) measured in a large number of 2\arcsec\ radius
  apertures.  The lower panels in each quadrant show the distribution
  of photometric uncertainties as a function of the magnitude in each
  IRAC band. The red and green lines indicate the median and 90$\%$ of
  the error distribution as a function of magnitude, respectively. The
  horizontal dashed line shows the SNR$\sim$5 limit; the intersection with
  the red line indicates the value quoted in Table~\ref{multiwav_data}
  for the IRAC bands. Note that these values are slightly lower than the
  5$\sigma$ sky-{\it rms}.}
\end{figure*}

We estimated the completeness of the IRAC catalog by analyzing the
recovery of simulated sources added in the mosaicked images. The
simulations were carried out in the central regions of the mosaic
where the coverage is uniform ($t_{exp}\geq$10ks). Artificial sources
spanning a wide range of sizes (from 1\arcsec\ to 6\arcsec\ ) and
brightnesses ([3.6]$=$16-25~mag) were created on the IRAC images at
random locations. The number of simulated sources was chosen to be
representative of the Poisson uncertainty in the observed number
densities. The source detection and photometry was performed again in
the simulated images keeping the same SExtractor parameters as in the
original frames. The success rate recovering the simulated sources
allow us to estimate the completeness level as a function of
magnitude. Figure~\ref{completeness} summarizes the completeness
analysis in the four channels.  For simplicity, we show only results
for point-like sources. The completeness for extended sources is
typically $\sim$10\% lower at faint magnitudes, however, these sources
represent a very small fraction of the catalog at these magnitudes.
Figure~\ref{completeness} also shows the source density as function of
the magnitude in each band.  We find that the catalog is 85\% complete
for point sources with [3.6,4.5]$=$23.75~mag, and 75\% complete at
[3.6,4.5]$=$24.75~mag. In the two other channels, the detection
efficiency is significantly lower, with 85$\%$ completeness at
[5.8,8.0]=22.25, and the completeness dropping rapidly beyond that
magnitude. Note that the forced photometric measurement in these bands
provides a significant number of $<$5$\sigma$ (vertical dashed lines
in Figure~\ref{completeness}) detections that would be missed
otherwise.

The lower panels of Figure~\ref{completeness} show the photometric
uncertainties as a function of magnitude in the four IRAC bands.  The
red and green lines indicate the median and the level enclosing 90$\%$
of the distribution, respectively. For the region with the deepest
coverage, we estimated a 3$\sigma$ limiting magnitude of
[3.6,4.5]$\sim$24.75 and [5.8,8.0]$\sim$22.90 from the median value of
the sky {\it rms} in our default photometric apertures (2\arcsec\
radius).

The 3.6+4.5~$\mu$m catalog contains 70,048 and 99,618 sources with
[3.6]$<$23.75 and [3.6]$<$24.75 respectively. The median magnitude of
the sample up to [3.6]$<$24.75 is [3.6]=23.14, and 75\% of the sources
present [3.6]$<$23.96. Note that these numbers correspond to the IRAC
3.6+4.5~$\mu$m catalog before applying the de-blending technique
discussed in \S~\ref{merged_properties}. Consequently, the number
  of sources quoted above is lower than in the final catalog (see
  Table~\ref{sample_summary}).

\subsection{Comparison to Barmby et al. IRAC selected catalog}
\label{merged_comparison}

Here we compare our IRAC photometric catalog to that published by
\citet[][hereafter BAR08]{2008ApJS..177..431B}. BAR08 used the same
dataset (except for the GO flanking regions) and obtained final
mosaics in all 4 IRAC bands with a very similar reduction to ours.
Concerning the source detection, our method is slightly different from
BAR08 since we detect galaxies in both the 3.6 and 4.5\mic\, channel
instead of only using the bluer band. Our dual detection technique
helps to alleviate the source confusion problems arising from the PSF
size and the remarkable depth of the IRAC data. In addition, we
measure fluxes in the four channels simultaneously, obtaining upper
limit values for undetected sources in the shallower bands ([5.8] and
[8.0]). Note also that we have increased the resolution of our catalog
by deconvolving blended sources using higher resolution information
from ground-based observations (\S\ref{rainbow_crossmatch}). However,
for the sake of clarity, we compare here the BAR08 catalog with ours
before carrying out the deblending procedure.

BAR08 measured aperture photometry with SExtractor. The publicly
available catalog\footnotemark[1] includes MAG\_AUTO and MAG\_ISO
measurements, jointly with aperture magnitudes for several radii
corrected to total magnitudes with empirical PSF corrections. We
compare our photometry to the magnitudes measured in the 3.5 pixel
aperture (2.1\arcsec) by BAR08. Their aperture corrections agree with
our measurements for the 4\arcsec\ diameter aperture within the errors
(due to alignment uncertainties).

\footnotetext[1]{http://www.cfa.harvard.edu/irac/egs/}

BAR08 and our catalog are cross-correlated in the region of highest
exposure (t$_{exp}$$>$4ks) using a 1\arcsec\ radius.  The comparison
of the WCS between the two mosaics is in very good agreement, with a
{\it rms} of $<$0.05\arcsec\ .  We find 41,514 and 52,130 sources in
common up to [3.6]$<$23.75 and [3.6]$<$24.75, respectively.

Attending to the density of sources per unit area, we find that our
catalog includes $11$\%$\pm$5\% and $18$\%$\pm$7\% more sources than
the catalog published by BAR08 at [3.6]$<$23.75~mag and
[3.6]$<$24.75~mag, respectively. The uncertainties in these
measurements were estimated by comparing number counts in 0.5~mag
bins, including Monte-Carlo simulations on the photometric errors.

The different source densities are a consequence of their more
conservative SExtractor detection threshold.  We have carefully chosen
the SExtractor parameters differentiating between the regions with
high/low coverage (\S\ref{irac_catalog}), trying to push down the
detection limits as much as possible without degrading the reliability
of the entire catalog. As a consequence, our catalog recovers a larger
number of sources at fainter magnitudes, and the completeness of our
catalog is larger than BAR08 for the same magnitude.  For example, they
quoted a $\leq50\%$ completeness for point-like sources at
[3.6]$=$23.75~mag, compared to our estimated $85\%$. Unfortunately,
lowering the detection threshold inevitably increases the number of
spurious detections. However, in the context of a merged multi-band
photometric catalog, spurious detections can be efficiently identified
as sources only detected by IRAC (cf. N(band)$<$5), and we will show
that the reliability is $>97$\%, with false detections located almost
uniquely around very bright sources (see \S~\ref{reliability_sect}).

\begin{figure*}
\includegraphics[width=8.5cm,angle=0.]{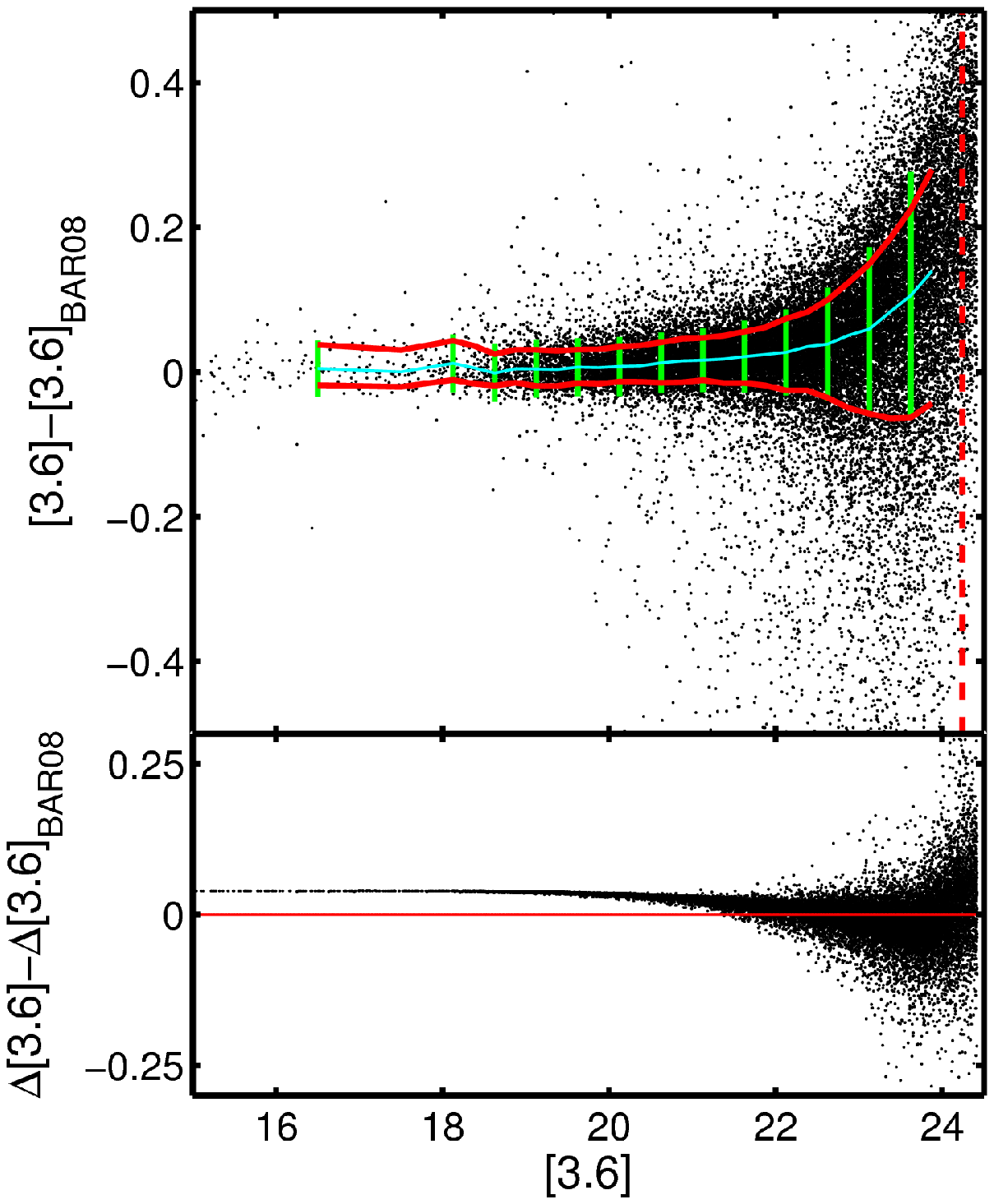}
\hspace{0.2cm}
\includegraphics[width=8.5cm,angle=0.]{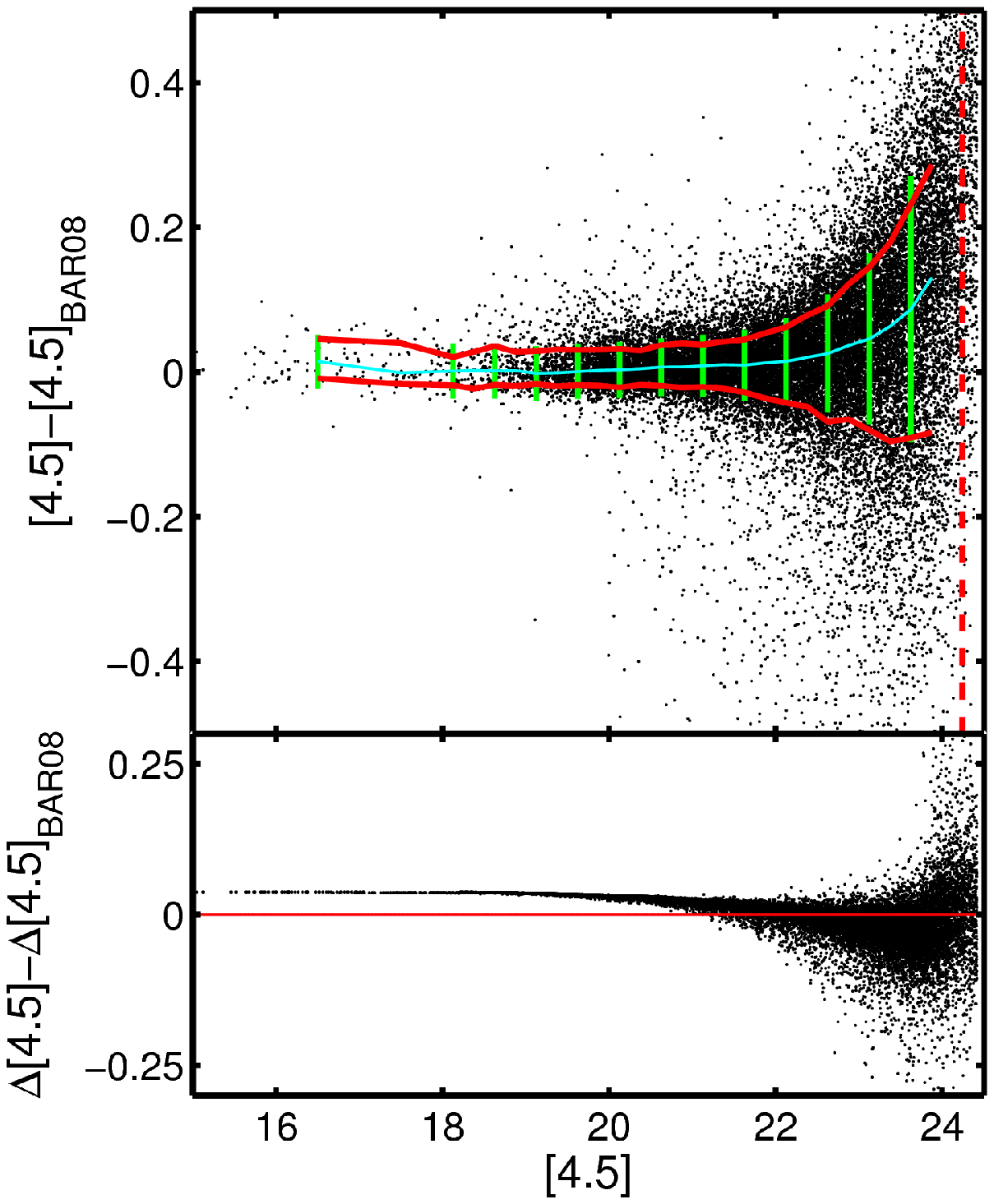}\\
\includegraphics[width=8.5cm,angle=0.]{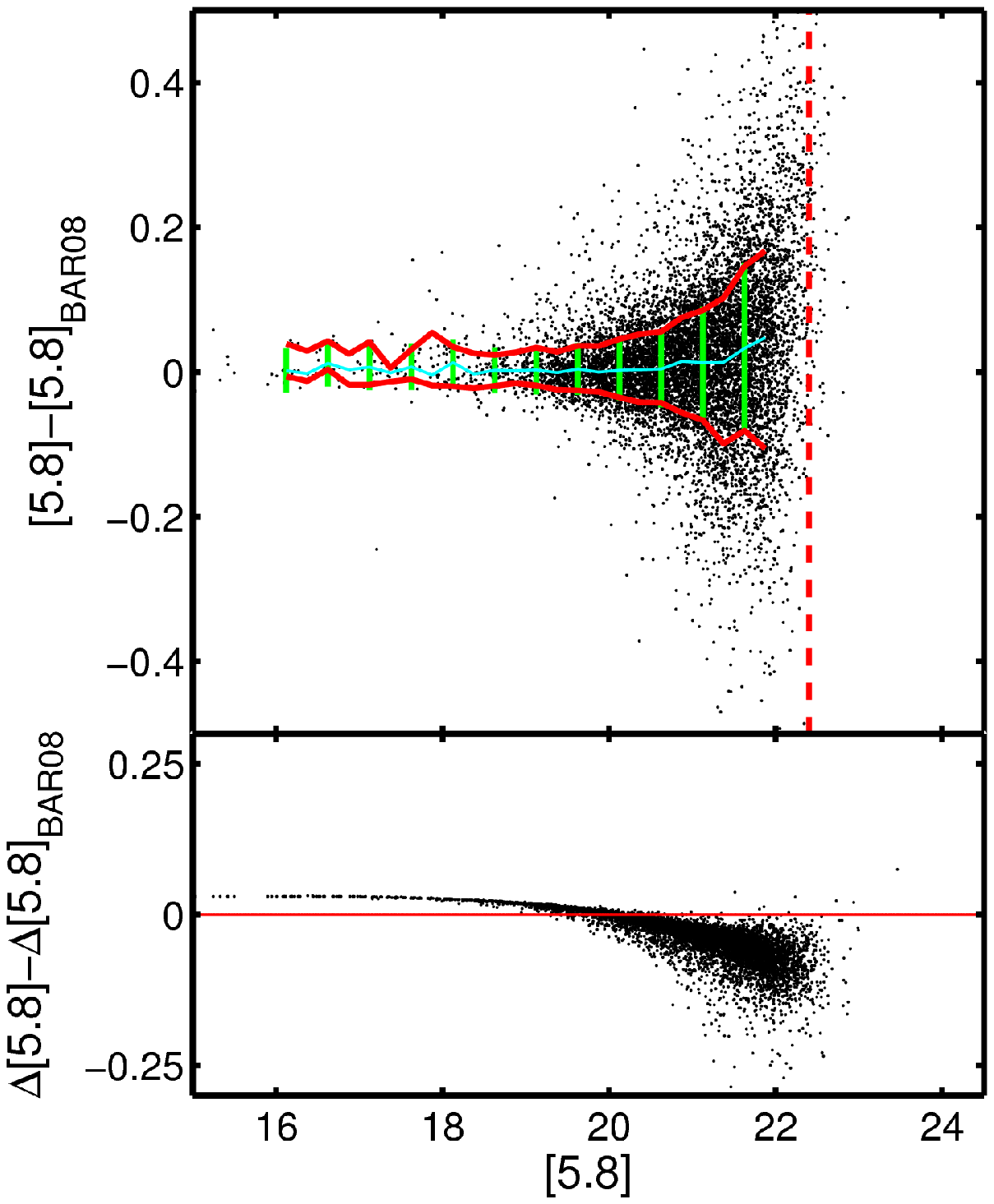}
\hspace{0.2cm}
\includegraphics[width=8.5cm,angle=0.]{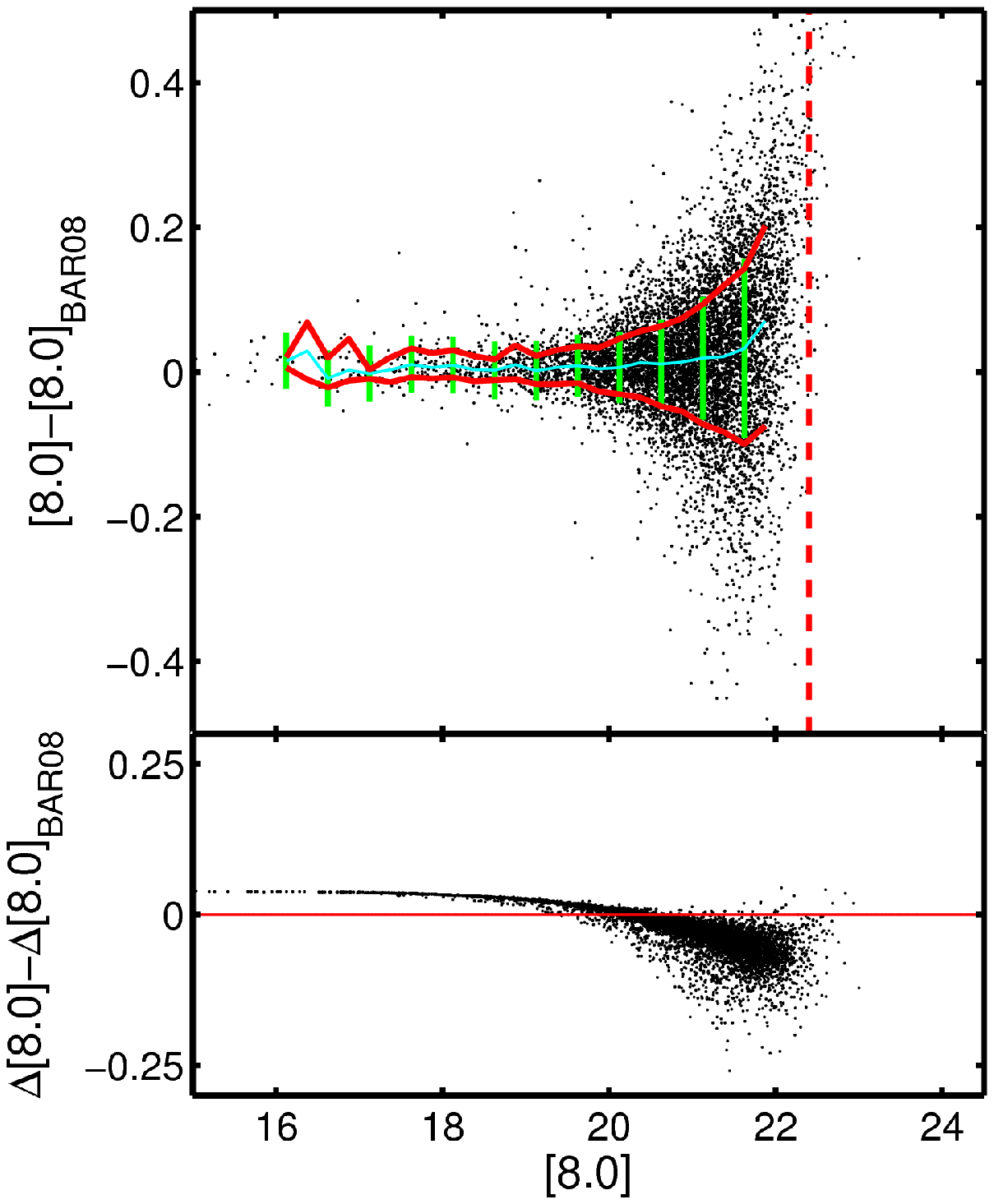}
\caption{\label{compare_barmby} Comparison of the observed magnitudes
  (upper panels) and the photometric errors (lower panels) in the four
  IRAC bands for the sources in common between our IRAC-3.6+4.5~$\mu$m
  catalog and BAR08 catalog. The cyan line shows the median value of
  the magnitude difference as a function of magnitude. We have
  corrected the comparisons by a constant value of -0.05~mag in [3.6]
  and [4.5], and by -0.04~mag and -0.03~mag in [5.8] and [8.0],
  respectively. Such small offsets can be attributed to slight
  differences in the data reduction and aperture corrections. The
  green bars indicate the average photometric errors per magnitude bin
  in our catalog.  The red lines enclose 1$\sigma$ of the distribution
  centered in the median value (cyan line). The vertical dashed line
  indicate the 5$\sigma$ limiting magnitude in our catalog. The
  photometry in the four bands is consistent up to the $\sim85\%$
  completeness limit. The uncertainties in BAR08 are 5$\%$ to 10$\%$
  larger at faint magnitudes, probably as a result of the slightly
  different procedure applied to measure the sky background.}
\end{figure*}

Figure~\ref{compare_barmby} shows the comparison of the photometric
magnitudes in the four bands for both catalogs.  In both cases the
flux was measured on circular apertures and corrected to total
magnitudes using aperture corrections.  We have corrected the
comparison by a constant value of -0.05~mag in [3.6] and [4.5], and by
-0.04~mag and -0.03~mag in [5.8] and [8.0], respectively. Such small
offsets are attributed to slight differences in the data reduction
(final absolute calibration, frame stacking, registering and
mosaicking) and in the aperture corrections.  Despite the small
offsets, the overall results in the four IRAC bands are in good
agreement. Note that the average photometric error in our catalog for
each magnitude bin (green bars) encloses 1$\sigma$ of the values
around the median difference (red and cyan lines).

The lower panel of each plot in Figure~\ref{compare_barmby} shows the
comparison of the photometric errors in BAR08 and in our catalog.  Our
quoted photometric uncertainties tend to differ from BAR08, specially at
faint magnitudes.  In contrast with that paper, we have considered
zero-point and WCS uncertainties, resulting on slightly larger
uncertainties ($\sim$0.05~mag) in our catalog for bright sources up to
[3.6][4.5]$\sim$21~mag and [5.8][8.0]$\sim$19~mag. At fainter
magnitudes, the photometric uncertainties increase with magnitude at a
faster rate in the catalog of BAR08. The cause for this difference is
the procedure to measure the background noise. Similarly to BAR08, we
estimated this value from the sky variance measured in circular
apertures at different locations of the images that are empty of
sources (\S\ref{merged_errors}). However, the definition of an empty
region depends on the limits of source detection. Therefore, given our
higher detection fraction, our {\it sky} regions would contain, in
principle, lower signal pixels effectively decreasing the {\it
  rms}. In addition, our photometric procedure estimates uncertainties
on a source-by-source basis studying the background around each object
in a independent way, while BAR08 relied on SExtractor photometric
errors and applied a correction to them based on the average
properties of the mosaic. Nevertheless, our estimates of the
photometric errors are consistent with the observed scatter of the
comparison between our photometry and that measured by BAR08.

\section{Multi-wavelength photometry: the {\it Rainbow} catalog}
\label{merged_technique}

Using the whole dataset available in the EGS field, we created a
multi-wavelength photometric catalog for the IRAC selected sample
described in the previous Section. For that purpose, we used the {\it
  Rainbow} software package, described in detail in PG05 and PG08.
This software was created to: (1) cross-correlate multi-band catalogs
and obtain consistent (aperture matched) photometry on the different
bands to build a UV-to-FIR SED; (2) estimate stellar parameters, such
as photometric redshifts, stellar masses, and SFRs from those SEDs. In
the rest of this Section we will describe the photometry procedure,
and in Paper II we will present the methods to estimate photometric
redshifts, stellar masses, and SFRs out the SEDs.

\subsection{Cross-correlation and source deblending}
\label{rainbow_crossmatch}

The {\it Rainbow} code starts from a primary selection catalog (in our
case, IRAC selected) and obtains merged photometry and spectroscopy in
other bands.  The first step is to identify the counterparts of the
IRAC sources in the other bands, where SExtractor catalogs have been
built following typical procedures. These catalogs are
cross-correlated to the 3.6+4.5~$\mu$m positions using a
2\arcsec\ search radius.  An exception to this rule are the MIPS,
radio and X-ray catalogs.  For the MIPS and radio bands, we used
a 2.5\arcsec\ and 3\arcsec\ radius, respectively, recognizing possible
alignment and center estimation problems of the order of one
pixel. For the latter, instead of using the WCS of the X-ray sources,
we cross-matched to the positions of the IRAC counterparts (given in
\citealt{2009ApJS..180..102L}) using a 1\arcsec\ radius. These authors
used as reference the IRAC catalog of BAR08, which cover a slightly
lower area than ours. Thus, for sources outside of the BAR08 mosaic we
cross-matched to the X-ray coordinates using a 2\arcsec\ radius. (see
\S~\ref{xray_radio} for more details).

The IRAC sources are identified in this way with objects in all the
other catalogs. One of the optical catalogs (typically the deepest; in
the EGS, the Subaru $R$-band data) is used as a reference to narrow
the following cross-correlations and alleviate confusion problems
present in the IRAC images. As the cross-correlation to the catalog of
spectroscopic redshifts is done to the coordinates of the counterpart
in the reference (optical) band, it is possible to choose a more
reliable 0.75\arcsec\ search radius for this catalog.

Before the cross-matching procedure is carried out between a pair of
images, these are re-aligned locally within a
4\arcmin$\times$4\arcmin\ square region using the positions of several
sources (typically more than 20) as reference.  The mean {\it rms}
between the central coordinates of matched sources in optical/NIR
images is typically $\leq$0.1\arcsec\, and $\leq$0.2\arcsec\, between
the IRAC and the ground-based images.  This procedure allows to
overcome small misalignment problems between the frames, and assures a
reliable identification of counterparts, and an accurate positioning
of the photometric aperture in all bands. It also allows to obtain
reliable photometry (in the appropriate aperture) even if the source
is very faint and/or undetected in an individual image.

\begin{figure}
\includegraphics[width=8.5cm,angle=0.]{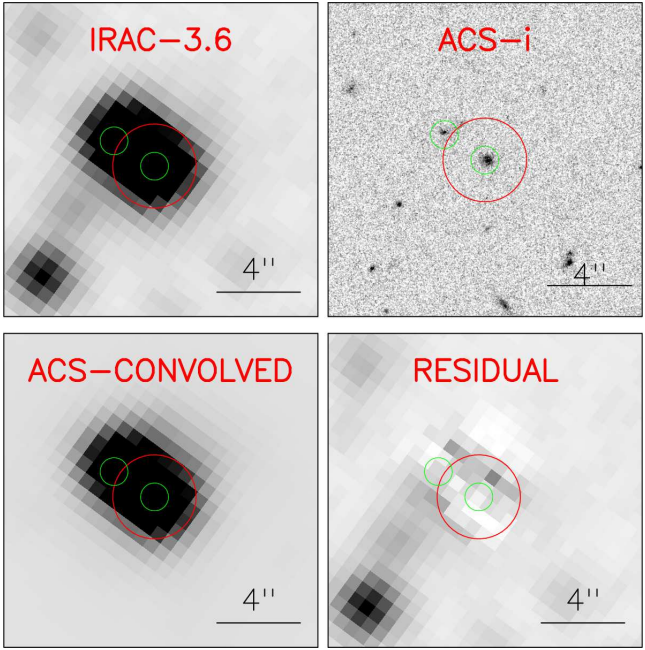}
\caption{\label{deblendplot} Example of the deconvolution
    procedure for unresolved sources in IRAC images. After the
    deblending, the unique irac source (irac164074) becomes two
    separate sources (irac164074 \_1 and \_2).{\it Top-left (A)}:
  IRAC-3.6$\mu$m image (0.61\arcsec/pixel) showing a blended source.
  The red circle depicts a 2\arcsec\ radius aperture (our default
  aperture for isolated sources). The green circles show
  0.9\arcsec\ apertures centered at the positions of sources in the
  reference image. {\it Top-right (B)}: \HST\,/ACS reference image
  (0.02\arcsec/px) after applying a 4\arcmin\,$\times$4\arcmin\ local
  WCS re-alignment, showing the individual sources. {\it Bottom-left
    (C)}: Model of the blended source obtained by convolving the PSF
  of \HST\,/ACS to the PSF of IRAC-3.6$\mu$m in (B), registering it to
  image (A) and scaling each source to the flux in the
  0.9\arcsec\ apertures (green) in (A). {\it Bottom-right (D)}
  Residual from the subtraction of the model and science images
  (C-A).}
\end{figure}

The combination of the remarkable depth and the $\sim$2\arcsec\ FHWM
of the IRAC observations inevitably leads to issues of source
confusion, specially around crowded environments. However, based on
the (ground-based) reference image, it is possible to deblend IRAC
sources which have not been separated by SExtractor in the original
IRAC images and lie at least 1\arcsec\ away (half the FWHM, chosen as
our resolution criterion). \HST\ images reveal that the multiplicity
is larger, but the deconvolution of sources separated by less than
1\arcsec\, is very uncertain. When multiple counterparts are found in
the optical/NIR images during the cross-matching, the IRAC photometry
is recomputed following a deconvolution method similar to that used in
\citet{2006A&A...449..951G}, \citet{2008ApJ...682..985W},
\citet{2009ApJ...691.1879W} or \citet{2010ApJS..187..251W}.

In our case, first, the coordinates of the photometric aperture are
re-positioned to that of the optical/NIR counterparts. Then, the PSF
of the higher resolution image is convolved to the IRAC PSF, and the
flux of each source is scaled to match that of the real IRAC sources
measured in 0.9\arcsec\ apertures (after re-centering to the positions
of the optical counterparts). Finally, total magnitudes are computed
applying an aperture correction of
[1.30$\pm$0.07,1.02$\pm$0.08,1.2$\pm$0.10,1.44$\pm$0.14]~mag in the
[3.6,4.5,5.8,8.0]~$\mu$m bands, respectively.
Figure~\ref{deblendplot} illustrates the deconvolution procedure using
a HST/ACS image as reference. The red and green apertures depict the
{\it standard} 2\arcsec\ aperture (for isolated sources) and the
0.9\arcsec\ aperture, respectively.  Pixel-by-pixel variations in the
residual from subtracting the model PSF do not exceed a 5$\%$ within
the 0.9\arcsec\ and 2\arcsec\ apertures. We also checked that the
average {\it rms} ($\sim$3$\%$) is well within the photometric error
of the sources.  The analysis of an average PSF, derived from observed
sources across the image, indicates that for the typical separation
between blended sources, $\sim$2.2\arcsec\ ($>$1.8\arcsec\ for 75$\%$
of them), the flux contamination from the nearby neighbor does not
exceed a 10$\%$ for sources with a flux ratio around 1:2-3.
Approximately 75$\%$ of the blended sources present flux ratios lower
than 1:3.5.

After applying the deblending method, our IRAC selected catalog
  contains 76,936 (113,023) sources to [3.6]$<$23.75 (24.75).  This
  means that we were able to deblend 8\% of the sources in the
  original IRAC catalog built with Sextractor (presented in
  \S~\ref{irac_catalog}), and 16\% of the final catalog of 76,936
  sources were deblended (typically, each blended sources was a
  combination of 2 sources).We find no significant difference in the
brightness distribution of the blended sources compared to the
resolved sources.

In the following sections, and in Paper II, we will analyze the SEDs
and physical properties of the IRAC sources, concentrating in the
sample with [3.6]$<$23.75~mag, which count with more accurate IRAC
photometry (SNR$\gtrsim$8). Therefore, this will be the working
  sample for the rest of the paper unless explicitly stated otherwise.
  Nonetheless, all the procedures discussed in the following are also
  applied to the sources down to [3.6]$<$24.75~mag. Although these
  objects are not included in the accompanying catalog (presented in
  \S~\ref{dataacess}, restricted to the most reliable detections), it
  is possible to retrieve their data through our online database (see
  \S\ref{rbnav}).

\subsection{Merged photometry}\label{rainbow_photometry}

The photometry is carried out in the same (Kron) elliptical aperture
in all bands.  The parameters of that aperture are obtained from a
reference image (the same as for the cross-matching procedure) whose
resolution is representative of the entire dataset.  Normally, this
reference image is a ground-based optical/NIR frame with a PSF of
approximately 1-1.5\arcsec, which is easy to translate to other
ground-based images avoiding aperture issues.

The bands are sorted according to depth to facilitate the cross-match
to the optical bands. The typical aperture band for the majority
  of the sources (83\% of the sample) is the Subaru $R$-band, followed
  by the CFHTLS $i'$-band (4\%), MMT -$i$,$-z$ (3$\%$ each) and
  MOIRC-$K$ (2\%). For the remaining 5\% of the sources, the aperture
  is computed from other bands (CFHTLS-r, $i_{814}$, V$_{505}$) that
  account for less than 1\% of the total each. The Subaru imaging was
preferred to the CFHTLS as primary aperture band due to the uniform
coverage of the whole IRAC mosaic (the CFHTLS frames cover only
$\sim$60\% of the IRAC survey). In order to ensure that the
  aperture radius is always large enough to enclose the full PSF
profile in all ground-based images, we established a minimum
  aperture radius equal to the worse value of the seeing
(1.5\arcsec). Once the best photometric aperture is defined, if no
counterpart is found in one band but data exist at that position, the
flux is measured within the same aperture.  Note that the local WCS
re-centering of the images allows an accurate positioning of that
aperture. In this way we recover fluxes for very faint sources whose
detection was missed by SExtractor. When the counter-part source is
too faint to be detected, the sky-{\it rms} value is stored to be used
as an upper limit in the SED analysis.

\begin{figure}
\includegraphics[width=8.5cm,angle=0.]{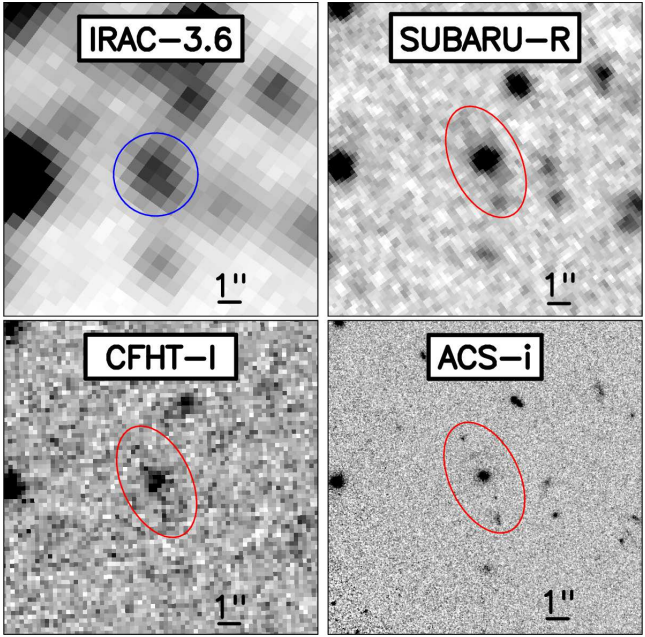}
\caption{\label{aperphot} Multi-band images of an IRAC selected source
  (irac096525) in the [3.6], R, I and $i_{814}$ bands (from left
  to right and top to bottom). The size of the frames is
  15\arcsec\,$\times$15\arcsec\,. In the three optical bands, the flux
  is measured in elliptical apertures (red) whose parameters are
  determined in the R-band image. Note that, despite the different
  image resolutions (and multiple \HST\ counterparts), the accurate
  WCS re-centering ({\it rms}$<$0.15\arcsec) allows to place the
  aperture correctly, recovering the flux of all sources. For the IRAC
  bands, we use fixed circular apertures of 2\arcsec\ radius (blue)
  and we apply a correction to the total magnitude.}
\end{figure}

Some bands are deliberately excluded from the merged photometric
measurement described above due to a significant difference of the
resolution compared with the reference band or because images are not
available. For all these bands, the photometry is obtained differently
and later incorporated in the merged catalog during the
cross-correlation procedure. For the four IRAC channels, the
optical/NIR Kron apertures are typically not large enough to enclose
the entire PSF. Therefore, we keep the values measured in
\S\ref{irac_catalog} with aperture photometry. Given the comparatively
large PSFs of the MIPS (24~$\mu m$, 70~$\mu m$) and GALEX (FUV, NUV)
images, the flux measurement in these bands was carried with
IRAF-DAOPHOT and SExtractor, respectively (see PG05 and
PG08). Figure~\ref{aperphot} illustrates the different resolutions and
aperture sizes involved in the photometric measurement (see also the
captions in
Figure~\ref{postages_navigator1},\ref{postages_navigator2},\ref{postages_navigator3}
and \ref{postages_navigator4}).

The \HST\,-ACS bands were included in the general photometric
procedure. Although the much higher resolution may lead to multiple
cross identifications even when compared with the optical ground based
images, the high spatial resolution (after the local re-alignment
described above) allows a reliable cross-matching within 0.15\arcsec\
({\it rms} of the local WCS solution for HST images).  As shown in
Figure~\ref{aperphot}, apertures were placed correctly even when
multiple counterparts are identified, and the photometry includes the
fluxes from all them.  We have conducted an additional test to check
the accuracy of the ACS photometry measured with this method. We
compared the observed photometry to synthetic magnitudes derived from
SED templates fitted to the multi-band photometry (except ACS) of
spectroscopically confirmed galaxies. We find a very small offset
($\leq0.02$~mag) and a scatter consistent with the typical photometric
errors in the bands (see \S3.1.3 of Paper II for more details).

The NIR source catalog of the Palomar/WIRC survey
\citep{2006ApJ...651..120B} is also included in the merged photometric
catalog, although no images are available to match apertures. In this
case, we use the SExtractor MAG\_AUTO value of the closest neighbor in
the $J$- and $K$-bands.

\begin{figure*}
\includegraphics[width=8.7cm,angle=0.]{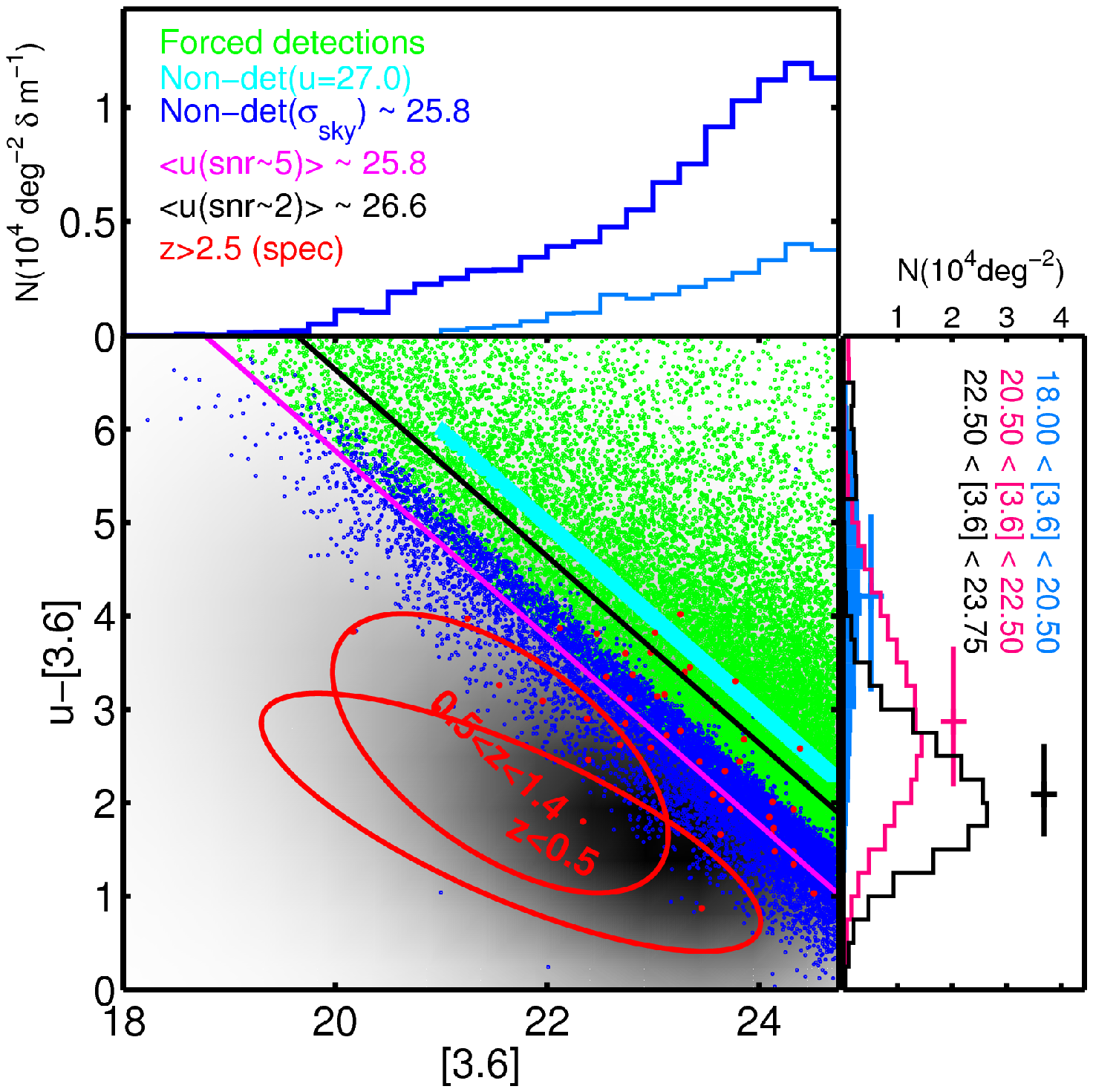}
\hspace{0.2cm}
\includegraphics[width=8.7cm,angle=0.]{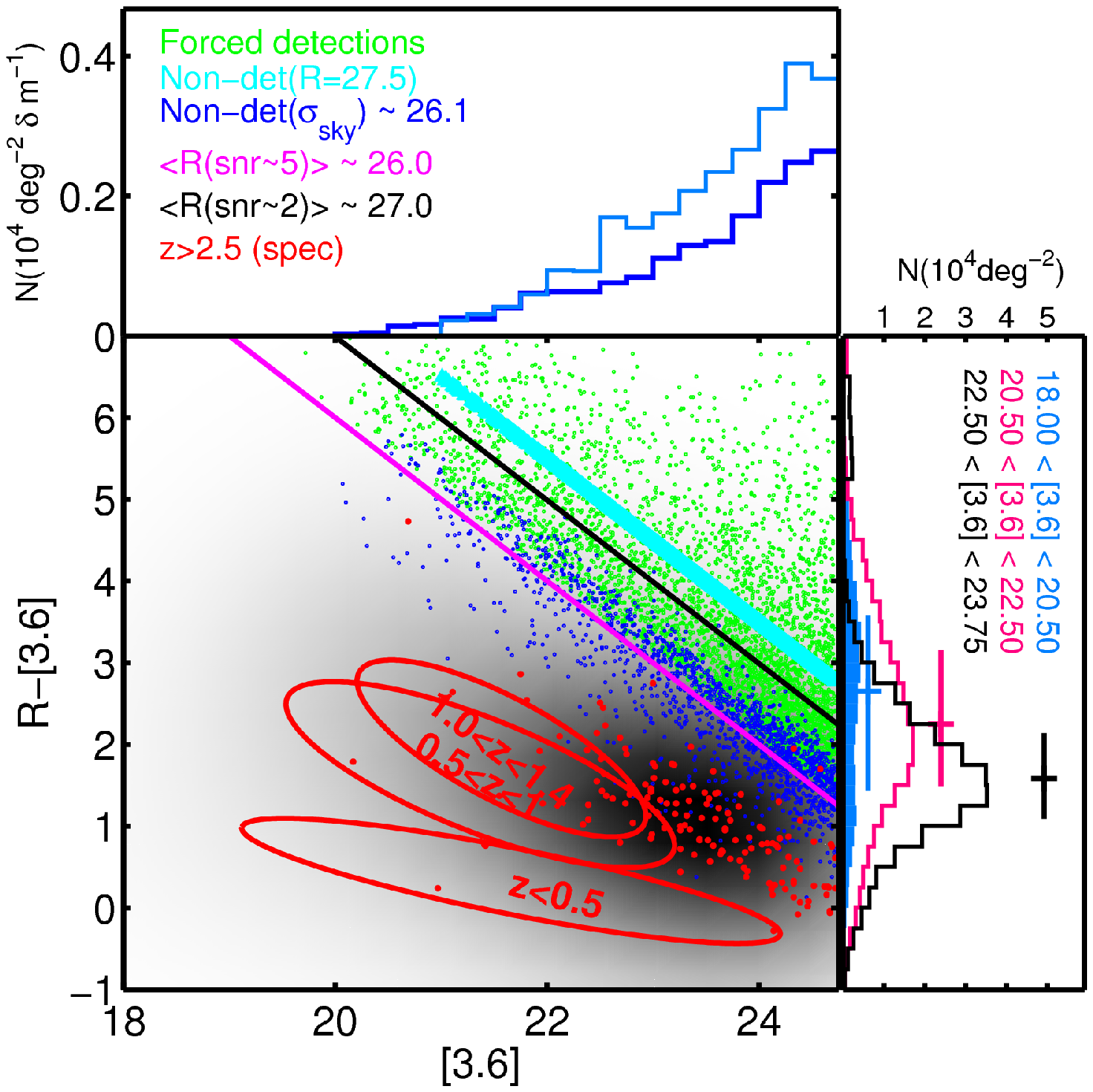}\\
\includegraphics[width=8.7cm,angle=0.]{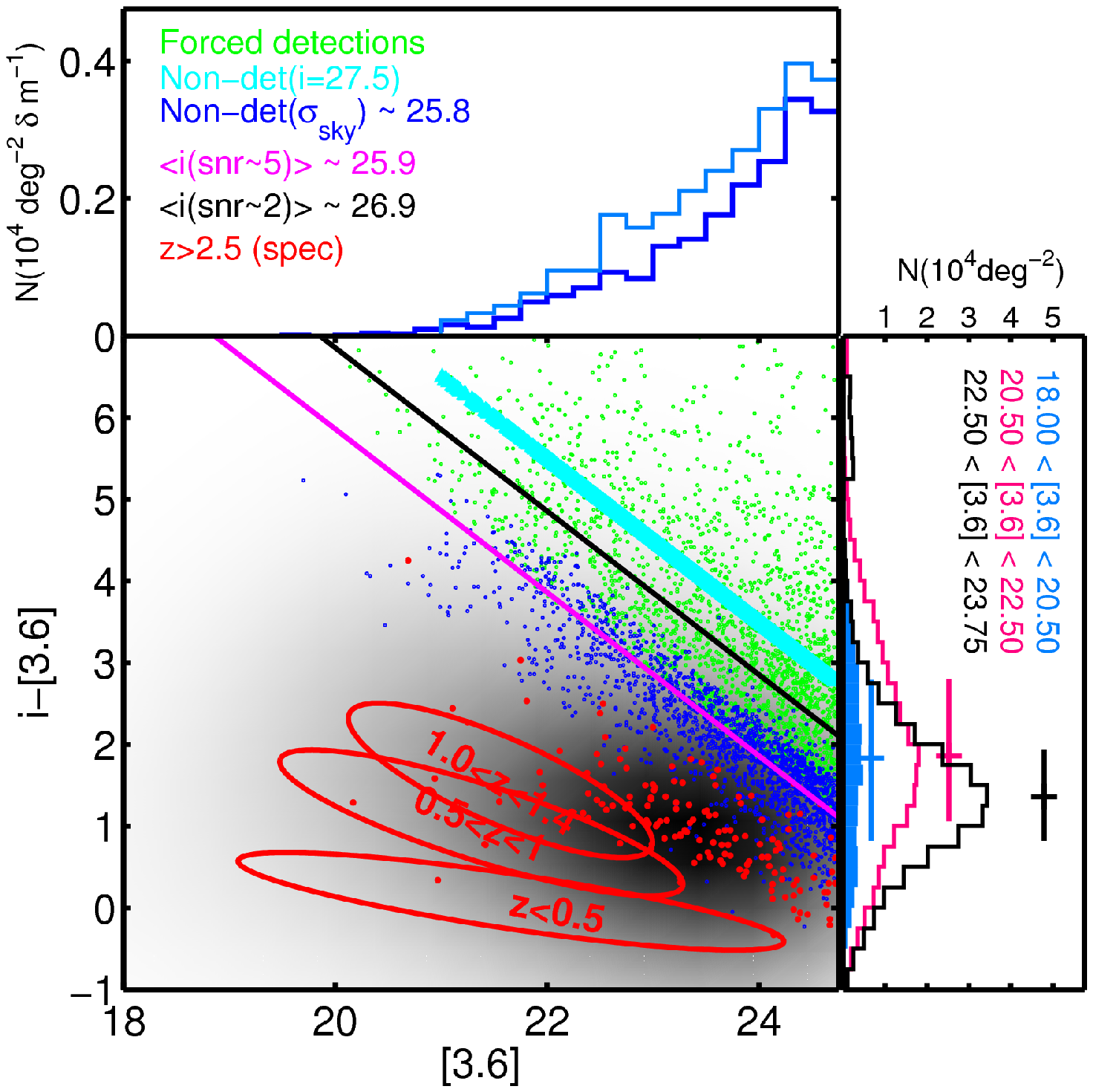}
\hspace{0.2cm}
\includegraphics[width=8.7cm,angle=0.]{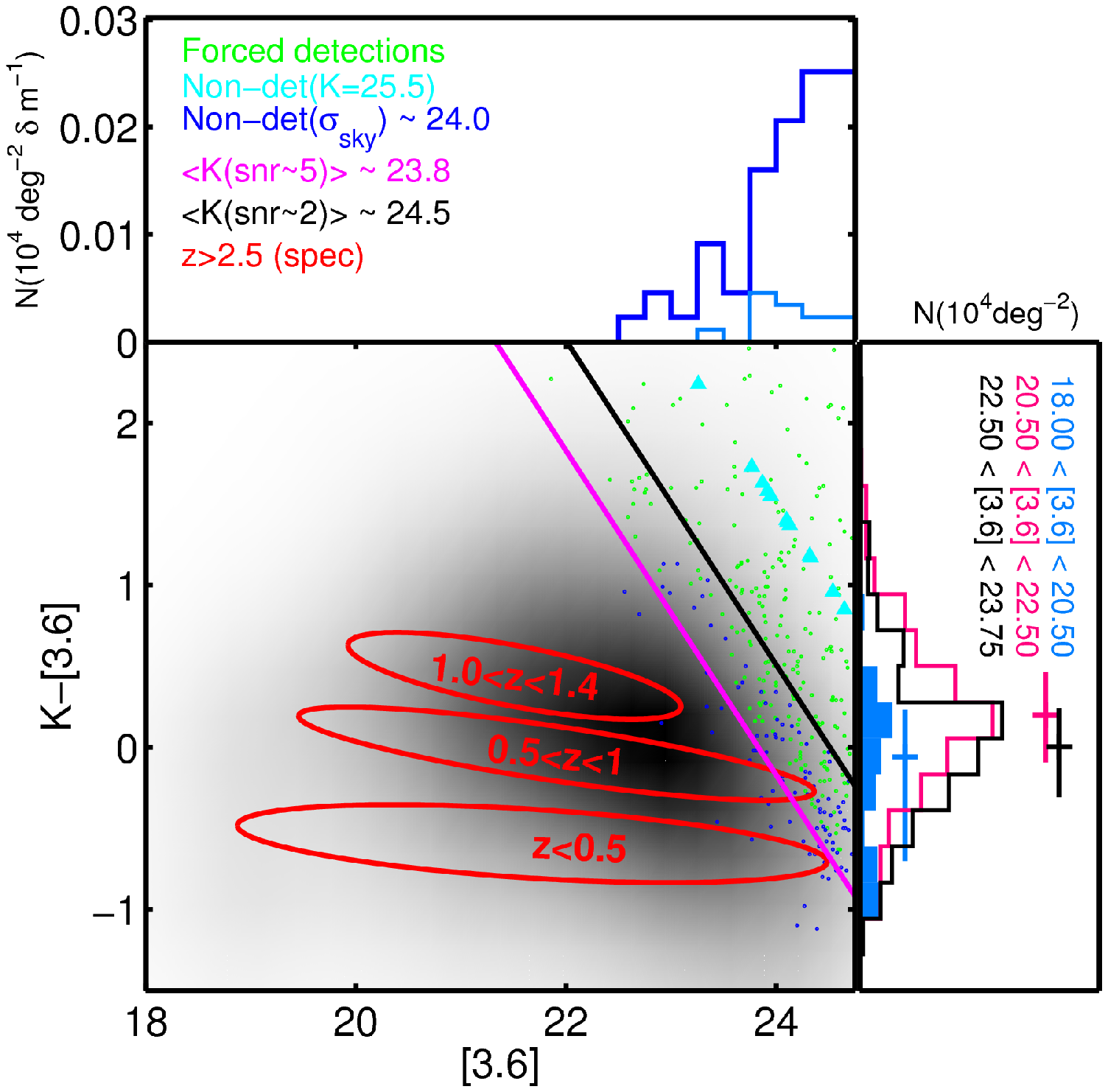}
\caption{\label{colorprops} The central panel in each plot shows the
  color magnitude diagrams in the $u^{*}$, R, $i'$ and $K$ bands with
  respect to [3.6] (top to bottom, left to right). The density map in
  grey scale shows the distribution of IRAC sources detected in each
  band.  For sources missing (not detected by Sextractor) in a given
  band but detected in any other, we force a flux measurement at the
  position of the existing source using the same Kron aperture. The
  green dots depict sources for which we are able to recover a
  positive flux in this forced measurement. The blue dots depict the
  value of the sky {\it rms} in the apertures where the forced
  measurement failed (i.e., the integrated flux was negative). The
  cyan triangles are sources undetected in any other band but
  IRAC. For these sources, we set an upper limiting magnitude
  $\sim$0.5-1~mag fainter than the typical magnitude for a source with
  SNR$=$2 (black line). The magenta solid line indicates the median
  magnitude of the sources with SNR$\sim$5. The red dots depict
  galaxies with spectroscopic redshift z$>$2.5. The red ellipses show
  the median and quartiles of the color-magnitude distribution for
  galaxies in different bins of redshift within 0$<$z$<$1.5.  The
  upper panel of each quadrant shows the [3.6] brightness distribution
  of undetected sources (blue and cyan markers in the central panel)
  in each band.  The right panel of each quadrant shows the color
  distribution of detected sources (histogram, median, and quartiles)
  in three bins of the [3.6] magnitude.}
\end{figure*}

\subsection{Photometric uncertainties}\label{merged_errors}

The uncertainties in the photometry are derived simultaneously to the
process of flux measurement in each individual band. As mentioned in
\S\ref{irac_catalog}, the photometric errors obtained using SExtractor
often underestimate the true background noise due to signal
correlation on adjacent pixels (\citealt{2003AJ....125.1107L},
\citealt{2006ApJS..162....1G}). In order to properly account for this
effect, we estimated the flux uncertainties in three different ways,
as described in PG08 (Appendix A). First, we measured the background
noise in a circular corona of 5~\arcsec\ width around each
source. This procedure is similar to that used by SExtractor and
provides uncertainties $\sigma_{\mathrm{AP}}\propto
N_{pixels}^{1/2}$. In addition, we measured the average sky signal on
non-connected artificial apertures built with random pixels around
each source, with the same size as the photometric aperture, and
containing only \textit{pure} sky pixels (rejecting pixels
$>$5$\sigma$ the {\it rms} obtained with the first method). Finally,
we also estimated the background noise following the method by
\citet{2003AJ....125.1107L}. The flux measured on several photometric
apertures around the source, identical to the one employed for the
photometry, is fitted to a Gaussian function to yield the {\it rms}
background fluctuation. The sky background is set to the resulting
average value of the three methods, and the final photometric
uncertainty is set to the largest estimate, typically, the one
measured with the second method.

\section{Properties of the merged photometric catalog}
\label{merged_properties}

The exposure time of the IRAC mosaics in the four channels is
remarkably homogeneous across the mosaic. However, the coverage of the
strip at other wavelengths is patchy and discontinuous (see
Figure~\ref{layout}). For the remaining of this Section and in Paper
II, we will differentiate between the region of the IRAC image covered
by the CFHTLS, and the rest. The CFHTLS/IRAC common region
(52.16$^{\circ}<\delta<$53.20$^{\circ}$ \&
214.04$^{\circ}$$<\alpha<$215.74$^{\circ}$) has been also surveyed
with Subaru, the CFHT12k instrument, MMT/Megacam, \HST\,/ACS,
\HST\,/NICMOS, and GALEX , so it constitutes the zone with the highest
data quality (hereafter, main region), where the SEDs are sampled with
the highest band coverage. The area of this region is 0.35~deg$^{2}$
($\sim$68\% of the total) and contains 53,030 (76,936) sources down to
[3.6]$<$23.75 (24.75). The 0.13~deg$^{2}$ outside the main region
(hereafter, flanking regions), also present a solid SED coverage,
relying mostly in the MMT and SUBARU bands. Nevertheless, it lacks
some of the best quality data (taken with ACS and MOIRCS), and the
fraction of surveyed area by the WIRC-$JK$ bands is $\sim$15\%
lower. There are 23,906 (35,416) sources in the flanking regions with
[3.6]$<$23.75 (24.75). Table~\ref{sample_summary} summarizes the
  number of sources in the catalog according to different limiting
  magnitudes and area constraints.

\placetable{summary}
\begin{deluxetable}{ccc}
\centering
\tabletypesize{\tiny}
\setlength{\tabcolsep}{0.01in} 
\tablewidth{0pt}
\tablecaption{\label{sample_summary}Number of sources in the IRAC-3.6+4.5$\mu$m catalog}
\tablehead{\colhead{} &  \colhead{[3.6]$\leq$23.75$^{a}$} & \colhead{[3.6]$\leq$24.75$^{b}$}}
\startdata
Prior to deblending       &     70,048  &  99,618\\
After deblending            &     76,936$^{\dagger_{1}}$  &  113,023$^{\dagger_{2}}$\\
Main$^{c}$/ Flanking regions         &    53,030/23,906  &  77,607/35,416
\enddata
\tablecomments{\\
$a$, 85\% completeness level.\\
$b$, 3$\sigma$ limiting magnitude. \\
$c$, The main region is defined as: 52.16$^{\circ}<\delta<$53.20$^{\circ}$ \&
214.04$^{\circ}$$<\alpha<$215.74$^{\circ}$. The Flanking regions consist of the remaining area. \\
$\dagger_{1}$, Only these sources have been included in the
final catalog presented in \S~\ref{dataacess}.\\
$\dagger_{2}$, A larger, although less complete, sample including all the sources down
to [3.6]$<$24.75 can be accessed through the web interface {\it Rainbow Navigator} (\S~\ref{rbnav}).}
\end{deluxetable}

Note that the bulk of the optical SED coverage is based on CFHTLS and
MMT data (the CFHT12k-BRI data are $\sim$2~mag shallower). On average
numbers, their respective filter sets cover a similar wavelength range
with comparable data quality (see Tables~\ref{multiwav_data} and
~\ref{efficiency}). Also, the subtle differences in the shape of the
$u$ and $z$ band filters of each survey improve the quality of the SED
coverage for the sources in the main region. The most remarkable
difference between the two surveys is found in the homogeneity of data;
whereas the CFHTLS present a nearly uniform data quality across the
mosaic, the MMT data show larger variations between the four pointings
(e.g., $\Delta$g(5$\sigma$)=26-25~mag, $\Delta$FWHM$=1.0-1.6$\arcsec\
; see \citealt{2009RAA.....9.1061Z} for more details). Unfortunately,
the lowest quality pointings (shallower data and highest seeing) are
precisely those covering the flanking regions (mainly the north
region, $\delta >$53.20$^{\circ}$). Furthermore, a small area in the
north and south regions of the MMT mosaic ($<$10\% of the total) is
not observed in the g and u bands, respectively.  Thus, comparatively,
the average data quality in the flanking regions is slightly lower
than in the main region.

\begin{figure*}
\includegraphics[width=9cm,angle=0.]{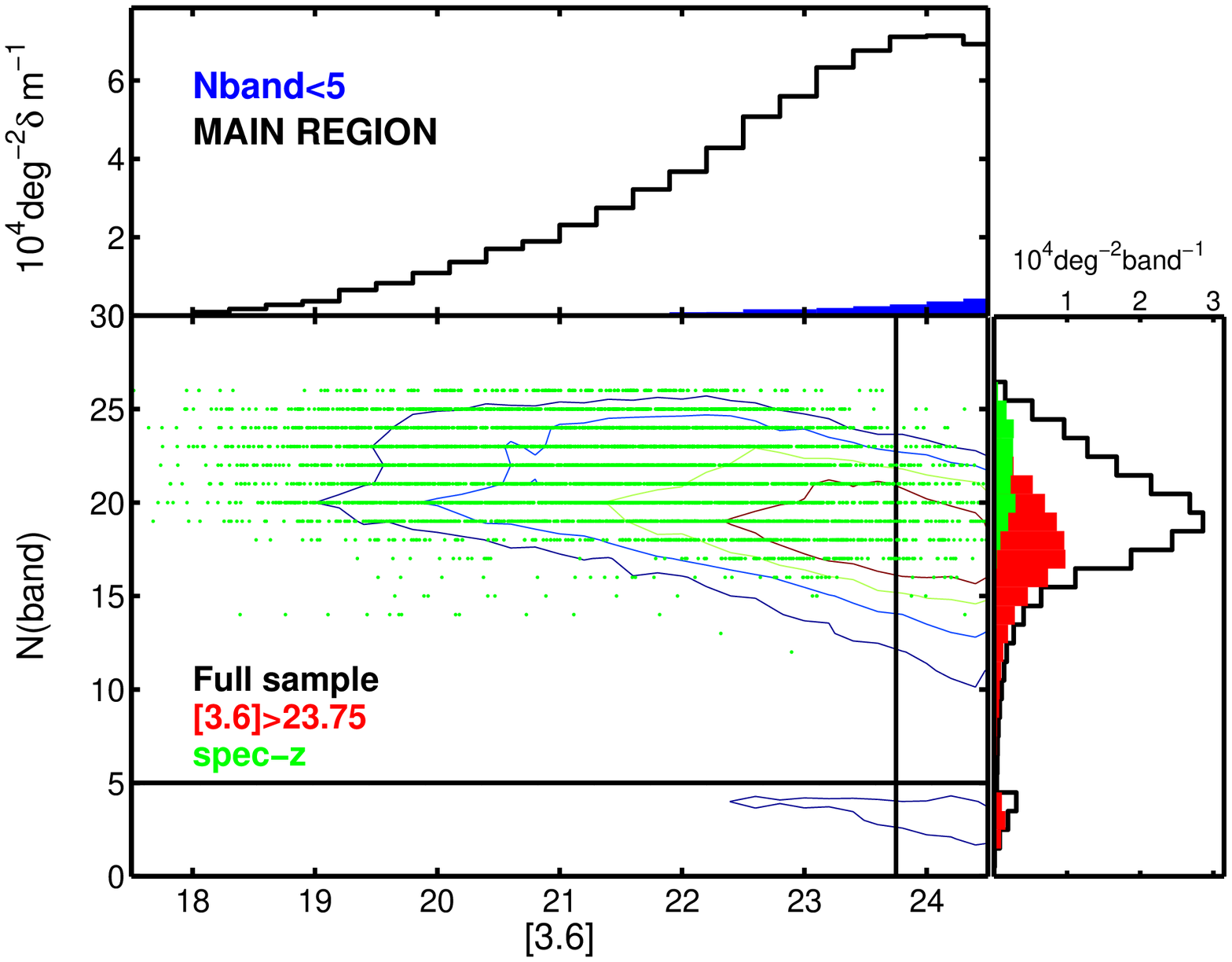}
\includegraphics[width=9cm,angle=0.]{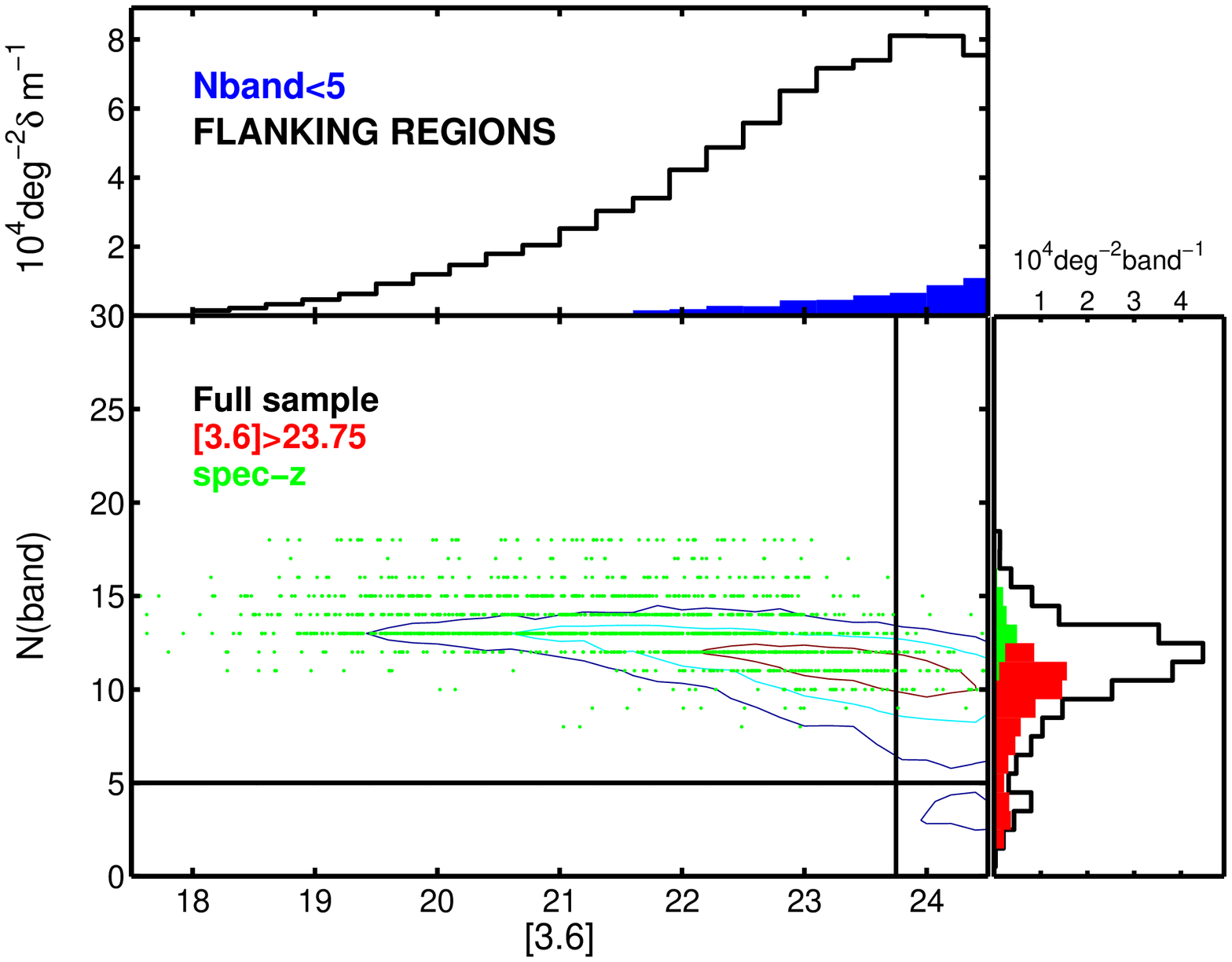}
\caption{\label{nbands_region} The central panel in each plot depicts
  the source density as function of the number of bands in which the
  source was detected (a positive flux was measured), N(band), and the
  magnitude in the [3.6] channel, for the main region (left) and the
  flanking regions (right). The color contours contain (from the
  inside out) approximately 25\%, 50\%, 75\% and 90\% of the
  sources. The green dots mark galaxies with a spectroscopic
  redshift. The black lines indicate the 85\% completeness limit of
  the catalog ([3.6]$<$23.75, vertical) and the band coverage limit
  N(band)$=$5 (horizontal), respectively. The histograms on top show
  the brightness distribution for the full sample (black) and for
  sources with N(band)$<$5 (blue).  The histograms to the right show
  the N(band) distribution of the full sample (black line), IRAC faint
  sources ([3.6]$>$23.75; red area) and galaxies with spectroscopic
  redshifts (green area).}
\end{figure*}

\subsection{Multi-band detection and color properties.}

Table \ref{efficiency} shows the fraction of the entire IRAC catalog
detected in the different bands compiled for this work. We give
specific values for different sub-samples divided as a function of
brightness in the selection band, detection level (SNR), and location
in the IRAC mosaic. The majority of sources are detected in the 4 IRAC
channels. However, while most sources ($>$90\%) have reliable
detections at 3.6 and 4.5~$\mu$m, only half of the sample is detected
with SNR$>$5 at the two longer wavelengths.  Note that the photometry
at 5.8 and 8.0~$\mu$m was not performed independently (see
\S\ref{irac_catalog}).

The fraction of IRAC sources detected in the optical bands is also
elevated ($\sim$95\%) for the majority of the bands. However, cutting
the measurements at SNR$>$5, the efficiency decreases to $\sim$85\%
for the deepest CFHTLS, MMT, Subaru and ACS bands, and to 50\%-60\%
for the $u$ and $z$ bands (in both CFHTLS and MMT data) and the
shallower CFHT $BRI$ bands. For the GALEX data, the detection is
typically lower than 10\% and 25\% in the FUV and NUV bands,
respectively. Around 12\% and 6\% of the sources are matched to high
quality spectroscopic redshift estimates in the main and flanking
regions, respectively. The lower efficiency in the flanking regions is
caused by the inhomogeneous DEEP2 coverage of the IRAC mosaic.

The NIR coverage of the strip surveyed by IRAC is also discontinuous.
The WIRC catalog provides the most uniform coverage, including 40\%
and $\sim$100\% of the area in the main region in the J and K bands,
respectively. The fraction of IRAC sources detected in each band is
very similar, $\sim$50-60\% ($\sim$20\% with SNR$>$5).  The deepest
NIR observations are those taken in CAHA-$J$ and Subaru-MOIRCS-$K_{s}$
bands. The latter covers $\sim$1/4 of the main region up to 1~mag
deeper than the WIRC-$K$ data, presenting a much higher detection
fraction, $\sim$90\% (70\% with SNR$>$5). The CAHA-$J$ data covers
40\% of the main region recovering $\sim$30\% (SNR$>$5) of the IRAC
detections, a slightly larger fraction than WIRC-$J$.

We note that for most bands, the fraction of sources detected with
SNR$<$5 can be significantly higher than the overall value. In the
shallowest bands, this is caused by the higher photometric
uncertainties (e.g., the WIRC J and K bands). However, for most bands
this is the result of the enhanced detection achieved with the forced
photometric measurement.  For each IRAC source undetected in a given
image, but having a counterpart in any other band, we still measure
the flux in the same aperture at the position of the counterpart.
With this method we increase the detection efficiency at fainter
magnitudes recovering sources that would be missed otherwise. However,
given the extreme faintness of these sources, some of the measurements
produce low significance detections (SNR$<$2).  As a precaution, to
preserve the overall quality of the SED, we do not include these
values (photometric uncertainties $>$0.4~mag) in the SED fitting
procedure, nor in the estimate of the physical parameters in Paper II.
Nevertheless, we keep these values as they can be useful as upper
limits. Another estimate of the limiting magnitude in each band is
obtained from the the value of the sky {\it rms} in failed forced
measurements (i.e., those for which the measured flux in the aperture
is negative).

Figure~\ref{colorprops} illustrates the characteristics of the
different flux measurements in four bands probing representative
wavelengths in the UV-to-NIR range. The central panel of each plot
shows a color-magnitude diagram in [3.6] vs $u^{*}$R$i'$ and K,
respectively. The grey-scale density map depicts the distribution of
{\it standard} (non-forced) photometric measurements, typically
detected with SNR$>$5 (magenta line).  Forced detections for which we
obtain a valid flux or a sky {\it rms} value are shown as green and
blue dots, respectively.  The cyan markers are sources detected in
IRAC only (for which we assign arbitrary magnitudes in each band).
For the deep R and $i'$ bands, which present the highest detection
efficiency, forced measurements account for less than 5\% of the flux
measurements, and present a median SNR$\lesssim$2 (black line).  In
the $u^{*}$ band, the source density for a similar limiting magnitude
is much lower, and consequently the fraction of forced detections is
higher, around 20\%.

The upper panels in each quadrant of Figure~\ref{colorprops} shows the
fraction of undetected sources in each band. In the $R$ and $i'$ bands
these sources make up for $<$4\% of the total sample. Most of them are
IRAC-only sources (plotted in cyan), whereas the rest are at least
marginally detected in one other (typically red) band, but the forced
measurement fails for the particular band shown in the plot (blue).
The fraction of undetected sources in the $u^{*}$ band (and other
shallower optical bands, e.g., CHFT12K-BRI) increases to 10\%, being
in this case dominated by failed forced measurements at the positions
of R or $i'$ detections. These kind of measurement are also
predominant for undetected sources in the $K$-band.

Note that some of the IRAC-only sources are relatively bright,
21$<$[3.6]$<$22. Most of them are detected in deep $K$-band
observations. However, there are a few galaxies detected only in the
IRAC bands (typically the faintest ones). The nature of these
interesting sources, potential candidates to massive high redshift
galaxies, will be explored in a future paper.

Figure~\ref{colorprops} also shows the color distribution of IRAC
sources in bins of [3.6]~mag (right panels in each quadrant), and
spectroscopically confirmed galaxies in several redshift bins (red
ellipses and dots in the central panels). The general trend with color
in the R and $i'$ bands is that faint IRAC sources are on average
bluer than brighter ones. The median colors for the faintest [3.6]
bin, 22.50$<$[3.6]$<$23.75, are R$-$[3.6]$=$1.6 and
$i'$$-$[3.6]$=$1.4.  These colors are similar to those of galaxies at
1$<$z$<$1.5. This is consistent with the fact that the median redshift
for the IRAC-3.6+4.5~$\mu$m magnitude limited sample ([3.6]$<$23.75)
is z$\sim$1 (See Paper II and PG08). Note also that most of the forced
detections in the R-band would qualify as IRAC extremely red objects
(by the criteria of \citealt{2004ApJ...616...63Y}, $R$-[3.6]$>$4.0),
that targets dusty starbursts or passively evolving galaxies at
z$>$1.5.

In general, all galaxies tend to become redder in optical$-$IRAC
colors with increasing redshifts. Indeed, for a typical galaxy SED,
the observed optical bands shift into the (fainter) UV whereas the
[3.6]~mag becomes brighter as it approaches the stellar bump at
1.6~$\mu$m rest-frame (for z$\lesssim$1.2). Therefore, it is not
surprising that the $u^{*}$$-$[3.6] median color is redder
($u^{*}$$-$[3.6]$=$2.1, typical of a z$\sim$1 galaxy) than the median
color involving the $Ri'K$ bands.  Note also that the observed UV
progressively shifts into the Lyman break, producing the large
fraction of z$>$2.5 (red dots in the top-left quadrant of
Figure~\ref{colorprops}) galaxies that are u-dropouts.  Finally, we
find that the $K$$-$[3.6] color presents an almost constant value as a
function of [3.6]~mag for a given redshift range, becoming
progressively redder as we move to higher redshifts: $K$$-$[3.6]$<$0
at z$<$0.5 and $K$$-$[3.6]$>$0 at z$>$1. Again, this is consistent
with the fact that both $K$ and [3.6] transit through the peak of the
stellar bump for z$<$1, changing their relative positions (see, e.g.,
\citealt{2004ApJS..154...44H}). An interesting consequence of the red
$K$$-$[3.6] colors of high-z galaxies is that IRAC-3.6~$\mu$m
observations are equivalent, in terms of source densities, to a
$K$-selected sample down to slightly deeper limiting magnitudes.

Figure~\ref{nbands_region} shows the density map (central panel) and
histograms (right panel) of the number of photometric bands with
measured fluxes, N(band), in the main (left) and flanking regions
(right). Typically, the SED of a galaxy in the main region has a
median of 19 photometric data points. The average spectral coverage is
$\sim$8 bands larger than in the flanking regions, mainly due to the
lack of data from the CFHTLS and NIR surveys. Galaxies with an
available spectroscopic redshift ($R$$\lesssim$24~mag; green dots)
typically present the highest band coverage ($\sim$22 and $\sim$14
bands in the main and flanking regions, respectively), since they are
relatively bright in the optical. Interestingly, the faintest IRAC
sources, [3.6]$>$23.75~mag (red histogram), present a relatively high
band coverage, N(band)$=$17 in the main region. In most cases, these
are low significance detections in the deepest optical/NIR bands where
the forced photometric measurement (see \S~\ref{rainbow_photometry})
is able to recover a flux (see, e.g.,
Figure~\ref{postages_navigator1}). On the opposite side, there is a
non-negligible number of IRAC-faint but optically-bright sources with
more than 19 photometric data points.  These constitute a population
of blue dwarf galaxies at intermediate redshift easily detected in the
optical but with faint IRAC counterparts (i.e., not very massive).

The upper panel of Figure~\ref{nbands_region} shows the [3.6]
magnitude distribution for the full sample (black) and for sources
with poor spectral coverage (N(band)$<$5; blue histogram). The latter
defines a clearly isolated group in the density contours (central
panel) and in the histograms (right panel) of
Figure~\ref{nbands_region}. These sources represent less than 3\%
(4\%) of the sample up to [3.6]$<$23.75 (24.75). Most of them are
clear IRAC-only detections (as the ones discussed in
Figure~\ref{colorprops}) relatively bright in [3.6,4.5]. However, some
of the faintest sources can be affected by contamination of spurious
sources (we give more details on these sources in
\S~\ref{reliability_sect}).

\subsection{FIR, X-Ray and radio counterparts}
\label{xray_radio}

\placetable{other_bands}
\begin{deluxetable}{cccc}
\tabletypesize{\tiny}
\tablewidth{0pt}
\tablecaption{\label{other_bands} Detection efficiency of X-ray, FIR and Radio sources}
\tablehead{\colhead{Source} & \colhead{Area} & \colhead{Fraction} &\colhead{Total}\\
(1)&(2)&(3)&(4)
}
\startdata
{\it Chandra}/ACIS$^{\dagger_{1}}$ & 0.67~deg$^{2}$  &  0.70& 1023 (77\%)           \\
MIPS-24~$\mu$m (f$>$60$\mu$Jy)$^{\dagger_{2}}$     &    0.53~deg$^{2}$  &  0.59          &     10771\\
MIPS-70~$\mu$m (f$>$3500$\mu$Jy)$^{\dagger_{2}}$    &   0.50~deg$^{2}$ &  0.61          &      868\\
VLA 20~cm$^{\dagger_{3}}$ & 0.73~deg$^{2}$  & 0.46  &  590 (52\%)\\  
\enddata
\tablecomments{\\
$\dagger_{1}$ Sources drawn from \citet{2009ApJS..180..102L}. These include, 815 high reliability identifications (based on a previous match to the IRAC catalog of BAR08) and 176 previously un-identified sources.\\
$\dagger_{2}$ The area of the survey refer to the GTO+FIDEL observations that reduced for this paper.\\
$\dagger_{3}$ Sources drawn from \citet{2007ApJ...660L..77I}.\\
(1) Name of the band.
(2) Total area of the survey. 
(3) Fraction of the survey area overlapping with the IRAC mosaic.
(4) Number of sources with IRAC counterparts ([3.6]$<$23.75) and fraction of recovered sources from the whole catalog.}
\end{deluxetable}

Table \ref{efficiency} shows the fraction of IRAC sources in our
  sample ([3.6]$<$23.75) detected in the X-ray, FIR and radio surveys
within the overlapping area. About 30\% (20\% with SNR$>$5 detections)
of the IRAC sources are detected at 24~$\mu$m (SNR$=$5 reached at
$\sim$60~$\mu$Jy), 10\% (2\% with SNR$>$5) at 70~$\mu$m (SNR$=$5
reached at $\sim$3.5~mJy), 1\% in the X-rays and $\leq$1\% at
20~cm. For the radio and X-ray surveys, we cross-matched our sample to
the catalogs of \citet{2007ApJ...660L..77I} and
\citet{2009ApJS..180..102L}, respectively. As explained in
\S~\ref{merged_technique}, the cross-match to the radio catalog was
performed using a 3\arcsec\ radius, whereas for the X-ray catalog we
followed a two step identification. First, we divided the X-ray sample
attending to the availability of pre-identified IRAC counterparts
(Laird et al. used the IRAC catalog of BAR08 as reference). For the
galaxies identified in IRAC, we used a 1\arcsec\ cross-matching
radius, whereas for the other we used a more conservative
2\arcsec\ radius.  With this procedure we found 848 out of the 882
galaxies with IRAC counterparts in their catalog (815 among the 830
with high reliability flag; see \citealt{2009ApJS..180..102L} for more
details).  The remaining galaxies were either outside of the
N(frame)$>$20 region or too close to a bright star. In addition, we
were able to identify 175 additional X-ray sources in the area of our
mosaic not covered by the data of BAR08, i.e., we recover a total of
1023 out of 1325 X-ray sources from the catalog of Laird et al. For
the the MIPS-24\mic\ and 70\mic\ and radio surveys, we identify 10758,
868 and 590 (out of 1122) sources, respectively. Noticeably, about
20\% and 54\% of the sources detected in MIPS-24\mic\ and
70\mic\ present a multiple IRAC counterparts (typically 2-3) and
$\sim$6\% in X-ray and Radio.

All of these surveys cover an area larger than the IRAC observations,
but none of them fully cover the IRAC strip (although the X-ray and
MIPS-24\mic\ data cover $>$90\% of the area).  Table~\ref{other_bands}
shows the total area covered by the X-ray, MIPS and radio surveys, the
fraction that area in common with the IRAC mosaic, and the number of
sources with an IRAC counterpart in our catalog. All the sources in
the different surveys located within the area covered by IRAC are
detected in our catalog (with the exception of a few objects too close
to bright stars).

Note that despite the larger area of the VLA observations,
this survey covers only $\sim$40\% of the region surveyed by IRAC. In
fact the data are limited only to the upper region of the IRAC mosaic
($\delta$$>$53.10$^{\circ}$; see Figure~\ref{layout}).

\subsection{Catalog reliability}
\label{reliability_sect}

In this Section we analyze in more detail the sources with a poor
spectral coverage to asses the reliability of the IRAC catalog as a
function of the magnitude in the [3.6] band.

First we test the reliability with the widely-used method of comparing
the number of detections at faint levels in the original and a
negative image. This test reveals that only $\sim$1\% of the sources
up to [3.6]$<$23.75~mag are spurious.

However, this procedure is conceived to detect faint spurious sources
based on the assumption of a symmetric noise, whereas most of the
spurious detections are concentrated around bright regions as a
consequence of PSF and saturation artifacts and an excessive
deblending. Therefore, we chose to follow a different approach to
analyze the reliability of the catalog based on the fact that faint
[3.6, 4.5] detections lacking a measurable counterpart in other bands
are potential candidates for spurious detections.

As discussed in \S~\ref{merged_properties}, we find a nearly isolated
group of sources detected in N(band)$\leq$4. Most of these sources are
IRAC-only detections (cyan histogram of Figure~\ref{colorprops}) that,
in the absence of low significance optical detections to validate
them, could be artifacts of the source extraction procedure.

The visual inspection of galaxies with N(band)$\leq$4 reveals a clear
dichotomy. We find that these sources are either strongly clustered
around bright stars and extended sources (typically low-z galaxies),
or uniformly distributed across the image. Most of the sources in the
first group are easily identified as spurious, whereas for the
isolated sources, their high SNR ($\sim$5-10) and the visual
inspection seem to favor that they are real. We dealt with the
reliability of sources with N(band)$<$4 by considering spurious all
detections in a 20\arcsec\ -30\arcsec\ radius region around the
brightest objects ([3.6]$=$16-15~mag). Up to [3.6]$\leq$23.75,
approximately half the N(band)$\leq$4 sources are spurious by this
criteria, accounting for less than a 2\% of the total sample.
Nevertheless, even for the more conservative scenario, the sum of all
N(band)$\leq$4 detections constitutes less than $\sim$4\% of the
sample in the main region.  Finally, we also check the reliability of
these sources by studying how many of them are simultaneously detected
at both 3.6 and 4.5~$\mu$m.  Unfortunately, we find that some real
detections are missed at 4.5~$\mu$m due to the slightly worse image
quality, while some of the spurious sources are detected in both
images in regions where the density of artifacts is larger (i.e.,
around bright stars).

Summarizing, we conclude that the overall reliability of the catalog
is very high ($>$97\%) and that the contamination by spurious sources
is strictly restricted to the surrounding areas of very bright
sources. For the sake of completeness, we do not remove the
  sources with N(band)$\leq$4 from the sample. Instead, we include in
  the data catalog (Table~\ref{photprop}) both the number of bands in
  which the source was detected (N(band)) and a flag indicating if the
  source is located in the vicinity of a bright object (see
  \S\ref{photpropsec}).

\subsection{Star-galaxy separation}
\label{STAR}

\begin{figure*}
\centering
\includegraphics[width=8.2cm,angle=0.]{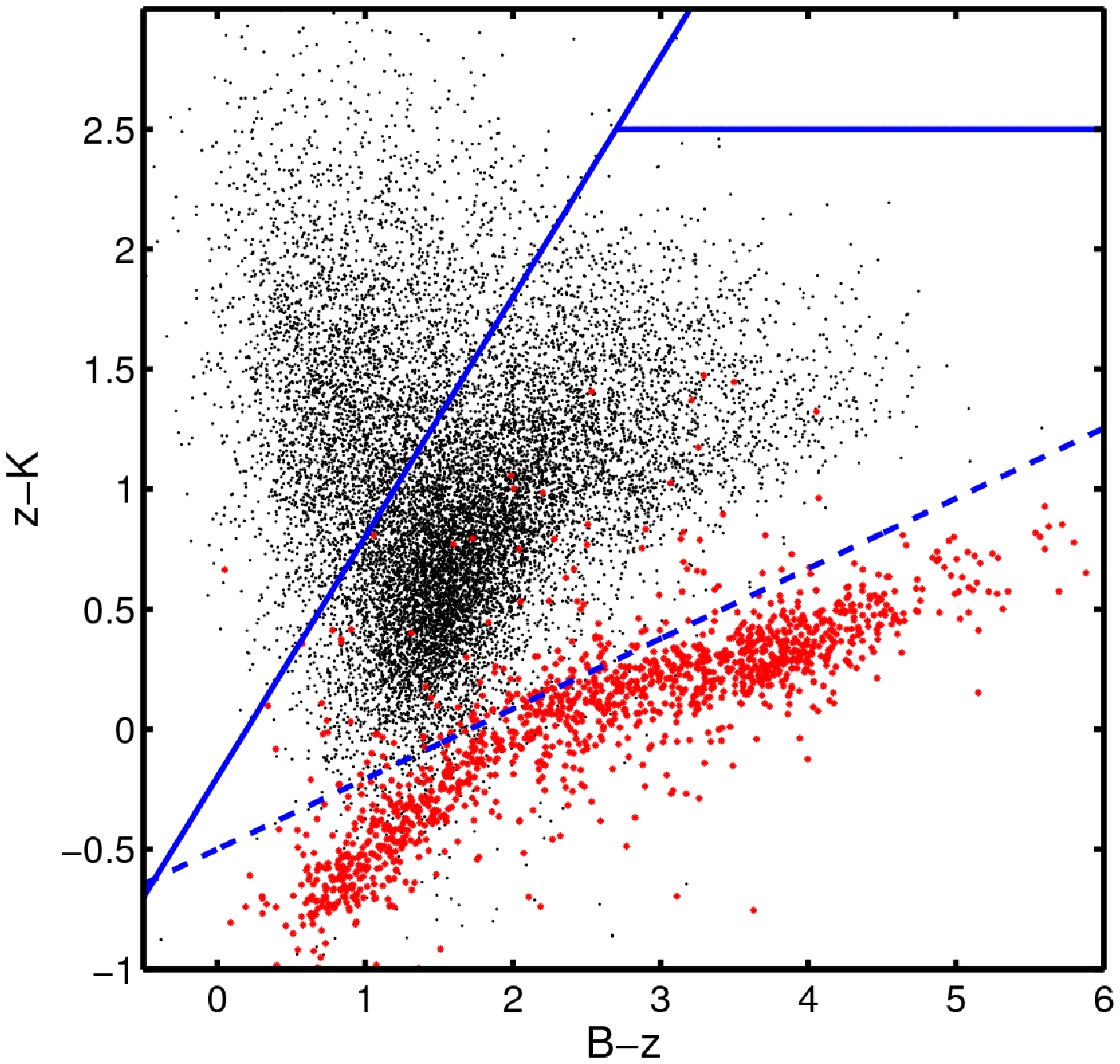}
\hspace{0.8cm}
\includegraphics[width=8.2cm,angle=0.]{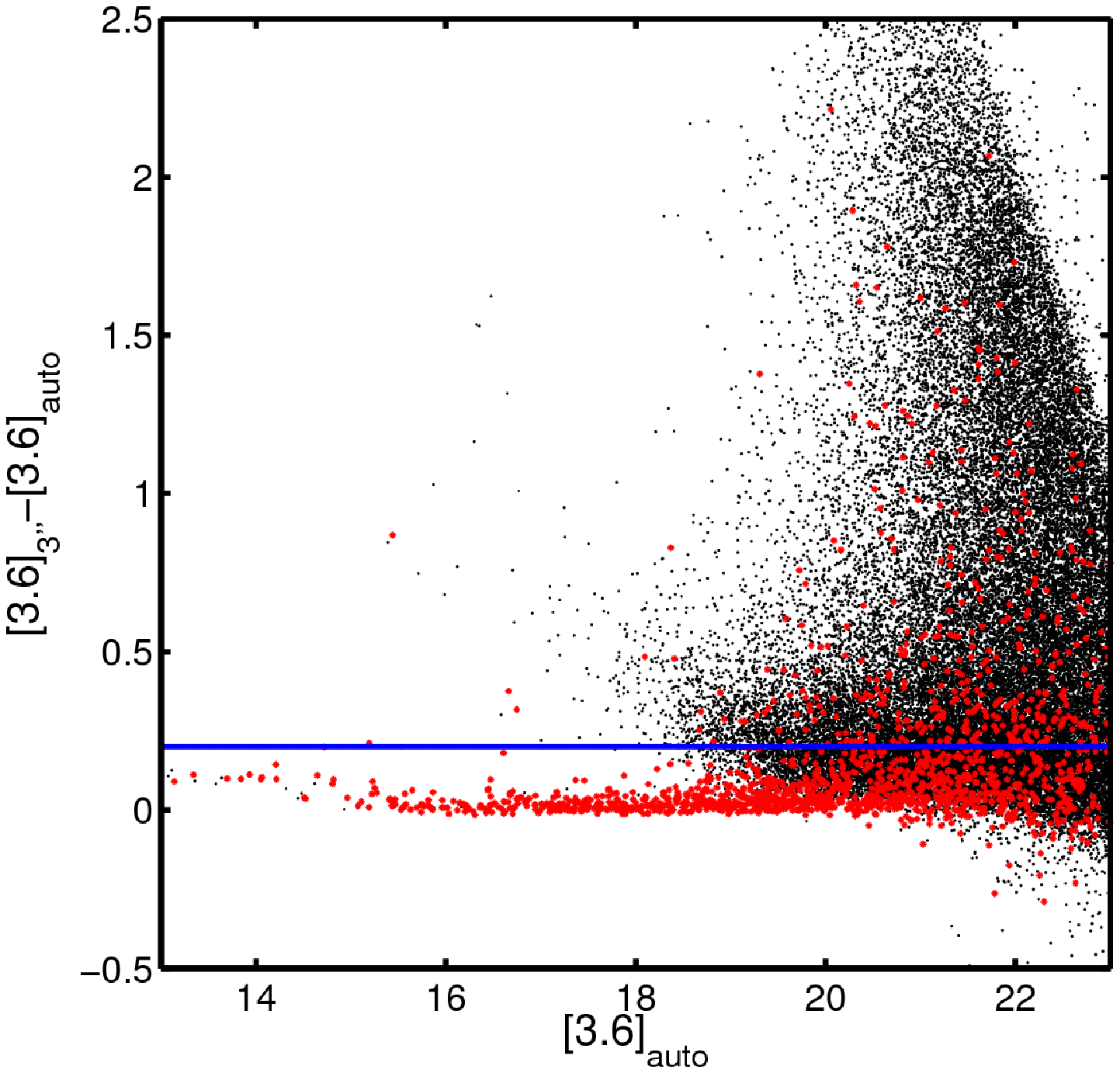}
\caption{\label{stellarity} Example of two of the stellarity criteria
  used in this paper to separate galaxies from stars.{\it Left}:
  $B$-$z$ versus $z$-$K$ for all sources in the catalog. The dashed
  line depicts the {\it BzK} criteria of \cite{2004ApJ...617..746D} to
  isolate stars. The solid lines delimit the regions where
  high-redshift galaxies are located. The red dots indicate sources
  satisfying 3 or more stellarity criteria. {\it Right}: Comparison of
  the magnitude in a 3\arcsec\ radius aperture ([3.6]$_{3\arcsec}$)
  and SExtractor MagAuto ([3.6]$_{\mathrm{auto}}$), i.e, the
  concentration parameter, versus [3.6]$_{\mathrm{auto}}$ for all
  sources in the catalog. The criterion of \cite{2004ApJS..154...48E}
  identifies stars with
  -0.25$<$[3.6]$_{3\arcsec}$-[3.6]$_{\mathrm{auto}}$$<$0.2 and
  [3.6]$_{\mathrm{auto}}$$<$17.80.}
\end{figure*}

Following PG08, we used eight different photometric and morphological
criteria to identify stars in the merged photometric catalog: (1) the
average of SExtractor star/galaxy separation parameter (stellarity)
for all bands where the source is detected; (2) when available, the
stellarity parameter and FWHM of the source in the {\it HST} images is
also considered as an independent, more reliable, criterion. (3,4) the
IRAC-based color-magnitude criteria criteria of
\cite{2004ApJS..154...48E} and \cite{2004ApJS..154...97B}. (5) the
concentration parameter (i.e., the difference of the [3.6]~mag
measured in a 3\arcsec\ radius aperture and the SExtractor MagAuto),
(6) the color-color and (7) color-magnitude criteria of \citet[][this
  time based on optical and NIR bands]{2004ApJS..154...48E}. (8) The
{\it BzK} criteria of \cite{2004ApJ...617..746D}.  In spite of using
multiple stellarity criteria, the heterogeneous band coverage makes
difficult the simultaneous application of the 8 criteria (e.g., the
{\it HST} data cover only $\sim$40\% of the full area). For this
reason, we chose the bulk of the criteria to be based on IRAC or
optical colors, which are available for the majority of the sources
independently of the region in which they are located. We verify that
at least 5 criteria can be estimated for 86\% of the sources with
[3.6]$<$23.75.

A source was identified as a star if 3 or more of the stellarity
criteria are satisfied.  Based on this method, we found 2913 stars.
Among them, we are able to identify 69$\%$ of the spectroscopically
confirmed stars. The rest of them satisfy at least 1 or 2 of the
stellarity criteria.  Figure~\ref{stellarity} illustrates the
efficiency of the IRAC morphological criteria by
\cite{2004ApJS..154...48E} and the {\it BzK} criteria of
\cite{2004ApJ...617..746D}. Red dots indicate sources with
N(criteria)$\geq$3. Clearly, the {\it BzK} criterion is more efficient
identifying stars over a wider magnitude range. Note that, as
discussed in PG08, an IRAC selected catalog down to $[3.6]$$<$23.75
only includes a minor fraction of stars (less than 5\% at all
magnitudes), the majority of them at bright magnitudes ($\sim$40$\%$
at [3.6]$<$18).At the faintest magnitudes, the overall dimming of the
objects could lead us to identify some stars as galaxies due to the
lack of applicable criteria. However, we find that both the galactic
and stellar number counts in the IRAC bands are in good agreement with
those presented in \citet{2004ApJS..154...39F} down to our limiting
magnitude. From the relative distribution of galaxies and stars we
also find that the fraction of stars makes up for only 3$\%$ of all
sources at magnitudes fainter than the the median of the sample
([3.6]=22.4).

\section{Data catalogs and database access}
\label{dataacess}

The catalog with the multi-band photometry for all the
IRAC-3.6+4.5~$\mu$m selected sample in the EGS is presented in
Table~\ref{dataphot}. Additional information regarding the photometry
of the sources and other properties described in the paper are given
in Table~\ref{photprop}. Furthermore, we also present here a web-based
interface to access our database containing all the results presented
in this and the companion paper. The web tool is publicly available
for the entire astronomical community.

 As explained in \S~\ref{merged_completeness}, the catalogs
  presented in this paper are restricted to the 76,936 sources with
  [3.6]$<$23.75, the $\sim$85$\%$ completeness level, which count with
  accurate IRAC photometry (SNR$>$8). As explained in
  \S~\ref{mergedcat}, our sample is extracted from the area of the
  IRAC mosaic counting with a frame coverage larger than 20
  ($t_{exp}\sim$4ks in the IRAC-3.6~$\mu$m band). Nevertheless, a
  deeper, although less complete, catalog without any magnitude
  restriction can be accessed through the online database
  (\S\ref{rbnav}).  In the following subsections we describe the
contents of Tables~\ref{dataphot} and \ref{photprop} and we present
{\it Rainbow Navigator}.

\subsection{Table~\ref{dataphot}: photometric catalog}

These are the fields included in Table~\ref{dataphot}:

\begin{itemize}
\item {\it Object}: Unique object identifier starting with irac000001.
  Objects labeled with an underscore plus a number (e.g,
  irac000356\_1) are those identified as a single source in the IRAC
  catalog built with Sextractor, but deblended during the photometric
  measurement carried out with the {\it Rainbow} software (see
  \S~\ref{rainbow_crossmatch}). Note that, although the catalog
    contains 76,936 elements, the identifiers do not follow the
    sequence irac000001 to irac076185. This is because the catalog is
    extrated from a larger reference set by imposing coordinate and
    magnitude constraints. The table is sorted according to this unique
    identifier.
\item $\alpha,\delta$: J2000.0 right ascension and declination in
  degrees.
\item {\it zspec}: Spectroscopic redshift determination drawn from the
  DEEP2 spectroscopic survey or the catalog of LBGs of
  \citet{2003ApJ...592..728S}.
\item {\it qflag}: Spectroscopic redshift quality flag from 1 to
  4. Sources with {\it qflag}$>$3 have a redshift reliability larger
  than 80\%.
\item FUV, NUV, $u'$,...: Effective wavelengths (in nanometers),
  magnitudes and uncertainties (in the AB system) for each of the 30
  photometric bands compiled for this paper.  The band-passes are
  GALEX~FUV and NUV, CFHTLS~$u^{*}$$g'$$r'$$i'$$z'$, MMT~$u'giz$,
  CFHT12k~BRI, ACS~$V_{606}$ and $i_{814}$, Subaru~$R$,
  NICMOS~$J_{110}$,$H_{160}$, MOIRCS~$K_{s}$, WIRC~$JK$, CAHA-$JK_{s}$
  IRAC~3.6-8.0, and MIPS-24~$\mu$m and 70~$\mu$m. The bands are sorted
  according to the effective wavelength of the filter. We refer to
  section \S3 for details on the photometric measurement and error
  calculation. The magnitudes do not include zero-point corrections
  (see Paper II, \S~3.1.3).  A value of the magnitude equal to -99.0
  with error equal to 0.0 indicates a not valid photometric
  measurement.  A negative error indicates that the source was
  undetected by Sextractor but a positive flux was obtained when
  forcing the measurement at that position using the appropriate
  aperture (see \S~\ref{rainbow_photometry}).  An error equal to 0.0
  indicates that the source was undetected and the forced measurement
  returned a negative flux. In this case, the value of the magnitude
  is an upper limit equal to the sky-{\it rms} (1$\sigma$) in the
  photometric aperture (see \S~\ref{rainbow_photometry}).
\end{itemize}
\noindent

\subsection{Table~\ref{photprop}: photometric properties catalog}
\label{photpropsec}
These are the fields included in Table~\ref{photprop}:

\begin{itemize}
\item {\it Object}: Unique object identifier (the same as in the
  photometric catalog in Table~\ref{dataphot}).
\item $\alpha,\delta$: J2000.0 right ascension and declination in
  degrees.
\item {\it N(bands,detect)}: Number of UV-to-NIR in which the source is detected.
\item {\it N(bands,forced)}: Number of UV-to-NIR in which the source is a priori
undetected, but the forced photometry is able to recover a valid flux.
\item {\it Flag}: Quality flag indicating that the source is located
  in the vicinity of a bright object. Sources detected only in the
  IRAC bands (N(band)$<$5) and close to a bright ([3.6]$>$16) source
  are likely to be spurious (see \S~\ref{reliability_sect}). The
  values of the flag indicate: (5) source within 70\arcsec-100\arcsec
  of the brightess saturated stars in the field, (4) source within
  30\arcsec of a [3.6]$<$15 source, (3) source within 20\arcsec of a
  15$<$[3.6]$<$16 source, (2) source within 15\arcsec of a
  16$<$[3.6]$<$17 source, (1) source within 10\arcsec of a
  17$<$[3.6]$<$18 source. (0) source un-flagged.
\item {\it Stellarity}: Total number of stellarity criteria satisfied
  (see \S~\ref{STAR}).  A source is classified as a star if it
  satisfies 3 or more criteria.  (\S~\ref{STAR}).
\item {\it Region}: Region of the field in which the source is located
  (\S~\ref{merged_properties}): A value of 1 or 0 indicates that the
  source is in the Main or Flanking region, respectively. The Main
  region is defined as the area of the IRAC mosaic within
  52.16$<$$\delta$$<$53.20 and $\alpha$$>$214.04, the Flanking regions
  are those containing the remaining area.

\end{itemize}

\subsection{Rainbow Database $\&$ Navigator}
\label{rbnav}

The photometric catalog presented here and the inferred stellar
parameters discussed in Paper II are obtained using the {\it Rainbow}
software (see PG05, PG08). The program includes different sub-routines
for each task, and the output of each step serves as the input for the
next. After the data processing, both the input (images, spectra,
templates) and the resulting catalogs are stored in a database with
individual sources as building blocks.  On doing so, we achieve
several goals: (1) each object is fully characterized with all the
available information; (2) the data can be easily sorted and retrieved
according to several criteria; (3) the data are homogeneously stored,
which allows a straightforward combination with data from other {\it
  Rainbow} fields and projects (such as those in PG05 or PG08).

In order to provide worldwide access to the data stored in our
database, we have developed a publicly available web interface, dubbed
{\it Rainbow Navigator}\footnotemark[8]. Here we briefly describe the
main features of the utility. A detailed description of its
capabilities can be found at the website.
\footnotetext[8]{http://rainbowx.fis.ucm.es}

{\it Rainbow Navigator} is essentially a user-friendly web interface
to a database containing all the data products resulting from the
process of creating and analyzing the IRAC-selected catalog presented
in this two papers, from the initial source detection to the estimate
of the stellar parameters.

As many other astronomical query interfaces (e.g, NED, SIMBAD), {\it
  Rainbow Navigator} allows to retrieve the information for single
sources searching for the source name or by coordinates. It also
allows to create subsets of the complete catalog based on multiple
constraints over the multi-band photometry, the redshifts or the
stellar parameters. In addition, we have incorporated a cross-matching
tool that allows to compare catalogs uploaded by the user to the
IRAC-selected sample stored in the database, returning the sources in
common.  {\it Rainbow Navigator} also has an on-the-fly utility to
create sky maps of a selected area, including point-and-click access
to the individual sources.

Furthermore, an interesting feature that we do not include in the
public catalog for the sake of simplicity is the possibility of
retrieving observed and rest-frame synthetic magnitudes over a
predefined grid of 52 different filters covering the whole spectral
range from the UV to the radio wavelengths. These values are computed
by convolving the best fitting template (see Paper II) for each source
with the appropriate filter transmission curve.

Each source of the catalog has its own data sheets that provides all
the available information including not only the photometry, the
redshifts or the stellar parameters, but also detailed information of
the stellarity, synthetic magnitudes, the multiplicity in other bands,
jointly with the unique identification and coordinates of the
counterparts in each given band. The tool also provides postage stamps
of the source in each of the available bands, that can be modified
on-the-fly or combined to create simple false color images.
Furthermore, each page includes a figure depicting the full SED
showing the fit of the optical and IR data jointly with the best
fitting templates.

\begin{figure}
\centering
\includegraphics[width=14.5cm,angle=0.]{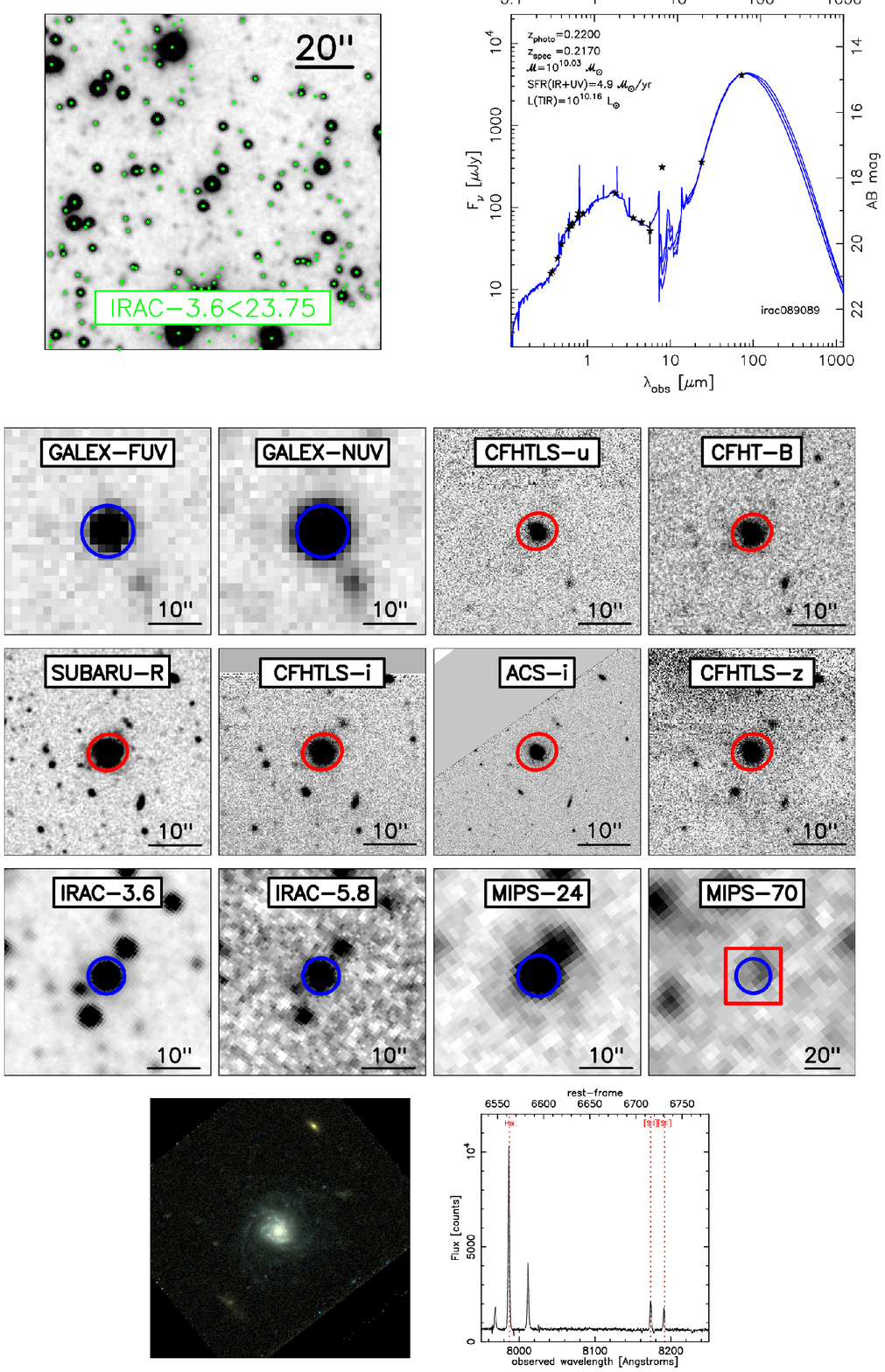}
\caption{\label{postages_navigator1} Example of the results available
  for each IRAC source in the {\it Rainbow} database, accessible
  through the {\it Rainbow Navigator} interface. This source is
    irac089089, a galaxy at z=0.2170. {\it Top-left}, Map
  (2\arcmin$\times$2\arcmin) of the sky area around the source. The
  green dots in the map depict sources in our IRAC catalog with
  [3.6]$<$23.75. By clicking on any of the sources, the interface
  presents an individual webpage with all the available information
  for that source. {\it Top-right}, Full UV-to-FIR SED of the central
  galaxy in the sky map, a source at z$=$0.21. The UV-to-NIR data is
  fitted to a stellar population model, while the IR-part of the SED
  is fitted to the models of CE01, DH02. A summary of the estimated
  stellar parameters, such us the stellar mass or the global SFR is
  shown in the upper-left corner of the Figure. {\it Middle panels}:
  grey-scale postage stamps (with size 40\arcsec$\times$40\arcsec,
  except for the MIPS-70 image, whose size is
  2\arcmin$\times$2\arcmin, same as the map in the top-left figure) of
  the galaxy in some of the available bands, covering different
  wavelength ranges. The Kron aperture used to measure consistent
  photometry in optical/NIR bands and the circular aperture used in
  bands with significantly lower resolution are shown in all panels
  (red and blue, respectively). {\it Bottom-left}: RGB color stamp
  obtained by combining images in the ACS-$V_{606},i_{814}$ bands. The
  {\it Rainbow Navigator} web interface allows to produce on-the-fly
  monochromatic and RGB images changing the cuts interactively.  {\it
    Bottom-right}: DEEP2 1-D spectra of the galaxy depicting some of
  the identified lines. The wavelength range and the redshift can be
  modified interactively.}
\end{figure}

\begin{figure*}
\centering
\includegraphics[width=16cm,angle=0.]{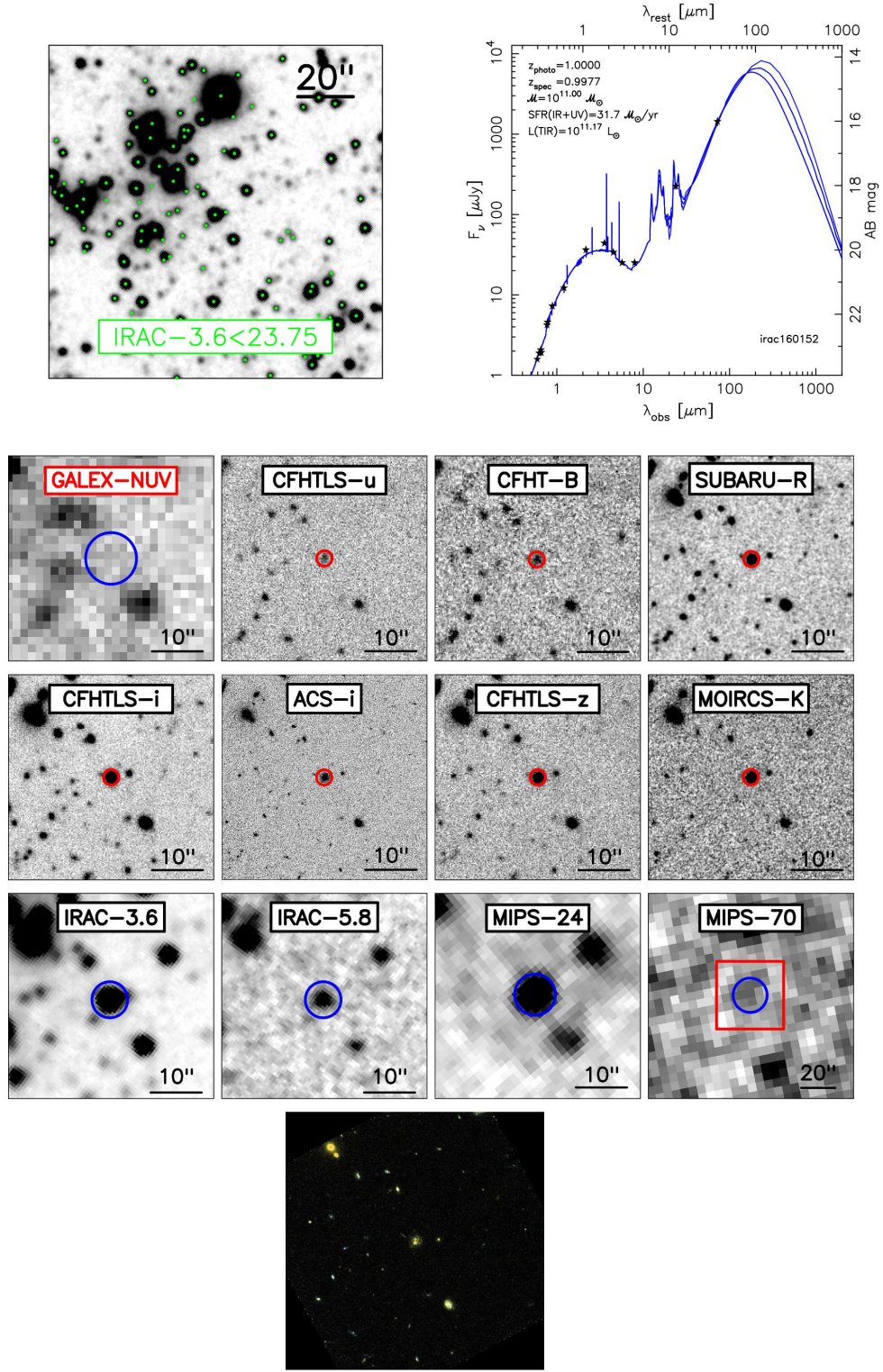}
\caption{\label{postages_navigator2} Same as
  Figure~\ref{postages_navigator1} for the galaxy irac160152 at
  z=0.99. The band label shown in red in the GALEX-NUV postage stamps
  indicates a non-detection. For each of these bands
  $\sigma_{\mathrm{sky}}$ is depicted in the UV-to-FIR SED as red
  arrow. Green name labels indicate forced detections (see
  \S~\ref{rainbow_photometry}). These bands are shown as green stars
  in the SED plot.}
\end{figure*}

\begin{figure*}
\centering
\includegraphics[width=16cm,angle=0.]{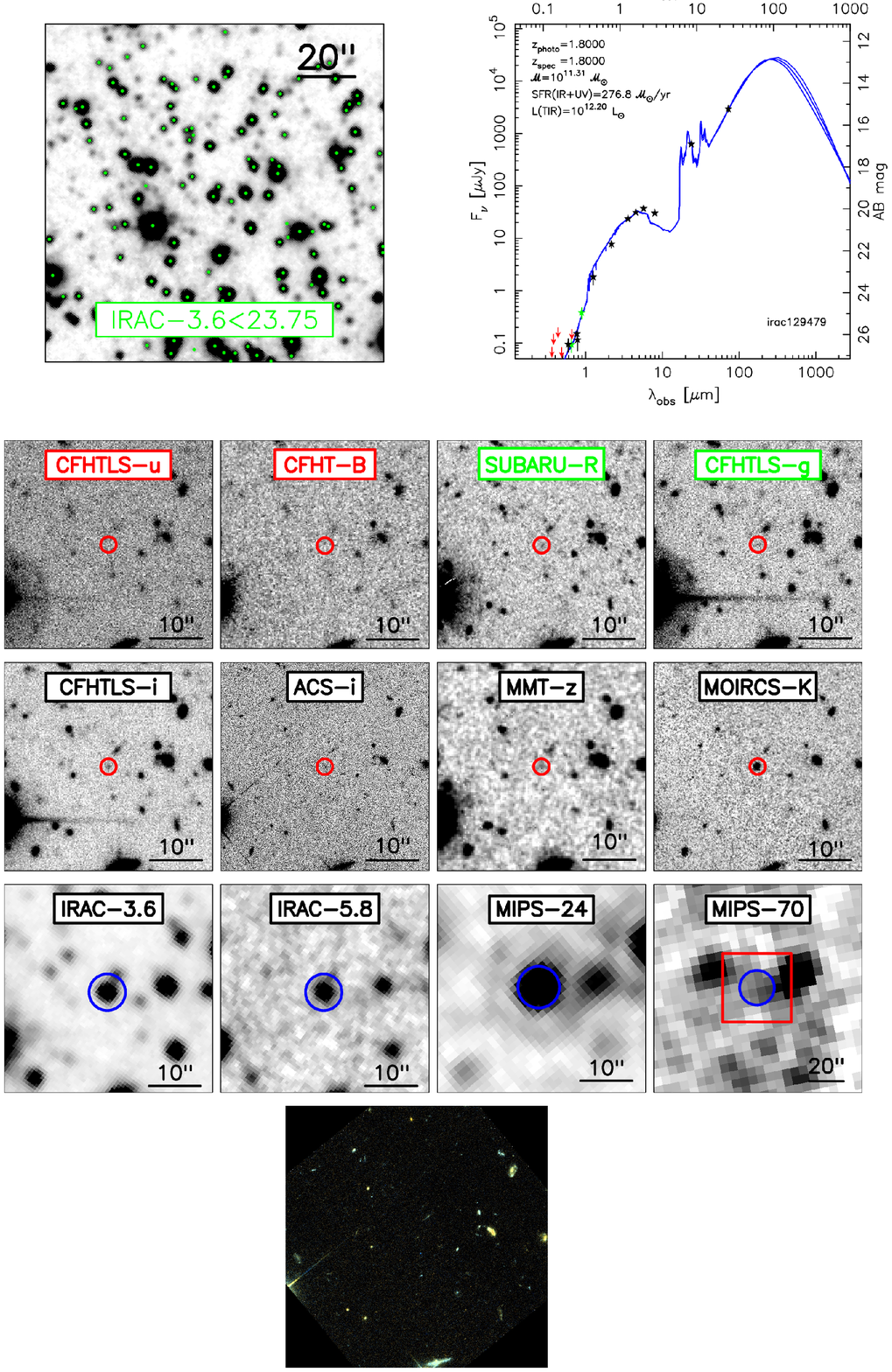}
\caption{\label{postages_navigator3} Same as
  Figure~\ref{postages_navigator1} for irac129479, an IR-bright
  galaxy at z$=$1.8.  This galaxy correspond to source EGS11 in the
  paper of \citet{2009ApJ...700..183H}.  Band labels in red in the
  grey-scale postage stamps indicate non-detections. For each of these
  bands $\sigma_{\mathrm{sky}}$ is depicted in the UV-to-FIR SED as
  red arrow. Band labels in green indicate forced photometric
  measurements (see \S~\ref{rainbow_photometry}). These bands are
  shown as green stars in the SED plot.}
\end{figure*}

\begin{figure*}
\centering
\includegraphics[width=16cm,angle=0.]{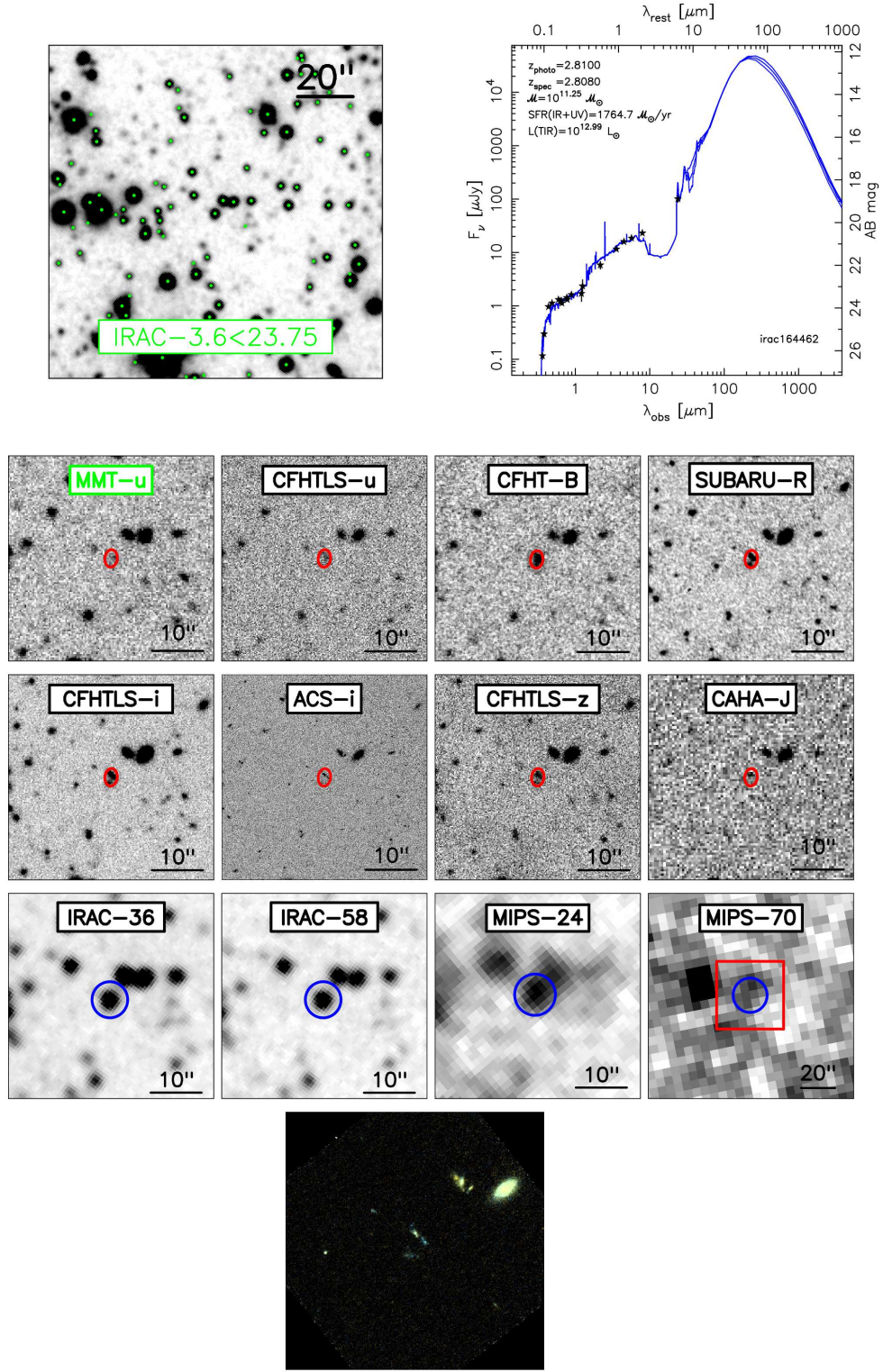}
\caption{\label{postages_navigator4} Same as
  Figure~\ref{postages_navigator3} for irac164462, an LBG at
  z$=$2.8. Band labels in red in the grey-scale postage stamps
  indicate non-detections. For each of these bands
  $\sigma_{\mathrm{sky}}$ is depicted in the UV-to-FIR SED as red
  arrow. Band labels in green indicate forced photometric measurements
  (see \S~\ref{rainbow_photometry}). These bands are shown as green
  stars in the SED plot.}
\end{figure*}

Figures~\ref{postages_navigator1}-\ref{postages_navigator4} show
examples of the multi-band postage stamps, the UV-to-FIR SED and
fitting templates, and the clickable map utility for a few galaxies at
different redshifts.  The postage stamps also illustrate the different
photometric measurements in each band (aperture matched and circular
apertures) and the forced detections for very faint sources (see,
e.g., the source in Figure~\ref{postages_navigator3}).

{\it Rainbow Navigator} has been conceived to serve as a permanent
repository for future versions of the catalogs containing improvements
over the previous results (the present version is data release
  1), and also to similar data products in other cosmological fields
(such as GOODS-N and GOODS-S, presented in PG08).  Currently, it
provides public access to the IRAC-selected catalog in the EGS
presented in this paper, and also a similar release of the data
described in PG08 for a small piece of the central region of the
GOODS-S region.

\section{Summary}

We presented an IRAC-3.6+4.5$\mu$m selected catalog in the EGS
characterized with multi-wavelength photometry. The sample
  contains 76,936 sources with [3.6]$<$23.75 (85$\%$ completeness of
  the sample) covering an area of 0.48~deg$^{2}$. The IRAC sources
are characterized with FUV
NUV~$u^{*}g'r'i'z'$~$u'gRiz$~BRI~$V_{606}$$i_{814}$~$J_{110}H_{160}$~$JK$~[3.6]-[8.0]
photometry.  In addition, we have cross-correlated the sample with
X-ray data \citep[][AEGIS-X]{2009ApJS..180..102L}, {\it Spitzer}/MIPS
24~$\mu$m and 70~$\mu$m FIR photometry, and VLA-20cm radio data
(AEGIS20; \citealt{2007ApJ...660L..77I}). Secure spectroscopic
redshifts are also included in the catalog for 7,636 sources with
[3.6]$<$23.75 obtained from the DEEP2 Survey and \citet[][LBGs at
  z$\gtrsim$3]{2003ApJ...592..728S}.  The data described in this paper
are publicly available, and will be part of future extended analysis
and projects. The main results of this work are summarized below.

\begin{itemize}

\item The extraction of the IRAC sample presented in this paper was
  limited to region with exposure times $>$4~ks. The average survey
  depth is $t_{exp}\sim$10~ks.  Aperture photometry was performed in
  the 4 IRAC bands simultaneously allowing us to obtain upper limit
  fluxes for undetected sources in the [5.8] and [8.0] bands. We
  removed spurious detections masking areas around bright stars. The
  estimated 85\% completeness level for point sources is
  [3.6,4.5]$\sim$23.75 and [4.5,5.8]$\sim$22.25. The 3$\sigma$
  limiting magnitude estimated from the sky-{\it rms} is
  [3.6,4.5]$\sim$24.75 and [4.5,5.8]$\sim$22.90. We also validated the
  quality of the photometry by comparing our results with the catalog
  of Barmby et al. (2008), finding good agreement in both magnitudes
  ($\lesssim$~0.05~mag) and uncertainties
  (\S~\ref{merged_comparison}). Some small systematics were found in
  this comparison, which can be attributed to the slightly different
  reduction versions and the limiting depth of the source extractions.

\item We described in detail our custom photometric procedure {\it
  Rainbow}, developed to measure photometry in multi-wavelength data
  in a consistent way. The main steps followed by our photometry
  software are: (1) re-calibration of the local
  (5\arcmin$\times$5\arcmin) astrometric solution for each pair of
  images, improving the accuracy of the cross-identification of
  sources to a limit below 0.1\arcsec\ (0.2\arcsec) between
  optical-NIR (IRAC-ground based) images.  (2) Deconvolution of
  blended IRAC sources ($\sim$16\% of the entire catalog) separated by
  more than 1\arcsec\ , using the positions of the optical/NIR
  counterparts as reference. The IRAC photometry of deblended sources
  is accurate to 0.03-0.10~mag, depending on the flux ratio between
  neighbors (\S~\ref{rainbow_crossmatch}). (3) Measurement of
  consistent aperture matched photometry for a wide range of ground$-$
  and space$-$ based observations, with different depths and
  resolutions (\S~\ref{rainbow_photometry}); (4) Obtaining (forced)
  flux measurements and upper limits for faint, undetected (in a
  direct analysis of each image) sources
  (\S~\ref{rainbow_photometry}). (5) Computing robust photometric
  errors that take into account variations in the sky {\it rms} and
  signal correlation (\S~\ref{merged_errors}).

\item The inhomogeneous multi-wavelength coverage of the region
  covered by IRAC and the differences in depth of the various datasets
  justifies the split of the complete sample in two complementary
  regions. The main region (covering 0.35~deg$^{2}$), delimited by the
  footprint of the CFHTLS image, constitutes the region with the
  highest data quality. Here the SEDs present a median coverage of 19
  bands, including high resolution imaging from \HST\ (for $\sim$50\%
  of the region) and deep NIR data from Subaru-MOIRCS (for $\sim$25\%
  of the region).  Nearly $\sim$70\% of the complete sample presented
  in this paper is located in the main region. In the flanking regions
  (0.13~deg$^{2}$), the optical-to-NIR coverage is still robust, but
  the median band coverage is reduced to 11 bands, lacking the high
  resolution HST data and the deepest NIR observations.

\item The overall detection efficiency of counterparts for the IRAC
  sources ([3.6]$<$23.75) in other bands is high: more than 85\% of
  the sources are detected (with SNR$>$5) in the deepest optical data
  ($R$- or $i'$-bands) and 70\% in the deepest (MOIRCS) $K_{s}$-band
  images. Our method to perform (forced) photometric measurements for
  a priory undetected sources allows us to recover 10--20\% additional
  sources in the shallowest images.  Despite the large fraction of
  IRAC sources detected in all other bands, we find that $\sim$2\% of
  the sample is detected in IRAC only, some of them at relatively
  bright magnitudes 21$<$[3.6]$<$22 and with high SNR.
  
\item Around 10\% of the sample counts with reliable spectroscopic
  redshifts. Nearly 20$\%$ and 2$\%$ of the sources are detected in
  the MIPS-24~$\mu$m and MIPS-70~$\mu$m data, respectively. This
  allows a detailed analysis of their IR-based SFRs. In addition, we
  recover 1023 of the X-ray sources in the catalog published by
  \citet[][77\% of their complete catalog]{2009ApJS..180..102L} and
  590 radio sources from the catalog published by \citet[][52\% of
    their entire sample]{2007ApJ...660L..77I}.

\item We presented the publicly available web-interface, {\it Rainbow
  Navigator}, to the database containing all the data products from
  this and the companion paper.  The interface allows the access to
  the data using customizable queries on the photometry and derived
  parameters.  Furthermore, it provides additional capabilities to
  inspect the data, such as the creation of on-the-fly clickable sky
  maps or the cross-match of the IRAC sample presented here to a
  user-provided catalog. We have made a significant effort to develop
  a useful and accessible tool that maximizes the legacy value of our
  catalogs in the EGS, and the future {\it Rainbow} based projects in
  other relevant fields (e.g., PG08). The online version of the
  catalog (IRAC-DR1) contains a deeper, although less complete, sample
  limited to [3.6]$<$25.

\item In Paper II, we make use of the very detailed UV-to-FIR SEDs presented
here to estimate photometric redshifts, stellar masses and SFRs.

\end{itemize}

\section*{Acknowledgments}

We acknowledge support from the Spanish Programa Nacional de
Astronom\'{\i}a y Astrof\'{\i}sica under grant AYA2009-10368. PGP-G
acknowledges support from the Ram\'on y Cajal Program financed by the
Spanish Government and the European Union.  Partially funded by the
Spanish MICINN under the Consolider-Ingenio 2010 Program grant
CSD2006-00070: First Science with the GTC.  Support was also provided
by NASA through Contract no. 1255094 issued by JPL/Caltech. This work
is based in part on observations made with the {\it Spitzer} Space
Telescope, which is operated by the Jet Propulsion Laboratory, Caltech
under NASA contract 1407. Observations reported here were obtained at
the MMT Observatory, a joint facility of the Smithsonian Institution
and the University of Arizona. GALEX is a NASA Small Explorer launched
in 2003 April. We gratefully acknowledge NASA's support for
construction, operation, and scientific analysis of the GALEX
mission. This research has made use of the NASA/IPAC Extragalactic
Database (NED) which is operated by the Jet Propulsion Laboratory,
California Institute of Technology, under contract with the National
Aeronautics and Space Administration. Based in part on data collected
at Subaru Telescope and obtained from the SMOKA, which is operated by
the Astronomy Data Center, National Astronomical Observatory of Japan.

\bibliographystyle{aa}
\bibliography{referencias}

\clearpage

\clearpage
\begin{landscape}
\begin{deluxetable}{llcccccccccccccccccccc}
\vspace{-0.4cm}
\tablecolumns{21}
\setlength{\tabcolsep}{0.05in} 
\tabletypesize{\tiny}
\tablewidth{0pt}
\tablecaption{\label{efficiency}The IRAC-3.6+4.5~$\mu$m sample: Detection efficiency in each band and region.}
\tablehead{  
\colhead{Band}& \colhead{$\lambda_{\mathrm{eff}}$}& & \multicolumn{9}{c}{MAIN REGION (0.35~deg$^{2}$)} & &\multicolumn{9}{c}{FLANKING REGIONS (0.13~deg$^{2}$)} \\
\cline{1-2} \cline{4-12} \cline{14-22}\\
& &  & \colhead{Area} & &\multicolumn{3}{c}{[3.6]$<$23.75 [53,030]} & & \multicolumn{3}{c}{[3.6]$<$24.75 [77,607]} &  & \colhead{Area} & &\multicolumn{3}{c}{[3.6]$<$23.75 [23,906]} & & \multicolumn{3}{c}{[3.6]$<$24.75 [35,416]} \\
\cline{4-4}\cline{6-8}\cline{10-12} \cline{14-14}\cline{16-18}\cline{20-22}\\
& & & &&\colhead{stats} & & \colhead{ALL/SNR$>$5(\%)} & & \colhead{stats} & & \colhead{ALL/SNR$>$5(\%)} & & &&\colhead{stats} & & \colhead{ALL/SNR$>$5(\%)} & & \colhead{stats} & &\colhead{ALL/SNR$>$5(\%)}\\
\cline{6-6}\cline{8-8}\cline{10-10}\cline{12-12}\cline{16-16}\cline{18-18}\cline{20-20}\cline{22-22}\\
\colhead{(1)} & \colhead{(2)} && \colhead{(3)} && \colhead{(4)} && \colhead{(5)} &&  \colhead{(6)} &&  \colhead{(7)} &&  \colhead{(8)} &&  \colhead{(9)}&&  \colhead{(10)} &&  \colhead{(11)} &&  \colhead{(12)}
}
\startdata
X-ray$^{\dagger}$   &0.31,1.24~nm& &0.97& & ... & & 717 (1.3\%)& &  ... & & 718 (1.3\%)& &0.93& & ... & & 307 (1.2\%)&  &  ... & & 307 (1.2\%)\\
FUV	 &0.154& &0.99& & 27.90$_{25.78}^{29.45}$& & 58.8/7.9 & &	28.03$_{26.19}^{29.55}$& & 57.4/6.1	 & &0.15& & 27.71$_{25.94}^{29.27}$& & 59.6/7.7 & &	27.83$_{26.27}^{29.34}$& & 58.3/6.8\\
NUV	 &0.232& &0.99& & 26.28$_{24.47}^{28.16}$& & 74.2/25.8 & &	26.60$_{24.78}^{28.35}$& & 70.8/20.2	 & &0.15& & 26.26$_{24.44}^{28.11}$& & 75.2/26.7  & &	26.57$_{24.81}^{28.33}$& & 72.5/23.7\\
$u'$	 &0.363& &1.00& & 25.53$_{24.29}^{27.01}$& & 88.4/52.8 & &	25.85$_{24.61}^{27.25}$& & 86.1/44.1	 & &1.00& & 25.39$_{24.15}^{26.78}$& & 89.6/64.3 & &	25.69$_{24.45}^{27.03}$& & 88.0/57.8\\
$u^{*}$	 &0.381& &1.00& & 25.30$_{24.13}^{26.68}$& & 88.2/53.9 & &	25.60$_{24.43}^{26.93}$& & 85.2/44.2	 & &...& &...& &...& &	...& &...\\
CFH-B	 &0.440& &1.00& & 25.03$_{23.82}^{26.26}$& & 89.6/55.2 & &	25.33$_{24.16}^{26.56}$& & 87.1/45.6	 & &1.00& & 24.94$_{23.73}^{26.20}$& & 87.3/53.1 & &	25.27$_{24.05}^{26.48}$& & 83.7/43.0\\
g	 &0.481& &1.00& & 25.08$_{23.85}^{26.31}$& & 96.1/85.1 & &	25.44$_{24.23}^{26.64}$& & 95.0/80.3	 & &1.00& & 25.12$_{23.76}^{26.40}$& & 96.6/84.9 & &	25.49$_{24.14}^{26.71}$& & 95.5/80.5\\
$g'$	 &0.486& &1.00& & 24.86$_{23.65}^{26.05}$& & 95.4/83.6 & &	25.20$_{24.01}^{26.34}$& & 94.0/78.2	 & &...& &...& &...& &	...& &...\\
$V_{606}$ &0.592& &0.56& & 24.58$_{23.23}^{25.71}$& & 94.7/83.5 & &	24.92$_{23.66}^{26.00}$& & 93.9/78.7	 & &...& &...& &...& &	...& &...\\
$r'$	 &0.626& &1.00& & 24.50$_{23.08}^{25.56}$& & 97.1/89.3 & &	24.85$_{23.52}^{25.93}$& & 96.0/84.2	 & &...& &...& &...& &	...& &...\\
R	 &0.652& &1.00& & 24.44$_{23.00}^{25.45}$& & 97.0/88.5 & &	24.80$_{23.44}^{25.82}$& & 95.8/82.5	 & &1.00& & 24.36$_{22.81}^{25.44}$& & 97.4/87.7 & &	24.74$_{23.27}^{25.80}$& & 96.5/81.9\\
CFH-R	 &0.660& &1.00& & 24.49$_{22.98}^{25.58}$& & 95.3/71.5 & &	24.85$_{23.45}^{25.96}$& & 93.1/60.2	 & &1.00& & 24.36$_{22.73}^{25.50}$& & 91.6/73.7 & &	24.76$_{23.22}^{25.90}$& & 88.4/63.0\\
$i'$	 &0.769& &1.00& & 24.00$_{22.43}^{25.14}$& & 95.3/85.4 & &	24.41$_{22.92}^{25.52}$& & 94.2/78.3	 & &...& &...& &...& &	...& &...\\
i	 &0.782& &1.00& & 24.09$_{22.52}^{25.15}$& & 97.4/90.3 & &	24.51$_{23.01}^{25.62}$& & 96.1/83.8	 & &1.00& & 24.13$_{22.25}^{25.31}$& & 96.4/81.1 & &	24.54$_{22.80}^{25.87}$& & 94.6/70.2\\
$i_{814}$ &0.807& &0.56& & 24.12$_{22.49}^{25.23}$& & 96.7/80.7 & &	24.52$_{23.00}^{25.68}$& & 94.7/69.3	 & &...& &...& &...& &	...& &...\\
CFH-I	 &0.813& &1.00& & 24.03$_{22.34}^{25.23}$& & 94.1/68.6 & &	24.45$_{22.83}^{25.70}$& & 90.1/54.5	 & &1.00& & 23.97$_{22.07}^{25.25}$& & 89.2/61.1 & &	24.43$_{22.58}^{25.70}$& & 84.1/47.5\\
$z'$	 &0.887& &1.00& & 23.79$_{22.17}^{24.98}$& & 93.3/66.2 & &	24.23$_{22.61}^{25.48}$& & 89.1/51.2	 & &...& &...& &...& &	...& &...\\
z	 &0.907& &1.00& & 23.96$_{22.30}^{25.18}$& & 96.0/77.7 & &	24.44$_{22.75}^{25.66}$& & 92.7/64.9	 & &1.00& & 23.95$_{22.15}^{25.25}$& & 97.5/85.5 & &	24.44$_{22.65}^{25.73}$& & 95.3/75.4\\
$J_{110}$ &1.10& &0.04& & 23.87$_{22.38}^{24.94}$& & 84.6/65.4 & &	24.25$_{22.79}^{25.48}$& & 88.9/55.6 & &...& &...& &...& &	...& &...\\
$\Omega2k$-$J$	 &1.21& &0.40& & 23.29$_{21.81}^{24.50}$& & 93.7/38.5 & &	23.74$_{22.21}^{25.01}$& & 88.4/27.1 & &...& &...& &...& &	...& &...\\
WIRC-$J$	 &1.24& &0.42& & 22.43$_{20.90}^{23.47}$& & 62.8/16.6 & &	22.49$_{20.96}^{23.59}$& & 46.1/11.7 & &0.32& & 22.31$_{20.69}^{23.41}$& & 57.9/20.4 & &	22.35$_{20.73}^{23.46}$& & 41.1/14.1\\
$H_{160}$	 &1.59& &0.04& & 24.71$_{23.19}^{25.84}$& & 92.3/80.8 & &	25.18$_{23.61}^{26.52}$& & 91.7/69.4 & &...& &...& &...& &	...& &...\\
$\Omega'-K_{s}$	 &2.11& &0.46& & 22.43$_{20.89}^{23.73}$& & 75.8/6.7  & &	22.74$_{21.38}^{23.98}$& & 67.0/4.6 & &0.26& & 22.17$_{20.72}^{23.46}$& & 67.3/7.3 & &	22.39$_{21.04}^{23.76}$& & 59.6/5.0\\
$Ks$	 &2.15& &0.26& & 24.64$_{23.03}^{25.77}$& & 95.5/73.8 & &	25.14$_{23.44}^{26.42}$& & 89.8/52.9	 & &...& &...& &...& &	...& &...\\
WIRC-$K$ &2.17& &0.97& & 21.76$_{20.21}^{22.74}$& & 62.6/27.1 & &	21.82$_{20.26}^{22.81}$& & 42.6/16.3	 & &0.80& & 21.42$_{19.86}^{22.50}$& & 45.3/18.4 & &	21.45$_{19.89}^{22.55}$& & 42.4/14.4\\
IRAC-36	 &3.56& &1.00& & 22.54$_{20.97}^{23.41}$& & 100.0/99.2 & &	23.18$_{21.43}^{24.25}$& & 100.0/84.5	 & &1.00& & 22.52$_{20.87}^{23.40}$& & 100.0/97.8 & &	23.17$_{21.39}^{24.24}$& & 100.0/80.6\\
IRAC-45	 &4.51& &1.00& & 22.62$_{21.05}^{23.55}$& & 99.8/95.3 & &	23.26$_{21.51}^{24.44}$& & 98.9/75.3	 & &1.00& & 22.63$_{21.00}^{23.61}$& & 99.3/91.1 & &	23.27$_{21.44}^{24.49}$& & 97.1/70.5\\
IRAC-58	 &5.69& &1.00& & 22.50$_{21.04}^{23.73}$& & 89.7/41.1 & &	22.89$_{21.33}^{24.28}$& & 78.4/28.0	 & &1.00& & 22.45$_{20.95}^{23.71}$& & 87.1/38.6 & &	22.80$_{21.22}^{24.20}$& & 75.5/26.4\\
IRAC-80	 &7.96& &1.00& & 22.62$_{21.14}^{23.89}$& & 85.8/34.4 & &	22.97$_{21.37}^{24.33}$& & 73.8/23.5	 & &1.00& & 22.51$_{21.05}^{23.83}$& & 82.6/32.9 & &	22.83$_{21.29}^{24.26}$& & 71.1/22.6\\
MIPS-24	 &23.84& &0.91& & 19.58$_{18.44}^{20.59}$& & 31.5/20.2 & &	19.71$_{18.50}^{20.72}$& & 29.7/18.1	 & &0.93& & 19.36$_{18.35}^{20.39}$& & 29.9/19.1 & &	19.49$_{18.42}^{20.53}$& & 25.2/15.1\\
MIPS-70	 &72.49& &0.80& & 15.79$_{15.23}^{16.37}$& & 13.1/1.5 & &	15.81$_{15.25}^{16.35}$& & 7.5/5.7	 & &0.86& & 15.57$_{14.86}^{16.24}$&         & 13.6/1.2   & &	15.57$_{14.87}^{16.24}$& & 6.7/4.7\\
VLA$^{\dagger}$   &2E06& &0.74& & ... & & 380 ($<$1\%)& &  ... & & 380 ($<$1\%)& &0.42& & ... & & 210 ($<$1\%)& &  ... & & 210($<$1\%)\\
\hline
\hline
\\
& &  & & &\multicolumn{3}{c}{[3.6]$<$23.75 [53,030]} & & \multicolumn{3}{c}{[3.6]$<$24.75 [77,607]}&  &  & &\multicolumn{3}{c}{[3.6]$<$23.75 [23,906]} & & \multicolumn{3}{c}{[3.6]$<$24.75 [35,416]} \\
\cline{1-2}\cline{4-4}\cline{6-8}\cline{10-12} \cline{14-14}\cline{16-18}\cline{20-22}\\
Redshifts$^{\dagger}$ & 0.64-0.91 &  & 0.94 &  & 6191 & & 12.0\% & & 6420 & & 8.4\% &  & 0.84 & & 1445 & & 6.0\% & & 1481 & & 4.2\%
\enddata
\tablecomments{
  Detection efficiency for the sources IRAC-3.6+4.5$\mu$m sample in the
  different bands compiled for this work. The analysis is divided in 2 zones:
  the main region, defined as the overlapping area between the IRAC and the CFHTLS mosaics
  (52.16$^{\circ}<\delta<$53.20$^{\circ}$ and $\alpha>$214.04$^{\circ}$), and the flanking regions.\\
 $\dagger$ For these catalogs, we quote only the number of sources detected in IRAC, and the percentage of total IRAC sample that they represent.\\
(1) Band name.\\
(2) Effective wavelength of the bands in microns.\\
(3) Fraction of the main region covered by the observations in each band.\\
(4,6) Median and quartiles of the magnitude distribution in each band for main region sources with [3.6]$<$23.75
and [3.6]$<$24.75, respectively.\\
(5,7) Percentage of IRAC sources in the main region detected in each band at any magnitude (ALL) and with SNR$>$5.
These values are computed in areas fully covered by both the IRAC mosaic and the observations in each band. \\
(8) Same as (3) in the Flanking regions.\\
(9,11) Same as (4,6) in the Flanking regions.\\
(10,12) Same as (5,7) in the Flanking regions.\\}
\end{deluxetable}
\clearpage
\end{landscape}

\clearpage
\placetable{dataphot}
\LongTables
\begin{landscape}
\begin{deluxetable}{cccccccccccccccccccc}
\setlength{\tabcolsep}{0.005in} 
\tabletypesize{\tiny}
\tablewidth{0pt}
\tablecaption{\label{dataphot} The IRAC-3.6+4.5\mic\ sample: Multi-band photometry}
\tablehead{ 
\colhead{Object}& \colhead{$\alpha$}& \colhead{$\delta$}&\colhead{zspec}& \colhead{qflag}&
\colhead{FUV} &
\colhead{NUV} &
\colhead{$u'$} &
\colhead{$u^{*}$} &
\colhead{B} &
\colhead{$g$} &
\colhead{$g'$} &
\colhead{$V_{606}$} &
\colhead{$r'$} &
\colhead{R} &
\colhead{CFH-R} &
\colhead{$i'$} &
\colhead{$i$} &
\colhead{$i_{814}$} &
\colhead{I} \\
&&&&&$\lambda_{\mathrm{eff}}$&$\lambda_{\mathrm{eff}}$&$\lambda_{\mathrm{eff}}$&$\lambda_{\mathrm{eff}}$&$\lambda_{\mathrm{eff}}$&$\lambda_{\mathrm{eff}}$&$\lambda_{\mathrm{eff}}$&$\lambda_{\mathrm{eff}}$&$\lambda_{\mathrm{eff}}$&$\lambda_{\mathrm{eff}}$&$\lambda_{\mathrm{eff}}$&$\lambda_{\mathrm{eff}}$&$\lambda_{\mathrm{eff}}$&$\lambda_{\mathrm{eff}}$&$\lambda_{\mathrm{eff}}$\\
&&&&&(mag)&(mag)&(mag)&(mag)&(mag)&(mag)&(mag)&(mag)&(mag)&(mag)&(mag)&(mag)&(mag)&(mag)&(mag)\\
&&&&&(err)&(err)&(err)&(err)&(err)&(err)&(err)&(err)&(err)&(err)&(err)&(err)&(err)&(err)&(err)\\
\\
&&&&&
\colhead{$z'$} &
\colhead{$z$} &
\colhead{J$_{110}$} &
\colhead{H$_{160}$} &
\colhead{$\Omega2k$-$J$} &
\colhead{WIRC-$J$} &
\colhead{$\Omega'$-$K_{s}$} &
\colhead{WIRC-$K$} &
\colhead{$Ks$} &
\colhead{[3.6]} &
\colhead{[4.5]} &
\colhead{[5.8]} &
\colhead{[8.0]} &
\colhead{[24]} &
\colhead{[70]}\\
&&&&&$\lambda_{\mathrm{eff}}$&$\lambda_{\mathrm{eff}}$&$\lambda_{\mathrm{eff}}$&$\lambda_{\mathrm{eff}}$&$\lambda_{\mathrm{eff}}$&$\lambda_{\mathrm{eff}}$&$\lambda_{\mathrm{eff}}$&$\lambda_{\mathrm{eff}}$&$\lambda_{\mathrm{eff}}$&$\lambda_{\mathrm{eff}}$&$\lambda_{\mathrm{eff}}$&$\lambda_{\mathrm{eff}}$&$\lambda_{\mathrm{eff}}$&$\lambda_{\mathrm{eff}}$&$\lambda_{\mathrm{eff}}$\\
&&&&&(mag)&(mag)&(mag)&(mag)&(mag)&(mag)&(mag)&(mag)&(mag)&(mag)&(mag)&(mag)&(mag)&(mag)&(mag)\\
&&&&&(err)&(err)&(err)&(err)&(err)&(err)&(err)&(err)&(err)&(err)&(err)&(err)&(err)&(err)&(err)\\
\\
(1)&(2)&(3)&(4)&(5)&(6)&(7)&(8)&(9)&(10)&(11)&(12)&(13)&(14)&(15)&(16)&(17)&(18)&(19)&(20)\\
&&&&&(21)&(22)&(23)&(24)&(25)&(26)&(27)&(28)&(29)&(30)&(31)&(32)&(33)&(34)&(35)\\
&&&&&(36)&(37)&(38)&(39)&(40)&(41)&(42)&(43)&(44)&(45)&(46)&(47)&(48)&(49)&(50)\\
} 
\startdata
irac003270\_1 &215.43892696 &53.08455063 &0.00000 &2 &153.9 &231.6 &362.6 &381.1 &439.0 &481.4 &486.3 &592.4 &625.8 &651.8 &660.0 &769.0 &781.5 &807.3 &813.3\\&&&&&26.370 &25.785 &24.068 &24.028 &23.629 &23.959 &23.712 &-99.000 &23.423 &23.404 &23.401 &22.905 &22.862 &-99.000 &22.824\\&&&&&0.000 &-0.127 &0.079 &0.033 &0.051 &0.039 &0.018 &0.000 &0.017 &0.036 &0.040 &0.018 &0.040 &0.000 &0.034\\&&&&&887.1 &907.0 &1103.3 &1593.2 &1209.4 &1235.5 &2114.6 &2145.3 &2161.3 &3561.2 &4509.6 &5689.4 &7957.6 &23844.0 &72493.7\\&&&&&22.467 &22.718 &-99.000 &-99.000 &-99.000 &-99.000 &-99.000 &-99.000 &22.197 &21.593 &21.853 &22.810 &22.378 &19.545 &-99.000\\&&&&&0.028 &0.045 &0.000 &0.000 &0.000 &0.000 &0.000 &0.000 &0.380 &0.051 &0.044 &0.263 &0.189 &0.238 &0.000\\ 
irac003278 &215.42614011 &53.09447161 &0.00000 &0 &153.9 &231.6 &362.6 &381.1 &439.0 &481.4 &486.3 &592.4 &625.8 &651.8 &660.0 &769.0 &781.5 &807.3 &813.3\\&&&&&27.016 &25.598 &23.354 &22.688 &21.342 &20.683 &20.461 &-99.000 &19.216 &19.434 &18.969 &18.041 &17.870 &-99.000 &17.928\\&&&&&-0.633 &-0.295 &0.071 &0.015 &0.041 &0.029 &0.010 &0.000 &0.011 &0.019 &0.030 &0.010 &0.029 &0.000 &0.029\\&&&&&887.1 &907.0 &1103.3 &1593.2 &1209.4 &1235.5 &2114.6 &2145.3 &2161.3 &3561.2 &4509.6 &5689.4 &7957.6 &23844.0 &72493.7\\&&&&&17.583 &17.661 &-99.000 &-99.000 &-99.000 &-99.000 &-99.000 &-99.000 &17.303 &18.011 &18.417 &18.841 &19.341 &-99.000 &-99.000\\&&&&&0.010 &0.029 &0.000 &0.000 &0.000 &0.000 &0.000 &0.000 &0.010 &0.038 &0.036 &0.033 &0.046 &0.000 &0.000\\ 
irac003291\_1 &215.44043562 &53.08128671 &0.85700 &4 &153.9 &231.6 &362.6 &381.1 &439.0 &481.4 &486.3 &592.4 &625.8 &651.8 &660.0 &769.0 &781.5 &807.3 &813.3\\&&&&&29.948 &27.232 &25.312 &25.157 &24.623 &24.680 &24.423 &-99.000 &23.579 &23.457 &23.350 &22.529 &22.463 &-99.000 &22.390\\&&&&&-4.101 &-0.624 &0.196 &0.092 &0.114 &0.055 &0.031 &0.000 &0.021 &0.034 &0.036 &0.014 &0.036 &0.000 &0.032\\&&&&&887.1 &907.0 &1103.3 &1593.2 &1209.4 &1235.5 &2114.6 &2145.3 &2161.3 &3561.2 &4509.6 &5689.4 &7957.6 &23844.0 &72493.7\\&&&&&22.130 &22.295 &-99.000 &-99.000 &-99.000 &-99.000 &-99.000 &-99.000 &20.449 &20.391 &20.721 &21.007 &21.247 &18.860 &-99.000\\&&&&&0.021 &0.041 &0.000 &0.000 &0.000 &0.000 &0.000 &0.000 &0.106 &0.050 &0.054 &0.054 &0.067 &0.118 &0.000\\ 
irac003310 &215.42129738 &53.09430607 &0.00000 &0 &153.9 &231.6 &362.6 &381.1 &439.0 &481.4 &486.3 &592.4 &625.8 &651.8 &660.0 &769.0 &781.5 &807.3 &813.3\\&&&&&22.626 &22.577 &21.787 &21.681 &21.104 &20.817 &20.706 &-99.000 &20.323 &20.344 &20.301 &20.033 &19.985 &-99.000 &19.978\\&&&&&0.026 &-0.017 &0.053 &0.012 &0.041 &0.029 &0.010 &0.000 &0.014 &0.020 &0.030 &0.013 &0.029 &0.000 &0.030\\&&&&&887.1 &907.0 &1103.3 &1593.2 &1209.4 &1235.5 &2114.6 &2145.3 &2161.3 &3561.2 &4509.6 &5689.4 &7957.6 &23844.0 &72493.7\\&&&&&19.983 &20.085 &-99.000 &-99.000 &-99.000 &-99.000 &-99.000 &-99.000 &19.779 &20.339 &20.757 &21.294 &19.695 &19.436 &-99.000\\&&&&&0.013 &0.030 &0.000 &0.000 &0.000 &0.000 &0.000 &0.000 &0.063 &0.039 &0.039 &0.102 &0.045 &0.165 &0.000\\ 
irac003313 &215.43553774 &53.08200958 &0.00000 &0 &153.9 &231.6 &362.6 &381.1 &439.0 &481.4 &486.3 &592.4 &625.8 &651.8 &660.0 &769.0 &781.5 &807.3 &813.3\\&&&&&25.415 &25.293 &24.750 &24.473 &24.155 &24.501 &24.269 &-99.000 &24.029 &23.998 &24.069 &23.517 &23.433 &-99.000 &23.539\\&&&&&0.000 &-0.147 &0.131 &0.067 &0.082 &0.044 &0.024 &0.000 &0.035 &0.056 &0.072 &0.023 &0.048 &0.000 &0.067\\&&&&&887.1 &907.0 &1103.3 &1593.2 &1209.4 &1235.5 &2114.6 &2145.3 &2161.3 &3561.2 &4509.6 &5689.4 &7957.6 &23844.0 &72493.7\\&&&&&23.198 &23.437 &-99.000 &-99.000 &-99.000 &-99.000 &-99.000 &-99.000 &-99.000 &21.783 &21.891 &22.385 &22.479 &-99.000 &16.266\\&&&&&0.043 &0.054 &0.000 &0.000 &0.000 &0.000 &0.000 &0.000 &0.000 &0.048 &0.057 &0.403 &0.383 &0.000 &1.282\\ 
\enddata
\tablecomments{
(1) Object unique identifier in the catalog. The catalog is sorted by this field.\\
(2,3) Right Ascension and Declination (J2000) in degrees.\\
(4) Spectroscopic redshift determination drawn from (\citealt{2007ApJ...660L...1D}; $\sim$8,000 galaxies) and (\citealt{2003ApJ...592..728S}; LBGs at z$\gtrsim$3).\\
(5) Quality flag of the spectroscopic redshift (4=$>$99.5\%, 3=$>$90\%, 2=uncertain, 1=bad quality). Only redshifts with qflag$>$2 have been used in the analysis.\\
(6-50) Effective wavelength of the  FUV
NUV~$u^{*}g'r'i'z'$~$u'gRiz$~BRI~$V_{606}$$i_{814}$~$J_{110}H_{160}$~$JK$~[3.6]-[8.0]~[24]~[70] bands in nanometer. Observed magnitude with the associated uncertainty as measured in
the apertured matched procedure (see \S~\ref{merged_technique}). All magnitudes refer to the AB photometric system.\\
Negative values of the photometric uncertainty are for forced photometric detections. Magnitudes $>$0 with uncertainty equal to 0.00 are non-detections; for these bands the photometry indicates the magnitude of $\sigma_{\mathrm{sky}}$ (see \S~\ref{rainbow_photometry}).
(This table is available in its entirety in a machine-readable in the online version. A portion is shown here for guidance.)}
\end{deluxetable}
\clearpage
\end{landscape}

\begin{deluxetable}{cccccccc}
\vspace{6cm}
\setlength{\tabcolsep}{0.001in} 
\tablewidth{0pt}
\tablecaption{\label{photprop} The IRAC-3.6+4.5~\mic\ sample: Photometric properties}
\tablehead{ 
\colhead{Object}& \colhead{$\alpha$}& \colhead{$\delta$}& 
 \colhead{N(bands,detect)} &
 \colhead{N(bands,forced)} &
 \colhead{Flag} &
 \colhead{Stellarity} &
 \colhead{Region} \\
(1)&(2)&(3)&(4)&(5)&(6)&(7)&(8)}
\startdata

irac003270\_1 &215.43910540 &53.08468920 &18 &1 &0 &0 &1\\ 
irac003278 &215.42614011 &53.09447161 &18 &2 &0 &7 &1\\ 
irac003291\_1 &215.44058360 &53.08123980 &18 &2 &0 &0 &1\\ 
irac003310 &215.42129738 &53.09430607 &19 &1 &1 &0 &1\\ 
irac003313 &215.43553774 &53.08200958 &17 &1 &0 &1 &1 
\enddata
\tablecomments{
(1) Object unique identifier in the catalog. The catalog is sorted by this field.\\
(2,3) Right Ascension and Declination (J2000) in degrees.\\
(4) Number of optical-to-NIR bands (with effective wavelengths below 8.0$\mu$m) in which the object is detected.\\
(5) Number of optical-to-NIR bands (with effective wavelengths below 8.0$\mu$m) in which the object is a priori un-detected, but the forced photometry (see \S~\ref{rainbow_photometry}) recovers a valid flux.\\
(6) Quality flag indicating the proximity of a very bright source in the vecinity of the source. Sources detected in very few bands (N(band)$<$5) and located nearby a bright source are likely to be spurious detections (see \S~\ref{merged_completeness}).\\
(7) Sum of all the stellarity criteria satisfied (see \S~\ref{STAR}). A source is classified as star for Stellarity$>$2.\\
(8) Region of the field in which the source is located: 1 for the Main region, 0 for the Flanking regions.\\ 
(This table is available in its entirety in a machine-readable in the online version. A portion is shown here for guidance.)}
\end{deluxetable}
\clearpage

\label{lastpage}
\end{document}